\documentclass[prd,%
preprint,%
tightenlines,showpacs,%
nofootinbib,eqsecnum,%
amsfonts,amsmath,%
]{revtex4}
\usepackage{graphicx}

\begin{document}

\title{Field-theoretical formulations of MOND-like gravity}

\author{Jean-Philippe Bruneton}

\author{Gilles \surname{Esposito-Far\`ese}}

\affiliation{${\mathcal{G}}{\mathbb{R}}
\varepsilon{\mathbb{C}}{\mathcal{O}}$, Institut d'Astrophysique
de Paris, UMR 7095-CNRS, Universit\'e Pierre et Marie
Curie-Paris6, 98bis boulevard Arago, F-75014 Paris, France}

\begin{abstract}
Modified Newtonian dynamics (MOND) is a possible way to
explain the flat galaxy rotation curves without invoking the
existence of dark matter. It is however quite difficult to
predict such a phenomenology in a consistent field theory,
free of instabilities and admitting a well-posed Cauchy
problem. We examine critically various proposals of the
literature, and underline their successes and failures both
from the experimental and the field-theoretical viewpoints.
We exhibit new difficulties in both cases, and point out the
hidden fine tuning of some models. On the other hand, we
show that several published no-go theorems are based on
hypotheses which may be unnecessary, so that the space of
possible models is \textit{a priori} larger. We examine a
new route to reproduce the MOND physics, in which the field
equations are particularly simple outside matter. However,
the analysis of the field equations \textit{within} matter
(a crucial point which is often forgotten in the literature)
exhibits a deadly problem, namely that they do not remain
always hyperbolic. Incidentally, we prove that the same
theoretical framework provides a stable and well-posed model
able to reproduce the Pioneer anomaly without spoiling any
of the precision tests of general relativity. Our conclusion
is that all MOND-like models proposed in the literature,
including the new ones examined in this paper, present
serious difficulties: Not only they are unnaturally fine
tuned, but they also fail to reproduce some experimental
facts or are unstable or inconsistent as field theories.
However, some frameworks, notably the tensor-vector-scalar
(TeVeS) one of Bekenstein and Sanders, seem more promising
than others, and our discussion underlines in which
directions one should try to improve them.
\end{abstract}

\date{May 25, 2007}

\maketitle
\section{Introduction}\label{Intro}

Although general relativity (GR) passes all precision tests with
flying colors \cite{Willbook, Willreview}, there remain some puzzling
experimental issues, notably the fact that the Universe seems to be
filled with about 72\% of dark energy and 24\% of dark matter, only
4\% of its energy content being made of ordinary baryonic matter
\cite{Spergel:2003cb,Spergel:2006hy}.
Dark energy, a fluid whose negative pressure is close to
the opposite of its energy density, would be responsible for the
present accelerated expansion of the Universe
\cite{Tonry:2003zg,Knop:2003iy,Riess:2004nr,Astier:2005qq}.
Dark matter, a pressureless and non\-interacting component of matter
detected only by its gravitational influence, is suggested by
several experimental data, notably the flat rotation curves of
clusters and galaxies~\cite{Rubin:1978}. Another experimental
issue is the anomalous extra acceleration $\delta a \approx
8.5\times 10^{-10}\, \text{m}.\text{s}^{-2}$ towards the Sun
that the two Pioneer spacecrafts exhibited between 30 and 70
AU \cite{Anderson:2001sg} (see also \cite{Lammerzahl:2006ex}).

Actually, none of these issues contradicts GR nor Newtonian
gravity directly. Indeed, dark energy may be understood as the
existence of a tiny cosmological constant $\Lambda \approx
3\times 10^{-122}\, c^3/(\hbar G)$, or as a scalar field (called
quintessence) slowly rolling down a potential
\cite{Ratra:1987rm,Caldwell:1997ii,Zlatev:1998tr}. Many
candidates of dark matter particles have also been predicted by
different theoretical models (notably the class of neutralinos,
either light or more massive, occurring in supersymmetric
theories; see e.g. \cite{Boehm:2003ha}), and numerical
simulations of structure formation have obtained great successes
while incorporating such a dark matter (see, e.g.,
\cite{Hayashi:2004}). [It remains however to explain why
$\Lambda$ is so small but nonzero, and why the dark energy and
dark matter densities happen to be today of the same order of
magnitude.] The Pioneer anomaly seems more problematic, but the
two spacecrafts are identical and were not built to test gravity,
therefore one must keep cautious in interpreting their data. A
dedicated mission would be necessary to confirm the existence of
such an anomalous acceleration.

Nevertheless, these various issues, considered simultaneously,
give us a hint that Newton's law might need to be modified at
large distances, instead of invoking the existence of several
dark fluids. To avoid the dark matter hypothesis,
Milgrom~\cite{Milgrom:1983ca} proposed in 1983 such a
phenomenological modification, which superbly accounts for galaxy
rotation curves~\cite{Sanders:2002pf} (although galaxy clusters
anyway require some amount of dark matter), and automatically
recovers the Tully-Fisher law~\cite{Tully:1977fu} $v_\infty^4
\propto M$ (where $M$ denotes the baryonic mass of a galaxy, and
$v_\infty$ the asymptotic circular velocity of visible matter in
its outer region). The norm $a$ of a particle's acceleration is
assumed to be given by its Newtonian value $a_N$ when it is
greater than a universal constant $a_0 \approx 1.2\times
10^{-10}\, \text{m}.\text{s}^{-2}$, but to read $a = \sqrt{a_N
a_0}$ in the small-acceleration regime $a < a_0$. In particular,
the gravitational acceleration should now read $a = \sqrt{GM
a_0}/r$ at large distances, instead of the usual $GM/r^2$ law.

Various attempts have been made to derive such a modified
Newtonian dynamics (MOND) from a consistent relativistic field
theory. The main aim of the present paper is to examine them
critically, by underlining their generic difficulties and how
some models managed to solve them. Three classes of difficulties
may actually be distinguished: (i)~theoretical ones, namely
whether the proposed model derives from an action principle, is
stable and admits a well-posed Cauchy problem; (ii)~experimental
ones, in particular whether solar-system and binary-pulsar tests
are passed, and whether the predicted light deflection by galaxies
or haloes is consistent with weak-lensing observations; and
(iii)~``esthetical'' ones, i.e., whether the proposed model is
natural enough to be considered as predictive, or so fine-tuned
that it almost becomes a fit of experimental data. These
different classes of difficulties are anyway related, because
fine-tuning is often necessary to avoid an experimental problem,
and because some tunings sometimes hide serious theoretical
inconsistencies.

We will not address below the problem of dark energy, although it
may also be tackled with a modified-gravity viewpoint. For
instance the DGP brane model \cite{Dvali:2000hr,Dvali:2000xg}
predicts an accelerated expansion of the Universe without invoking a
cosmological constant \cite{Deffayet:2000uy,Deffayet:2001pu}, and the
k-essence models
\cite{Armendariz-Picon:1999rj,Chiba:1999ka,Armendariz-Picon:2000ah}
also reproduce many cosmological features in a somewhat more natural
way than quintessence theories. Some of our theoretical discussions
below, notably about stability and causality, are nevertheless
directly relevant to such models. Indeed, k-essence theories are
characterized by a non-linear function of a scalar field's kinetic
term, precisely in the same way relativistic aquadraric Lagrangians
(RAQUAL) were devised to reproduce the MOND phenomenology in a
consistent field theory
\cite{Bekenstein:1984tv}. Our aim is not to discuss the Pioneer
anomaly in depth either. Although the numerical value of these
spacecrafts' extra acceleration $\delta a$ is of the same order
of magnitude as the MOND parameter $a_0$, there indeed exist
important differences between their behavior and the MOND
dynamics. However, we will come back to the Pioneer anomaly at
the very end of the present paper. While analyzing a new class of
models \textit{a priori} devised to reproduce the MOND
phenomenology, we will show that it can account for the Pioneer
anomaly without spoiling any of the precision tests of general
relativity, while being stable and admitting a well-posed Cauchy
problem.

Our paper is organized as follows. In Sec.~\ref{Field}, we
examine the various ways one may try to modify the laws of
gravity in a consistent relativistic field theory, and give a
critical review of several models proposed in the literature. We
notably underline that an action should not depend on the mass of
a galaxy, otherwise one is defining a different theory for each
galaxy. We also recall why higher-order gravity is generically
unstable, contrary to scalar-tensor theories. When the dynamics
of the scalar field is defined by an aquadratic kinetic term
(RAQUAL or k-essence theories), we discuss the different
consistency conditions it must satisfy. We point out that such
conditions suffice for local causality to be satisfied, although
some modes may propagate faster than light or gravitational
waves. We finally recall how Ref.~\cite{Bekenstein:1984tv}
reproduced the MOND gravitational force thanks to a RAQUAL model,
but we exhibit a serious fine-tuning problem which does not seem
to have been discussed before --- or at least was not pointed out
so clearly.

In Sec.~\ref{Light}, we recall that many models fail to reproduce
the observed light deflection by ``dark matter'' haloes. We
exhibit a counter-example to an erroneous claim in the literature,
although this counter-example does not reproduce the correct MOND
phenomenology. We also recall how a ``disformal'' (i.e.,
non-conformal) coupling of the scalar field to matter allowed
Refs.~\cite{Bekenstein:1992pj,Bekenstein:1993fs} to predict the
right light deflection, but that this class of models has been
discarded too quickly, because of the existence of superluminal
gravitons. Actually, the consistency and causality of such models
is clear when analyzed in the Einstein frame. On the other hand,
we show that the consistency of the field equations
\textit{within} matter implies new conditions which must be
satisfied by the functions defining the theory. As far as we are
aware, these crucial conditions were not derived previously in
the literature. We finally discuss how the best present model,
called
TeVeS~\cite{Bekenstein:2004ne,Bekenstein:2004ca,Sanders:1996wk,
Sanders:2005vd}, solved the problem of light deflection by
considering a stratified theory, involving a unit vector field in
addition to a metric tensor and one or several scalar fields.

In Sec.~\ref{Section4}, we discuss the various difficulties which
anyway remain in this TeVeS model, including some which have
already been discussed in the literature and that we merely
summarize. Reference \cite{Clayton:2001vy} proved that the TeVeS
Hamiltonian is not bounded by below, and therefore that the model
is unstable. We also point out other instabilities in related models.
Many papers underlined that stratified theories define a preferred
frame, and that they are \textit{a priori} inconsistent with
solar-system tests of local Lorentz invariance of gravity. However,
the vector field is assumed to be dynamical in TeVeS, and it has been
argued in \cite{Bekenstein:2004ne,Bekenstein:2004ca} that such tests
are now passed. Although we do not perform a full analysis ourselves,
our discussion suggests that preferred-frame effects are actually
expected in TeVeS, and that they may be avoided only at the
expense of an unnatural fine-tuning. The reason why disformal
models were discarded in favor of the stratified TeVeS theory was
mainly to avoid superluminal gravitons. However, the opposite
phenomenon occurs in TeVeS: photons (and high-energy matter
particles) propagate faster than gravitons. References
\cite{Moore:2001bv,Elliott:2005va} actually proved that the
observation of high-energy cosmic rays imposes tight constraints
on such a behavior, because these rays should have lost their
energy by Cherenkov radiation of gravitational waves. After
recalling this argument, we conclude that the sign of one of the
TeVeS parameters should be flipped, implying that gravitons now
propagate faster than light, like in the RAQUAL models. We
finally discuss briefly binary-pulsar constraints, still without
performing a full analysis, but underlining that a large amount
of energy should be emitted as dipolar waves of the scalar field.
The matter-scalar coupling constant should thus be small enough
in TeVeS to pass binary-pulsar tests, and this implies an
unnatural fine-tuning of its Lagrangian.

Section~\ref{Nonmin} is devoted to new models whose point of view
differs significantly from those of the literature. They are
extremely simple in vacuum, reducing respectively to general
relativity and to Brans-Dicke theory. The MOND phenomenology is
obtained via a non-minimal coupling of matter to the
gravitational field(s). However, the first model exhibits a
subtle instability, similar to the one occurring in higher-order
gravity (although no tachyon nor ghost degree of freedom may be
identified in vacuum). The second model avoids this instability
due to higher derivatives, and reproduces the MOND phenomenology
(including light deflection) quite simply, as compared to the
literature. However, our analysis of the field equations shows
that they do not remain hyperbolic within the dilute gas in outer
regions of a galaxy. Therefore, although promising, this model is
inconsistent. But the same framework provides a simple model to
reproduce consistently the Pioneer anomaly, without spoiling the
precision tests of general relativity.

Finally, we present our conclusions in Sec.~\ref{Concl}. We notably
mention other experimental constraints that any complete MOND-like
theory of gravity should also satisfy, besides those which are
discussed in the present paper.

\section{Looking for MOND-like field theories}\label{Field}

\subsection{General relativity}\label{GR}

Einstein's general relativity is based on two independent
hypotheses, which are most conveniently described by
decomposing its action as $S = S_\text{gravity} +
S_\text{matter}$. First, it assumes that all matter
fields, including gauge bosons, are minimally coupled
to a single metric tensor, that we will denote
$\tilde g_{\mu\nu}$ throughout the present paper.
This metric defines the lengths and times measured
by laboratory rods and clocks (made of matter),
and is thereby called
the ``physical metric'' (the name ``Jordan metric''
is also often used in the literature). The matter
action may thus be written as
$S_\text{matter}[\psi;\tilde g_{\mu\nu}]$, where
$\psi$ denotes globally all matter fields, and
where the angular brackets indicate a functional
dependence. This so-called ``metric coupling'' implies
the weak equivalence principle, i.e., the fact that
laboratory-size objects fall with the same acceleration
in an external gravitational field, which is experimentally
verified to a few parts in
$10^{13}$~\cite{Baessler:1999iv,Williams:2004qb}.
For instance, the action of a point-particle reads
$S_\text{pp} = -\int mc \sqrt{-\tilde g_{\mu\nu}(x) v^\mu
v^\nu}\, dt$, and depends both on the spacetime position $x^\mu$ of
the particle and its velocity $v^\mu \equiv dx^\mu/dt$.
The second building block of GR is the Einstein-Hilbert action
\begin{equation}
S_\text{gravity} = \frac{c^4}{16 \pi G}
\int\frac{d^4 x}{c}\sqrt{-g_*}\, R^*,
\label{EH}
\end{equation}
which defines the dynamics of a spin-2 field $g^*_{\mu\nu}$,
called the ``Einstein metric''.
We use the sign conventions of~\cite{MTW}, notably the
mostly-plus signature, and always indicate with a tilde
or a star (either upper or lower) which metric is used
to construct the corresponding quantity. For instance,
$g_* \equiv \det(g^*_{\mu\nu})$ is the determinant of
the Einstein metric and $R^*$ its scalar curvature.
Similarly, we will denote as $\tilde\nabla_\mu$
and $\nabla^*_\mu$ the covariant derivatives corresponding
respectively to the Jordan and Einstein metrics, and as
$\tilde g^{\mu\nu}$ and $g_*^{\mu\nu}$ the inverses of
these two metrics. This
rather heavy notation will allow us to be always sure
of what we are talking about in the following. Einstein's
second hypothesis is that both metrics coincide:
$\tilde g_{\mu\nu} = g^*_{\mu\nu}$.

Milgrom analyzed in~\cite{Milgrom:1992hr,Milgrom:1998sy} the
consequences of modifying the matter action, that he called
``modified inertia''. He focused on a point particle in an external
gravitational field, and assumed that its action could depend also
on the acceleration ${\bf a} \equiv d{\bf v}/dt$ and its higher
time-derivatives. However, he proved that to obtain both the
Newtonian and the MOND limits and satisfy Galileo invariance, the
action must depend on all time-derivatives $d^n{\bf v}/dt^n$ to any
order, i.e., that the action is necessarily nonlocal. This does not
necessarily violate causality (see the counter-example
in~\cite{Soussa:2003vv}), and is actually a good feature for the
stability of the theory (see Sec.~\ref{R2} below), but the actual
computation of the predictions is quite involved. We refer the
reader to the detailed paper~\cite{Milgrom:1992hr} for more
information about this interesting viewpoint, but we focus below on
metric field theories, i.e., such that matter is minimally coupled
to $\tilde g_{\mu\nu}$ as in GR, with only first derivatives of the
matter fields entering the action.

Another way to modify GR is to assume that this physical
metric does not propagate as a pure spin-2 field, i.e.,
that its dynamics is no longer described by the
Einstein-Hilbert action (\ref{EH}), and that $\tilde g_{\mu\nu}$
is actually a combination of various fields. Since the
weak equivalence principle is very well tested experimentally,
and implied by a metric coupling, most MOND-like models in the
literature focused on such a ``modified gravity'' viewpoint, and we
will examine them in the present paper. The fact that they involve
extra fields, besides the usual graviton (fluctuation of
$g^*_{\mu\nu}$) and the various matter fields entering the Standard
Model of particle physics, makes the distinction with dark matter
models rather subtle. The crucial difference is that, in the
dark-matter paradigm, the amount of dark matter is imposed by initial
conditions, and its clustering generates gravitational wells in which
baryonic matter fall to form galaxies and large-scale
structures. On the other hand, in the modified-gravity
viewpoint, baryonic matter generates itself an effective
dark-matter halo $M_\text{dark} \propto \sqrt{M_\text{baryon}}$.
Such a halo may just be an artifact of the way we interpret the
gravitational field of baryonic matter alone at large distances.
But it may also be a real dark-matter halo,
made of the extra gravitational fields, and generating itself
a Newtonian potential. In such a case, the difference with
standard dark-matter models would be that its mass
$M_\text{dark}$ is imposed as above by the baryonic one,
and modified gravity could thus be considered as a constrained
class of dark-matter models. Most of the theories that we will
mention in the following predict that the energy density of the
extra gravitational degrees of freedom is negligible with respect
to baryonic matter: $|T_{\mu\nu}^\text{extra fields}| \ll
|T_{\mu\nu}^\text{matter}|$.
Therefore, the effective dark-matter haloes will be most of the
time artifacts of our interpretation of observations. However, we
will also consider in Sec.~\ref{LightRAQUAL} a model of the
``constrained dark matter'' type ($|T_{\mu\nu}^\text{extra
fields}| \gg |T_{\mu\nu}^\text{matter}|$), as a counter-example
to some claims in the literature about light
deflection.\footnote{Let us also mention that the recent
reinterpretation of MOND proposed in
Refs.~\cite{Blanchet:2006yt,Blanchet:2006sc} is also of the
constrained dark matter type: The MOND force $\propto 1/r$ is
caused by a background of gravitational dipoles contributing
directly to the Newtonian potential.}

As suggested by Milgrom, one may also consider models modifying
\textit{both} inertia and gravity. Paradoxically, as we will see
in Sec.~\ref{MatterSpin2} below, there also exists a non-trivial
(non-GR) possibility in which \textit{neither} of them is
modified in the above sense. Matter will be universally coupled
to a second-rank symmetric tensor $\tilde g_{\mu\nu}$, i.e.,
still described by an action $S_\text{matter}[\psi; \tilde
g_{\mu\nu}]$, and the dynamics of gravity will be described by
the pure Einstein-Hilbert action (\ref{EH}) in vacuum, but the
two metrics $\tilde g_{\mu\nu}$ and $g^*_{\mu\nu}$ will
nevertheless differ within matter.

\subsection{Various ideas in the literature}\label{Lit}

Although interesting from a phenomenological point of view, some
models proposed in the literature write field equations which do
not derive from an action, and cannot be obtained within a
consistent field theory. For instance, actually not to reproduce a
MOND-like behavior but anyway as a model of dark matter,
Ref.~\cite{Piazza:2003ri} proposes to couple differently dark matter
and baryonic matter to gravity (an idea explored in the cosmological
context by various authors, notably
\cite{Damour:1990tw,Damour:1990eh,Casas:1991ky,Anderson:1997un,
Amendola:1999er,Comelli:2003cv,Farrar:2003uw,Amendola:2006dg}).
The authors underline themselves that it cannot be considered as
a fundamental theory (and they actually need some negative energy
to obtain a repulsive force), but it is also instructive to
stress that it cannot derive from an action. Indeed, the scalar
field entering this model is assumed not to be generated by
baryons, which implies that the scalar-baryon vertex vanishes. On
the other hand, the assumed equation of motion for this baryonic
matter does depend on the scalar field, thereby implying that the
scalar-baryon vertex does not vanish. In conclusion, although the
behavior of such a model might be mimicked as an effective theory
of a more fundamental one, it cannot be described itself as a
field theory, and notably does not satisfy conservation laws. In
the present paper, our discussion will be restricted to
consistent field theories.

Some models in the literature do write actions, but they depend
on the (baryonic or luminous) mass $M$ of the galaxy
\cite{Cadoni:2003nz,Sobouti:2006rd}. For instance
Ref.~\cite{Sobouti:2006rd} proposes a gravitational action in
which the kinetic term is some power of the Ricci scalar
$R^{1-n/2}$. Since this implies, in general, an asymptotic
velocity of the form $v^2 \propto n c^2 + \mathcal{O}(GM/r)$,
Ref.~\cite{Sobouti:2006rd} then proposes to replace $n$ by
$\sqrt{G M a_0 / c^4}$ in order to recover the Tully-Fisher law.
However, without an additional prescription (which must be
non-local in nature) specifying which mass $M$ enters the action, this
theory is ill defined. We will come back to such $f(R)$ theories of
gravity in Sec.~\ref{R2} and in Sec.~\ref{ST}. It should be also
underlined that a gravitational force $\propto 1/r$ is not difficult
to obtain by choosing an appropriate scalar field potential, for
instance, as we will illustrate in Sec.~\ref{ST} below. However, the
second crucial feature of MOND, which arises from the Tully-Fisher law,
is \textit{also} the factor $\sqrt{G M a_0} \propto \sqrt{M}$
multiplying this $1/r$. Some consistent field theories actually
obtain a flattening of rotation curves but do not predict the
right amplitude of the asymptotic velocity
\cite{Capozziello:2006ph,Kao:2005ji}.

In \cite{Einstein,EinsteinStraus}, Einstein and Straus studied an
extension of general relativity in which the metric tensor is
nonsymmetric, $g_{\mu\nu}\neq g_{\nu\mu}$. In modern language,
this corresponds to considering an antisymmetric tensor field
$B_{[\mu\nu]}$ as a partner to the usual graviton $g_{(\mu\nu)}$.
Moffat analyzed the phenomenology of this class of models in many
papers (starting with \cite{Moffat:1978tr}), and it was shown to
define a consistent and stable field theory provided
$B_{[\mu\nu]}$ is massive \cite{Damour:1992bt}. More recently,
Moffat showed that it may reproduce the flat rotation curves of
clusters and galaxies
\cite{Moffat:2004nw,Moffat:2004ba,Moffat:2004bm,Moffat:2005si,
Brownstein:2005zz,Brownstein:2005dr}, probably even better that
the original MOND proposal. However, several assumptions are
needed to obtain such a prediction, and it cannot yet be
considered as a predictive field theory, in our opinion. Indeed,
the author points out himself that a constant entering his action
must take three different values at the solar system, galaxy, and
cluster scales. It must therefore be considered as a running
(renormalized) coupling constant instead of a pure number imposed
in the action, and a complete theory should be able to predict
the three needed values. A related difficulty is the fact that
the force generated by the antisymmetric tensor field
$B_{[\mu\nu]}$ is repulsive, whereas galaxy rotation curves need
a gravitational attraction greater that the Newtonian one. Here
again, the author invokes renormalization of the constants to
obtain the right behavior, but does not provide a full
derivation. But the most instructive difficulty of
Ref.~\cite{Moffat:2004ba} is that the extra force is predicted to
be $\propto k M^2/r$ instead of the needed $\propto \sqrt{M}/r$
behavior. The author first assumed in \cite{Moffat:2004ba} that
an unknown mechanism could tune the proportionality constant to
be $k \propto M^{-3/2}$, so that the right MOND phenomenology
would be recovered. More recently \cite{Moffat:2005si}, he
managed to derive this relation by assuming a particular form of
a potential. However, this potential does depend explicitly on
the mass $M$ of the galaxy, and the predictivity of the model is
thus lost. In conclusion, although the framework of nonsymmetric
gravity seems promising, it is not yet formulated as a closed
consistent field theory, and it should only be considered at
present as a phenomenological fit of observational data.

In the following, we will focus on field theories whose actions do
not depend on the mass or scale of the considered objects. They may
involve fine-tuned numbers, but they will be fixed once for all.
This is necessary to define a predictive model, but we will
underline that it does not suffice to prove its consistency. In
particular, the stability needs to be analyzed carefully.

\subsection{Higher-order gravity}\label{R2}

One-loop divergences of quantized GR \cite{'tHooft:1974bx} are well
known to generate terms proportional to\footnote{To simplify, we
do not write the tildes which should decorate all quantities in
this subsection, indicating that $g_{\mu\nu} = \tilde g_{\mu\nu}$
will be later assumed to be minimally coupled to matter.} $R^2$,
$R_{\mu\nu}^2$ and $R_{\mu\nu\rho\sigma}^2$. It is thus natural to
consider extensions of GR already involving such higher powers of
the curvature tensor in the classical action. At quadratic order,
it is always possible to write them as
\begin{equation}
S_\text{gravity} = \frac{c^4}{16 \pi G}
\int\frac{d^4 x}{c}\sqrt{-g}\, \left[R
+ \alpha\, C_{\mu\nu\rho\sigma}^2 + \beta\, R^2 +
\gamma\, \text{GB}\right],
\label{R+R2}
\end{equation}
where $C_{\mu\nu\rho\sigma}$ denotes the Weyl (fully traceless)
tensor, $\text{GB}\equiv
R_{\mu\nu\rho\sigma}^2-4R_{\mu\nu}^2+R^2$ is the Gauss-Bonnet
topological invariant, and $\alpha,\beta,\gamma$ are constants
having the dimension of a length squared (i.e., $(\hbar/mc)^2$ if
$m$ denotes the corresponding mass scale). The topological
invariant $\text{GB}$ does not contribute to the local field
equations and may thus be discarded. In his famous thesis
\cite{Stelle:1976gc}, Stelle proved that such an action gives a
renormalizable quantum theory, to all orders, provided both
$\alpha$ and $\beta$ are nonzero. However, he also underlined
that this cannot be the ultimate answer to quantum gravity
because such a theory contains a ghost, i.e., a degree of freedom
whose kinetic energy is negative. To understand this intuitively,
it suffices to decompose the schematic propagator in irreducible
fractions:
\begin{equation}
\frac{1}{p^2+\alpha p^4} = \frac{1}{p^2} - \frac{1}{p^2+1/\alpha},
\label{propag}
\end{equation}
where the first term, $1/p^2$, corresponds to the propagator of
the usual massless graviton, whereas the second term corresponds
to a massive degree of freedom such that $m^2 = 1/\alpha$. The
negative sign of this second term indicates that it carries
negative energy. Note that one cannot change its sign by playing
with that of $\alpha$: If $\alpha < 0$, this extra degree of
freedom is even a tachyon (negative mass squared), but it anyway
remains a ghost (negative kinetic energy). The problem with such a
ghost degree of freedom is that the theory is violently unstable.
Indeed, the vacuum can disintegrate in an arbitrary amount of
positive-energy usual gravitons whose energy is balanced by
negative-energy ghosts. Even at the classical level, although
such an instability is difficult to exhibit explicitly on a
toy model, one expects anyway the perturbations of a given background
to generically diverge, by creating growing gravitational waves
containing both positive-energy (massless) modes and negative-energy
(massive) ones.

The actual calculation, taking into account all contracted indices
in the propagators, confirms the above schematic reasoning for the
$\alpha\, C_{\mu\nu\rho\sigma}^2$ term in action (\ref{R+R2}), and
shows that the extra massive mode has also a spin 2, like the usual
massless graviton. On the other hand, the $\beta\, R^2$ term
generates in fact a positive-energy massive scalar degree of
freedom. We will recall a well-known and simple derivation in
Sec.~\ref{ST} below. The intuitive explanation is that the scalar
mode corresponds in GR to the negative gravitational binding
energy (which is not an actual degree of freedom because it is
constrained by the field equations), so that the negative sign
entering the right-hand side of Eq.~(\ref{propag}) in fact
multiplies an already negative term. The resulting extra degree of
freedom thereby appears as a positive-energy scalar field
\cite{Woodard:2006nt}.

More general actions $f(R,R_{\mu\nu},R_{\mu\nu\rho\sigma})$ have
been considered several times in the literature, notably in
\cite{Hindawi:1995cu} and \cite{Tomboulis:1996cy}. The conclusion
is that they generically contain the massive spin-2 ghost
exhibited above, which ruins the stability of the theory.
The only allowed models, within this class, are functions of the
scalar curvature alone, $f(R)$, that we will study in
Sec.~\ref{ST} below.

The fact that such functions of the curvature
\textit{generically} involve a ghost tells us that some
particular models may avoid it. One may for instance consider
a Lagrangian of the form $R + \gamma \text{GB}
+ \alpha (R_{\mu\nu\rho\sigma}^2)^n$, $n \geq 2$, whose
quadratic term reduces to the Gauss-Bonnet topological invariant
$\text{GB}\equiv R_{\mu\nu\rho\sigma}^2-4R_{\mu\nu}^2+R^2$.
Around a flat Minkowski background, the only quadratic kinetic term
$\mathcal{O}(g_{\mu\nu}-\eta_{\mu\nu})^2$ is thus the one coming
from the standard Einstein-Hilbert term $R$, so that no
propagator can be defined for any ghost degree of freedom.
However, Ref.~\cite{Hindawi:1995cu} underlined that flat
spacetime is generically not a solution of such higher-order
gravity theories. Around a curved background, the second-order
expansion of higher-order terms like $(R_{\mu\nu\rho\sigma}^2)^n$
thereby generates a nonzero kinetic term for the spin-2 ghost
(see Fig.~\ref{fig6} in Sec.~\ref{MatterSpin2} below for a
diagrammatic illustration).

One may still try to devise higher-order gravity models such that
their second-order expansion around \textit{any} background never
generates a negative-energy kinetic term. For instance,
Refs.~\cite{Navarro:2005da,Cognola:2006eg,DeFelice:2006pg}
consider Lagrangians of the form $R+f(\text{GB})$, and show
that the two degrees of freedom corresponding to the generic
spin-2 ghost are not excited. A similar trick as the one presented
in Sec.~\ref{ST} below for $f(R)$ models indeed proves that
$R+f(\text{GB})$ theories only involve a scalar degree of freedom
in addition to the usual massless spin-2 graviton. However, the
kinetic terms take nonstandard expressions in this framework, and at
present there is no proof that the full Hamiltonian is bounded by
below. One can even show that ghost modes do exist in particular
backgrounds (see, e.g., Eq.~(22) of Ref.~\cite{DeFelice:2006pg}),
although this does not prove either that the Hamiltonian is
necessarily unbounded by below. In our opinion, the clearest hint that
$R+f(\text{GB})$ models are probably unstable comes from the schematic
reasoning of Eq.~(\ref{propag}) above. In $f(R)$ models, the scalar
degree of freedom happened to have positive kinetic energy because it
corresponded to a higher-order excitation of the negative-energy
Newtonian potential (constrained by the field equations). In
$R+f(\text{GB})$ models, the scalar degree of freedom \textit{a
priori} corresponds to a higher-order excitation of \textit{another}
mode present in GR, generically dynamical and therefore carrying
positive energy. The crucial minus sign entering Eq.~(\ref{propag})
then shows that the new scalar mode should probably be a ghost.

More generally, deadly instabilities can exist in higher-order
field theories even if no ghost degree of freedom can be identified
in a perturbative way. Indeed, as recalled notably in
\cite{Woodard:2006nt}, their Hamiltonian is generically unbounded by
below because it is linear in at least one of the canonical momenta.
As in \cite{Woodard:2006nt}, let us illustrate this on the simplest
example of a Lagrangian depending on a variable $q$ and its first
two time derivatives, say $\mathcal{L}(q,\dot q, \ddot q)$. If
$\ddot q$ can be eliminated from $\mathcal{L}$ by partial
integration, we are in the standard case of a theory depending
only on first derivatives, and we know that some models do give a
Hamiltonian which is bounded by below. This is the case when
$\ddot q$ appears only linearly in $\mathcal{L}$, even multiplied
by a function of $q$ and $\dot q$ since $\int dt\, q^n \dot q^m
\ddot q = -n/(m+1)\int dt\, q^{n-1} \dot q^{m+2} +$ boundary
terms. Let us thus consider only the case of a ``non-degenerate''
Lagrangian,\footnote{Note that the $R+f(\text{GB})$ models
studied in Refs.~\cite{Navarro:2005da,Cognola:2006eg,DeFelice:2006pg}
are degenerate, so that their Hamiltonian cannot be written in the
Ostrogradski form (\ref{H}), but this does not suffice to prove
their stability.} i.e., such that the definition $p_2\equiv
\partial\mathcal{L}/\partial\ddot q$ can be inverted to express
$\ddot q$ as a function of $q$, $\dot q$ and $p_2$:
\begin{equation}
\ddot q = f(q,\dot q, p_2).
\label{ddotq}
\end{equation}
In such a case, Ostrogradski showed in 1850 \cite{Ostrogradski}
that the initial data must be specified by two pairs of conjugate
momenta, that he defined as
\begin{subequations}
\label{qp}
\begin{eqnarray}
q_1 \equiv q\,, &\qquad& p_1\equiv
\frac{\partial\mathcal{L}}{\partial\dot q} - \frac{d}{dt}\left(
\frac{\partial\mathcal{L}}{\partial\ddot q}\right),
\label{q1p1}\\
q_2 \equiv \dot q\,, &\qquad& p_2\equiv
\frac{\partial\mathcal{L}}{\partial\ddot q}\,, \label{q2p2}
\end{eqnarray}
\end{subequations}
and he proved that the following definition of the Hamiltonian
\begin{equation}
\mathcal{H} \equiv p_1 \dot q_1 + p_2 \dot q_2
- \mathcal{L}(q,\dot q,\ddot q)
\label{H}
\end{equation}
does generate time translations. Indeed, $\dot q_i = \partial
\mathcal{H}/\partial p_i$ and $\dot p_i = -\partial
\mathcal{H}/\partial q_i$ reproduce the Euler-Lagrange equations of
motion deriving from the original Lagrangian $\mathcal{L}$.
However, this Hamiltonian must be expressed in terms of the momenta
$q_i$ and $p_i$ defined in Eqs.~(\ref{qp}) above. Recalling that
Eq.~(\ref{ddotq}) allows us to write $\ddot q = f(q_1,q_2, p_2)$,
one gets
\begin{equation}
\mathcal{H} = p_1 q_2 + p_2 f(q_1,q_2, p_2)
- \mathcal{L}\left(q_1, q_2,f(q_1,q_2, p_2)\right).
\label{Hbis}
\end{equation}
The crucial problem of this expression is that it is linear in
$p_1$, and therefore unbounded by below \cite{Woodard:2006nt}. In
other words, the theory is necessarily unstable, even if one does
not identify any explicit contribution $-p_i^2$ defining a ghost
degree of freedom. If one diagonalizes the kinetic terms in
Eq.~(\ref{Hbis}) at quadratic order, say in terms of new momenta
$p'_i$, then the standard positive-energy degree of freedom should
appear as $+p_1'^2$. Since we know that $\mathcal{H}$ is unbounded by
below, the other momentum $p_2'$ must have a negative contribution,
but it may be for instance of the form $-p_2'^4$, and therefore absent
at quadratic order.

This result can be straightforwardly extended to (non-degenerate)
models depending on even higher time-derivatives of $q$, say up to
$d^n q/dt^n$. In that case, Ref.~\cite{Ostrogradski} shows that
the Hamiltonian is linear in $n-1$ of the momenta $p_i$, and
thereby unbounded by below. On the other hand, Ostrogradski's
construction of the Hamiltonian cannot be used if $\mathcal{L}$
depends on an \textit{infinite} number of time derivatives, i.e.,
if it defines a \textit{nonlocal} theory. In such a case, the
theory may actually be stable although its expansion looks
pathological \cite{Simon:1990ic}. This is notably the case in the
effective low-energy models defined by string theory, which do
involve quadratic curvature terms like Eq.~(\ref{R+R2}) above, but
also any higher derivative of the curvature tensor (whose
phenomenological effects occur at the same order as the quadratic
terms). Nonlocal models of MOND have been studied in the
literature \cite{Soussa:2003vv}, and they can be proved to satisfy
anyway causality, but they remain difficult to study from a
phenomenological point of view. In the following, we will focus on
local field theories.

\subsection{Scalar-tensor theories}\label{ST}
As mentioned in the previous section, gravity models whose
Lagrangians are given by functions of the scalar curvature,
$f(R)$, do not exhibit any ghost degree of freedom, and avoid the
generic ``Ostrogradskian'' instability of higher-derivative
theories \cite{Woodard:2006nt}. Let us recall here the simplest
way to prove it (see
\cite{Higgs,Bicknell,Teyssandier,Whitt:1984pd,
Bel,Schmidt:2001ac,Wands:1993uu,Hindawi:1995cu,Tomboulis:1996cy},
and note that the following derivation assumes that the scalar
curvature $R$ is a function of the metric $g_{\mu\nu}$ and its
derivatives alone; the derivation and the result differ in the
first-order Palatini formalism, where the scalar field does not
acquire any kinetic term \cite{Flanagan:2003rb,Flanagan:2003iw}).
We start from an action
\begin{equation}
S_\text{gravity} = \int d^4 x\,\sqrt{-\tilde g}\, f(\tilde R),
\label{fR}
\end{equation}
where the global factor $c^3/16\pi G$ is temporarily set to 1, to
simplify. We now introduce a Lagrange parameter $\phi$
to rewrite this action as
\begin{equation}
S_\text{gravity} = \int d^4 x\,\sqrt{-\tilde g} \left\{f(\phi) +
\left(\tilde R-\phi\right)f'(\phi)\right\}.
\label{fphi}
\end{equation}
The field equation for $\phi$ reads $\left(\tilde
R-\phi\right)f''(\phi) = 0$, and implies $\phi = \tilde R$ within
each spacetime domain where $f''(\phi)\neq 0$. [If there exist
hypersurfaces or more general domains where $f''(\phi)=0$, then
either $f$ is constant and there is no gravitational degree of
freedom in such domains, or $f(\tilde R) \propto \tilde R-2\Lambda$
and the theory reduces to GR plus a possible cosmological constant.
In both cases, the scalar degree of freedom that we will exhibit
below cannot propagate within such domains. A consistent field
theory, without any discontinuity of its degrees of freedom, can
thus be defined only within the domains, possibly infinite, where
$f''(\tilde R)\neq 0$.] The field equations for the metric $\tilde
g_{\mu\nu}$ may now be derived from action (\ref{fphi}), and one
finds that they reduce to those deriving from action (\ref{fR})
when $\phi$ is replaced by $\tilde R$. This can also be seen by the
\textit{a priori} illicit use of the field equation $\phi = \tilde
R$ directly within action (\ref{fphi}), which reduces trivially to
(\ref{fR}). In other words, the theory defined by action (\ref{fR})
is equivalent to the scalar-tensor one
\begin{equation}
S_\text{gravity} = \int d^4 x\,\sqrt{-\tilde g}
\left\{f'(\phi) \tilde R - 0
\left(\partial_\mu \phi\right)^2 -
\left[\phi f'(\phi)-f(\phi)\right]\right\},
\label{fphiBis}
\end{equation}
within each spacetime domain where $f''(\tilde R)$ never vanishes.
If one redefines the scalar field as $\Phi \equiv f'(\phi)$, this
action takes the form of the famous (Jordan-Fierz)-Brans-Dicke
theory \cite{Brans-Dicke,Jordan,Fierz}
with no explicit kinetic term for $\Phi$, i.e., with a
vanishing $\omega_\text{BD}$ parameter, and with a potential
defined by the last term within square brackets. It should be
stressed that solar-system experiments \cite{Bertotti:cassini}
impose the bound $\omega_\text{BD} > 4000$ when the scalar-field
potential vanishes. Therefore, if matter is assumed to be coupled
to $\tilde g_{\mu\nu}$, the resulting scalar degree of freedom
needs to be massive enough to have a negligible effect in the
solar-system gravitational physics. The shape of the initial
function $f(\tilde R)$ is thus constrained by experiment.

Although action (\ref{fphiBis}) does not involve any explicit
kinetic term for $\phi$ (or the Brans-Dicke scalar $\Phi$), this
scalar field does propagate. Indeed, the first contribution
$f'(\phi) \tilde R$ contains terms of the form $\phi \partial^2
\tilde g$, i.e., cross terms $-\partial \phi \partial \tilde g$
after partial integration. A redefinition of the fields then allows
us to diagonalize their kinetic terms. This is achieved with the
new variables
\begin{subequations}
\label{vars}
\begin{eqnarray}
g^*_{\mu\nu} &\equiv& f'(\phi) \tilde g_{\mu\nu},
\label{gstar}\\
\varphi &\equiv& \frac{\sqrt{3}}{2}\,\ln f'(\phi),
\label{varphi}\\
V(\varphi) &\equiv& \frac{\phi f'(\phi)-f(\phi)}{4f'^2(\phi)},
\label{V}
\end{eqnarray}
\end{subequations}
and the full action, Eq.~(\ref{fphiBis}) plus the matter part
$S_\text{matter}[\psi; \tilde g_{\mu\nu}]$, now reads
\begin{equation}
S = \frac{c^4}{4\pi G}\int \frac{d^4 x}{c}\sqrt{-g_*} \left\{
\frac{R^*}{4} - \frac{1}{2} g_*^{\mu\nu} \partial_\mu\varphi
\partial_\nu\varphi - V(\varphi)\right\}
+ S_\text{matter}[\psi; \tilde g_{\mu\nu} =
A^2(\varphi)g^*_{\mu\nu}],
\label{TS1}
\end{equation}
where we have put back the global factor $c^3/16\pi G$
multiplying the gravitational action, and where $A(\varphi) =
e^{\varphi/\sqrt{3}}$. This is a particular case of scalar-tensor
theory \cite{DEF92}, where the precise matter-scalar coupling
function $A(\varphi) = e^{\varphi/\sqrt{3}}$ comes from our
initial hypothesis of a $f(\tilde R)$ theory. More general models
may involve an arbitrary nonvanishing function $A(\varphi)$
relating the Einstein metric $g^*_{\mu\nu}$ (whose fluctuations
define the spin-2 degree of freedom) to the physical one $\tilde
g_{\mu\nu} = A^2(\varphi) g^*_{\mu\nu}$. Action (\ref{TS1})
clearly shows that the spin-0 degree of freedom $\varphi$ does
propagate and carries positive kinetic energy. Of course, this
does not suffice to guarantee the stability of the theory: The
potential $V(\varphi)$, Eq.~(\ref{V}), also needs to be bounded by
below, and this imposes constraints on the initial function
$f(\tilde R)$.

Although such tensor-mono-scalar models are perfectly
well-defined field theories, they do not reproduce the MOND
phenomenology, at least not in the most natural cases. Indeed,
when the potential $V(\varphi)$ has a negligible influence, the
scalar-field contribution to the gravitational interaction is
proportional to the Newtonian one, at lowest order, i.e., of the
form $\varphi \propto G M /rc^2$ where $M$ is the baryonic mass
of the body generating the field. Therefore, it does not give the
MOND potential $\sqrt{GM a_0} \ln r$ we are looking for. It
has been shown in \cite{DEF93,DEF96b,DEF98} that Gaussian
matter-scalar coupling functions $A^2(\varphi) = e^{\beta_0
\varphi^2}$, with $\beta_0 < 0$, generate nonperturbative effects
such that $\varphi$ is no longer strictly proportional to $M$,
but still almost so for two different mass ranges. Therefore, the
factor $\sqrt{M}$ we are looking for cannot be obtained that way
either, besides the fact that the radial dependence of the scalar
field is still $\propto 1/r$ instead of being logarithmic. When
the potential $V(\varphi)$ has a minimum, its second derivative
defines the mass $m$ of the scalar degree of freedom, and its
contribution to the gravitational interaction is of the Yukawa
type, $\varphi \propto GM e^{-mr}/rc^2$, still not of the MONDian
form, notably because its global factor is $M$ instead of
$\sqrt{M}$. [Note that the Yukawa force is proportional to
$\partial_r \varphi\propto GM e^{-mr} (1/r^2+m/r)$, and therefore
includes a MOND-like $1/r$ contribution. However, it dominates
the main $1/r^2$ term only for $mr \gg 1$, i.e., precisely when
the exponential $e^{-mr}$ makes the whole force quickly tend
towards zero.] A ``quintessence''-like potential, whose minimum
occurs at $\varphi \rightarrow \infty$, seems better because it
allows us to build a logarithmic gravitational potential. Indeed,
if $V(\varphi) = -2 a^2 e^{-b\varphi}$, where $a$ and $b$ are two
constants, then $\varphi = (2/b)\ln(abr)$ is a solution of the
vacuum field equation $\Delta \varphi = V'(\varphi)$. However,
not only this potential is unbounded by below, thereby spoiling
the stability of the model, but $\varphi$ is also multiplied by a
constant $2/b$ (independent of the matter source) instead of
being proportional to $\sqrt{M}$.

A generalization of $f(R)$ theories has been considered in
Refs.~\cite{Gottlober:1989ww,Wands:1993uu}, where the Lagrangian
density is given by some function of the scalar curvature and its
iterated d'Alembertian, $f(R,\Box R, \dots,
\Box^n R)$. The theory can be shown to be generically equivalent
to a tensor-multi-scalar theory \cite{DEF92} involving $n+1$
scalar fields. The signs of their respective kinetic terms
depends on the numerical constants entering the function
$f(R,\Box R, \dots, \Box^n R)$, and there is thus no guarantee
that they all carry positive energy in the general case. However,
well chosen functions $f$ do define stable tensor-multi-scalar
theories.

Although it seems difficult to reproduce MOND dynamics with
tensor-mono-scalar models, general tensor-multi-scalar theories
might actually succeed. Indeed a particular tensor-bi-scalar
model, called ``phase coupling gravity'' (PCG), has been shown to
reproduce the Tully-Fisher law in some appropriate regime (there
exists a critical mass above which MOND phenomenology arises)
\cite{Bekenstein:1988zy,Sanders:1988}. However this model is
marginally ruled out experimentally \cite{Bekenstein:2004ne}, and
moreover needs a potential which is unbounded by below, thereby
spoiling the stability of the model. This kind of theory has then
been promoted to a ``tensor-vector-bi-scalar'' model
\cite{Sanders:2005vd}, that we shall examine in more detail in
Sec.~\ref{Stabilite}.

Among stable scalar-tensor theories, let us finally mention those
involving a coupling of the scalar field to the Gauss-Bonnet
topological invariant, i.e., involving a term of the form
$f(\varphi)\times \text{GB}$ in the gravitational action.
[Note that the function of $\varphi$ is \textit{a priori} free, but
that the Gauss-Bonnet term must appear linearly.] Contrary to the
higher-order theories discussed in Sec.~\ref{R2} above, such models
do not always contain ghost degrees of freedom. They have been
studied in various contexts (see e.g.
\cite{Boulware:1986dr,Gef-Semboloni,Esposito-Farese:2004cc,
Amendola:2005cr, Amendola:2007ni}), but not with the aim of
reproducing the MOND phenomenology. Our own investigations showed
that it is \textit{a priori} promising, because the Riemann
tensor $\tilde R_{\mu\nu\rho\sigma}$ generated by a body of
baryonic mass $M$ does not vanish outside it, and therefore gives
us a local access to $M$. However, a second information is also
needed to separate $M$ from the radial dependence, and we did not
find any simple way to generate a potential $\sqrt{GM a_0} \ln
r$. [It is actually possible if one defines a nonstandard kinetic
term for the scalar field, as in Sec.~\ref{SecRAQUAL} below, but
the scalar-Gauss-Bonnet coupling is not necessary in such a
case.] A related but much simpler idea is explored in
Sec.~\ref{Nonmin} below.

\subsection{K-essence or RAQUAL models}\label{SecRAQUAL}
The above scalar-tensor models may involve three functions of the
scalar field, namely the potential $V(\varphi)$, the
matter-scalar coupling function $A(\varphi)$, and the possible
scalar-Gauss-Bonnet coupling $f(\varphi)\times \text{GB}$. Their
generalization to $n$ scalar fields $\varphi^a$ \cite{DEF92} may
also involve a $n\times n$ symmetric matrix $\gamma_{ab}(\varphi^c)$
depending on them and defining their kinetic term $g_*^{\mu\nu}
\gamma_{ab}(\varphi^c) \partial_\mu \varphi^a \partial_\nu
\varphi^b$. Their phenomenology is however similar to the single
scalar case, at least if all of them carry positive energy and
their potential is bounded by below, to ensure the stability of
the theory \cite{DEF92}. On the other hand, a significantly
different physics arises if one considers more general kinetic
terms of the form $f(s, \varphi)$, where $s \equiv g_*^{\mu\nu}
\partial_\mu \varphi \partial_\nu \varphi$ is the standard one.
[Beware that the literature often uses the notation $X \equiv
s/2$, which actually does not change the following discussion.]
Such models, without any matter-scalar coupling, have been
studied in the cosmological context under the name of
``k-inflation'' \cite{Armendariz-Picon:1999rj} or ``k-essence''
\cite{Chiba:1999ka,Armendariz-Picon:2000ah}, the letter k meaning
that their dynamics is kinetic dominated, as opposed to
quintessence models in which the potential $V(\varphi)$ plays a
crucial role. When a matter-scalar coupling $A(\varphi)$ is also
assumed, such models have been studied to reproduce the MOND
dynamics under the name of ``Relativistic AQUAdratic
Lagrangians'' (RAQUAL) \cite{Bekenstein:1984tv}. Their action is
thus of the form
\begin{equation}
S = \frac{c^4}{4\pi G}\int \frac{d^4 x}{c}\sqrt{-g_*} \left\{
\frac{R^*}{4} - \frac{1}{2} f(s,\varphi) - V(\varphi)\right\}
+ S_\text{matter}[\psi; \tilde g_{\mu\nu} =
A^2(\varphi) g^*_{\mu\nu}],
\label{RAQUAL}
\end{equation}
with $s \equiv g_*^{\mu\nu} \partial_\mu \varphi \partial_\nu
\varphi$, and they can be considered as natural generalizations of
the simplest $f(\tilde R)$ model, Eq.~(\ref{TS1}). Note that the
potential $V(\varphi)$ may be reabsorbed within the general
function $f(s,\varphi)$, and that it is therefore unnecessary.
However, since one often considers functions $f(s)$ of the kinetic
term alone, it remains convenient to keep an explicit potential
$V(\varphi)$. Reference \cite{Bekenstein:1984tv} proved that an
appropriate nonlinear function $f(s,\varphi)$ allows us to
reproduce the MOND gravitational potential $\sqrt{GMa_0} \ln r$,
but that some difficulties remain, as discussed below and in
Secs.~\ref{MONDRAQUAL} and \ref{Light}.

Several conditions must be imposed on this function $f(s,\varphi)$ to
guarantee the consistency of the field theory. For any real value of
$s \in [-\infty,+\infty]$, one must have
\begin{enumerate}
\item[(a)]$f'(s,\varphi) > 0$, \item[(b)]$2s f''(s,\varphi) +
f'(s,\varphi) > 0$,
\end{enumerate}
where a prime denotes derivation with respect to $s$, i.e., $f' =
\partial f(s,\varphi)/\partial s$. The condition $f'(s) \geq 0$,
implied by (a), is \textit{necessary} for the Hamiltonian to be
bounded by below. Indeed, up to a global factor $c^4/8\pi G$ that we
do not write to simplify the discussion, the contribution of the
scalar field to this Hamiltonian reads $H = 2 (\partial_0 \varphi)^2
f'(s) + f(s)$ in a locally inertial frame. If there existed a value
$s_-$ such that
$f'(s_-) < 0$, then $H$ could be made arbitrarily large and
negative by choosing diverging values of $(\partial_0 \varphi)^2$
and $(\partial_i \varphi)^2$ such that $-(\partial_0 \varphi)^2 +
(\partial_i \varphi)^2 = s_-$. On the other hand, conditions (a)
and (b) are necessary and sufficient for the field equation to be
always hyperbolic, so that the Cauchy problem is well posed for
the scalar field. Indeed, it reads
\begin{equation}
G^{\mu\nu} \nabla^*_\mu\nabla^*_\nu\varphi
= \frac{1}{2}\,\frac{\partial f}{\partial\varphi}
- s \frac{\partial f'}{\partial \varphi} +
\frac{\partial V}{\partial\varphi} -
\frac{4\pi G}{c^3\sqrt{-g_*}}\,\frac{\delta
S_\text{matter}}{\delta\varphi},
\label{eqPhiRAQUAL}
\end{equation}
where $G^{\mu\nu} \equiv f' g_*^{\mu\nu} + 2 f''
\nabla_*^\mu\varphi\nabla_*^\nu\varphi$ plays the role of an
effective metric in which $\varphi$ propagates. The lowest
eigenvalue of $G^{\mu\nu}$ is negative (i.e., defining a
consistent time) only if condition (b) is satisfied, and it is
easy to check\footnote{The simplest calculation consist in
diagonalizing the matrix $G^{\mu\rho}g^*_{\rho\nu}$, instead of
the tensor $G^{\mu\nu}$, and impose that its four eigenvalues are
positive. This also ensures that there exists a coordinate system
in which the time directions defined by $G_{\mu\nu}$ and
$g^*_{\mu\nu}$ coincide.} that the three others are then positive
(defining spatial dimensions) if (a) is also
satisfied.\footnote{One could also consider the degenerate case
where $f'(s)$ vanishes on some hypersurfaces. On them, the
effective metric can then be put in the form
$\textrm{diag}(-1,0,0,0)$ in some appropriate basis, and thus
define a consistent time, if and only if $f''(s)<0$. Strictly
speaking, the scalar field equation is not hyperbolic in that
case but it has nevertheless a well-posed Cauchy problem. Using
condition (b), we can conclude that $f'(s)$ may vanish only at
negative values of $s$.} As a quick check of condition (b), one
may for instance consider the particular case of a homogeneous
scalar field ($\partial_i\varphi = 0$) in a locally inertial frame
($g^*_{\mu\nu} = \eta_{\mu\nu}$): The fact that we need $G^{00} =
-f'+2f''\times(-s) < 0$ then immediately yields inequality (b).
Finally, conditions (a) and (b) together \textit{suffice} for the
Hamiltonian to be bounded by below, provided the function
$f(s,\varphi)$ is analytic and $f(s=0,\varphi)$ is itself bounded
by below for any $\varphi$. Indeed, still in a locally inertial frame,
we know that $s = -(\partial_0 \varphi)^2+(\partial_i \varphi)^2
\geq -(\partial_0 \varphi)^2$. Together with condition (a),
we can thus conclude that the Hamiltonian $H =
2 (\partial_0 \varphi)^2 f' + f$ is greater than $-2 s f'+f$,
which is known to be a decreasing
function of $s$ because of condition (b). Therefore, for any
$s \leq 0$, the Hamiltonian is necessarily greater than
$f(s=0,\varphi)$. On the other hand, for $s \geq 0$, we
also know that $H \geq f \geq f(s=0,\varphi)$, first because
$(\partial_0 \varphi)^2 \geq 0$ and then because
condition (a) implies that $f$ is an increasing function of $s$.
Although we used the hyperbolicity condition (b) to derive that
the Hamiltonian is bounded by below, note that condition (b) is
\textit{not} implied by this boundedness, on the contrary.
Nevertheless, it reappears when considering the propagation
velocity of perturbations around a background
\cite{Armendariz-Picon:1999rj}, $c_s^2 = f'/(2sf''+f')$. Such
perturbations are unstable if $c_s^2 < 0$, and one may thus think
that condition (b) should be implied by the boundedness by below
of the Hamiltonian. However, the energy of the perturbations is
only part of the total Hamiltonian, which also includes the
energies of the background and its interactions with the
perturbations. A total Hamiltonian bounded by below therefore
does not suffice to guarantee the stability of perturbations; the
hyperbolicity of the field equations, condition (b), is
\textit{also} crucial.

\begin{figure}
\includegraphics[scale=.5]{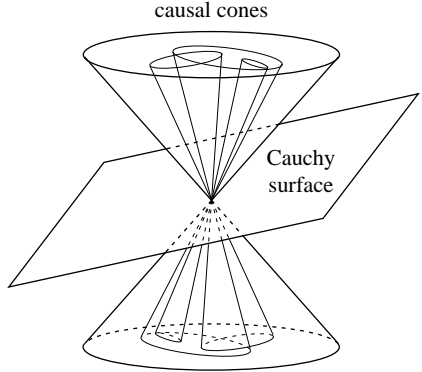}
\caption{In a theory where different fields have different
causal cones, it suffices that their union be embedded in
a wider cone for local causality to be satisfied. Initial
data for all the fields simultaneously may be specified
on a surface exterior to the wider cone, i.e., spacelike
with respect to each cone. If the topology of spacetime
is such that there does not exist any CTC with respect
to the wider cone, then causality is preserved, although
some fields may propagate faster than light (i.e.,
faster than electromagnetic waves, a mere particular
case of matter field).}
\label{fig1}
\end{figure}

In Ref.~\cite{Aharonov:1969vu} (see also \cite{Adams:2006sv}), a
third condition was underlined for such k-essence models, besides
(a) and (b) above:
\begin{enumerate}
\item[(c)]$f''(s) \leq 0$.
\end{enumerate}
This extra condition is necessary for the causal cone of the
scalar field to remain inside the light cone defined by the
spacetime metric $g^*_{\mu\nu}$, i.e., to avoid superluminal
propagation. The simplest way to derive this inequality is to
consider a null vector $k^\mu$ with respect to the metric
$G_{\mu\nu} = (1/f')[g^*_{\mu\nu} - 2f''\partial_\mu\varphi
\partial_\nu\varphi/(2sf''+f')]$, defined as the inverse of the
effective metric $G^{\mu\nu}$ entering Eq.~(\ref{eqPhiRAQUAL})
above. [We assume here that $f'$ never vanishes.] The equation
$G_{\mu\nu} k^\mu k^\nu = 0$ then gives $g^*_{\mu\nu} k^\mu k^\nu
= 2f'' (k^\mu\partial_\mu\varphi)^2/(2sf''+f')$, and obviously
implies $g^*_{\mu\nu} k^\mu k^\nu \leq 0$ if conditions (b) and
(c) are satisfied. The causal cone of the scalar field is thus
timelike or null with respect to the Einstein metric
$g^*_{\mu\nu}$. Such a condition (c) would be extremely
constraining, since it would rule out notably any monomial $f(s)
= s^n$, which would violate either (a) or (c) depending on the
parity of $n$. However, let us underline that (c) is actually
\textit{not} required for the consistency of the model. Indeed,
even if $\varphi$ propagates superluminally, this does not ruin
the causality of the model. Thanks to condition (b), the causal
cone of $\varphi$ never opens totally, i.e., always admits a class
of Cauchy surfaces which can be defined in its exterior. In the
general situation where different fields have different causal
cones, it suffices that their union still admits a nonvanishing
exterior where one may consistently define initial data for all
of them simultaneously; see Fig.~\ref{fig1}.
This becomes quite clear by shifting our viewpoint. Let us assume
that the union of all causal cones may be embedded in the
interior of a wider cone which never totally opens, and let us
define space and time with respect to this exterior cone. Since
all fields propagate within this cone, there is no more causal
problem than in special relativity where particle worldlines lie
within the standard lightcone. Causal pathologies are still
possible, like in GR itself which admits solutions containing
closed timelike curves (CTC), like the G\"odel universe or the
interior of the Kerr solution.\footnote{It has even been proven that
CTCs can form in GR from smooth initial data, while assuming
asymptotic spatial flatness and various energy conditions; see
\cite{Ori:2007kk} and references therein.} Since the global topology
of the Universe is not imposed by our local field equations, it is
necessary to \textit{assume} that it does not involve any CTC to
ensure causality. In our example of different fields having different
causal cones, one thus just needs to assume that spacetime does not
admit any CTC with respect to the wider cone surrounding all of them.
Causal paradoxes due to superluminal propagation, like the nice one
exhibited in Sec.~III. 2 of \cite{Adams:2006sv}, are thus more a
matter of global assumptions than a problem with local field
equations.\footnote{In Ref.~\cite{Adams:2006sv}, a CTC is exhibited by
considering a background involving two bubbles of a scalar field with
a fast relative motion. If one assumes that all clocks can be globally
synchronized by using only electromagnetic waves (for instance that
spacetime is Minkowskian with respect to $g^*_{\mu\nu} =
\eta_{\mu\nu}$, like in \cite{Adams:2006sv}), then such a CTC
does exist. However, although this hypothesis sounds extremely
natural, there \textit{also} exist spacetimes in which there are
no CTCs for any field, and in such a case, light within a given
bubble of scalar field does not define the same synchronization
as within the other bubble. The CTC constructed in \cite{Adams:2006sv}
would then become an artifact of gluing independent coordinate systems
at the edges of the bubbles.} We refer to Ref.~\cite{Bruneton:2006gf}
for a more detailed discussion of this subtle issue (see also
Sec.~\ref{Disformal} below and the very recent article
\cite{Babichev:2007dw}). Reference \cite{Ellis:2007ic} claims that
microscopically Lorentz-invariant particles cannot give rise to
superluminal signals, but this conclusion does not take into account
their possible self-interactions via their kinetic term, like $f(s)$
in our present case of k-essence models (\ref{RAQUAL}). As shown
above, superluminal signals do occur when $f''(s) > 0$, although the
theory is microscopically Lorentz-invariant --- and causal when the
hyperbolicity conditions (a) and (b) are satisfied. The causal cone of
the scalar field $\varphi$ is either interior or exterior to the light
cone defined by $g^*_{\mu\nu}$ (depending on the local value of $s$),
but it always remains a cone thanks to these conditions. At each
spacetime point, any surface exterior to the wider of these two cones
may be used as a Cauchy surface to impose initial data. Of course, a
surface lying \textit{between} the two cones behaves as a spacelike one
for the thinner cone but as a timelike one for the thicker. Some causal
paradoxes discussed in the literature (including in
\cite{Aharonov:1969vu}) are actually based on the improper use of such
intermediate surfaces. They are clearly not consistent Cauchy surfaces
for the fastest field (i.e., the wider cone), so that data are
necessarily constrained on them. The conclusion of such
paradoxes, namely that initial data are constrained, is thus
actually hidden in the use of these intermediate surfaces.

\subsection{MOND as a RAQUAL model}\label{MONDRAQUAL}

As mentioned above, Ref.~\cite{Bekenstein:1984tv} proved that RAQUAL
models (\ref{RAQUAL}) may reproduce the MOND gravitational
potential. Indeed, if $f(s)$ is assumed not to depend explicitly
on $\varphi$, if the potential $V(\varphi)$ vanishes, and if one
chooses an exponential (Brans-Dicke-like) matter-scalar coupling
function $A(\varphi) = \exp(\alpha \varphi)$,
Eq.~(\ref{eqPhiRAQUAL}) reduces to
\begin{equation}
\nabla^*_\mu\left[f'(s)\nabla_*^\mu\varphi\right] = -\frac{4\pi
G}{c^4}\, \alpha\, T^*,
\label{eqPhiRAQUAL2}
\end{equation}
where $T^* \equiv g^*_{\mu\nu} T_*^{\mu\nu}$ denotes the trace of
the matter energy-momentum tensor $T_*^{\mu\nu}\equiv
\left(2c/\sqrt{-g_*}\right) \delta S_\text{matter}/\delta
g^*_{\mu\nu}$. In order to recover the right Newtonian limit for
large accelerations $a> a_0$, one may follow
Ref.~\cite{Bekenstein:1984tv} and impose\footnote{Note that imposing
$f'(s) \rightarrow \text{const.}$ is strictly equivalent, since
the arbitrary constant can be absorbed in a redefinition of the
matter-scalar coupling constant $\alpha$. In the following, the
experimental constraints on $\alpha$ that we quote correspond to
the choice $f'(s) \rightarrow 1$ in the Newtonian regime.} $f'(s)
\rightarrow 1$ for large positive values of $s$. In such a case,
the scalar field reads $\varphi \approx -\alpha GM/rc^2$ near a
body of mass $M$, and test particles feel an extra potential
$\alpha\varphi c^2\approx -\alpha^2 GM/r$ adding up to the
standard gravitational potential $-GM/r$ mediated by the spin-2
interaction (i.e., by the Einstein metric $g^*_{\mu\nu}$). The
total potential $-G(1+\alpha^2)M/r$ is thus of the Newtonian type,
with a renormalized effective gravitational constant
$G_\text{\textrm{eff}} = G(1+\alpha^2)$ (see for instance
\cite{DEF92} and line (a) of Fig.~\ref{fig5} in Sec.~\ref{Light}
below). Precision tests in the solar system are however sensitive
to post-Newtonian corrections, and they prove that the scalar
contribution must be negligible. Indeed, the parametrized
post-Newtonian (PPN)
parameter $\gamma^{\textrm{PPN}}$ assumes the value
$\gamma^{\textrm{PPN}}= 1-2 \alpha^2/(1+\alpha^2)$
in the present (conformally-coupled) scalar-tensor framework
[see Refs.~\cite{Willbook,DEF92} and
Eq.~(\ref{defl2}) below], and the impressive experimental bound
$|\gamma^{\textrm{PPN}}-1|<2
\times 10^{-5}$ obtained in \cite{Bertotti:cassini} from
the observation of the Cassini spacecraft implies therefore
\begin{equation}
\alpha^2 < 10^{-5}.
\label{alpha2}
\end{equation}
Instead of imposing $f'(s) \rightarrow 1$ for large values of $s$,
one may also recover the right Newtonian limit by choosing a
decreasing function $f'(s)$, such that the scalar contribution is
even smaller. Condition (b) should however still be satisfied,
thereby constraining the slope of $f'(s)$.

\begin{figure}
\includegraphics[scale=0.75]{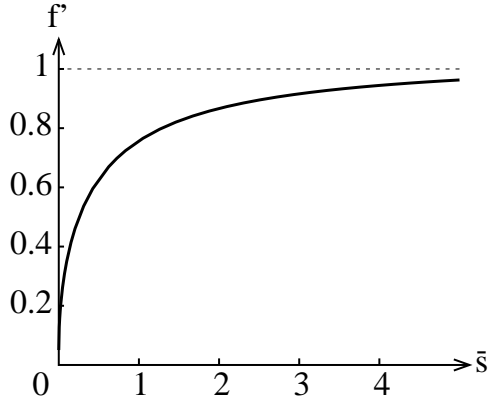}
\caption{A simple function $f'(s)$ reproducing the MOND dynamics
for small $s$ (i.e., large distances) and the Newtonian one for
large $s$ (i.e., small distances).} \label{fig2}
\end{figure}

But the crucial feature of Eq.~(\ref{eqPhiRAQUAL2}) is that it
also allows us to reproduce the MOND potential for small
accelerations $a < a_0$. Indeed, if $f'(s) \approx \ell_0
\sqrt{s}$ for small and positive values of $s$, where $\ell_0$ is
a constant having the dimension of a length, one gets
$(\partial_r\varphi)^2 \approx \alpha GM/\ell_0c^2r^2$ near
a spherical body of mass $M$. Test particles therefore feel, in
addition to the usual Newtonian potential $-G M /r$, an extra
potential $\alpha\varphi c^2\approx
\sqrt{\alpha^3 GMc^2/\ell_0}\ln r$ which reduces to the MOND one
$\sqrt{GMa_0} \ln r$ for $\ell_0 = \alpha^3c^2/a_0$. A simple way
to connect this MOND limit to the above Newtonian one is to choose
for instance $f'(s) = \sqrt{\bar s}/\sqrt{1+\bar s}$, where $\bar
s \equiv \alpha^6 c^4 s/a_0^2$ is dimensionless. The shape of this
function is illustrated in Fig.~\ref{fig2}.
By integrating this expression, one can thus conclude that the
precise function $f(s) = (a_0^2/\alpha^6 c^4)\left[\sqrt{\bar
s(1+\bar s)} - \sinh^{-1}(\sqrt{\bar s})\right]$, or any other one
having the same asymptotic behaviors for large and small positive
values of $s$, allows us to reproduce the MOND dynamics.

However, the authors of Ref.~\cite{Bekenstein:1984tv} noticed
that the scalar field propagates superluminally in the MOND
(small positive $s$) regime, since $f''(s) \approx
\frac{1}{2}\ell_0/\sqrt{s} > 0$ contradicts condition (c) above.
This is the reason why they discarded this model, although we
underlined above that such a superluminal propagation actually
does not threaten causality, provided the hyperbolicity
conditions (a) and (b) are satisfied. In fact, the only
experimental constraints we have on the existence of several
causal cones is in favor of superluminal fields! As we will
discuss in Sec.~\ref{Cherenkov}, if light (and thus matter)
travelled faster than some field to which it couples, then it
would emit Cherenkov radiation of this field. High-energy cosmic
rays would thus be significantly suppressed.

It thus seems that the above RAQUAL model reproducing the MOND
dynamics is \textit{a priori} consistent. However, it presents
several other difficulties. The main one has been immediately
recognized and addressed in the literature: Because of the
specific form of the matter-scalar coupling assumed in action
(\ref{RAQUAL}), this model does not reproduce the observed light
deflection by ``dark matter''. We will devote Sec.~\ref{Light}
below to this crucial problem. Let us here mention the other
problems that we noticed, and which do not seem to have been
discussed in the literature.

First of all, the above function $f(s)$ is clearly not defined for
negative values of $s$. One may try to replace $s$ by its absolute
value $|s|$, and also multiply globally $f(s)$ by the sign of $s$
in order to still satisfy conditions (a) and (b). However, this
would not cure the serious problem occurring at $s = 0$. Indeed,
since $f(s) \rightarrow (2\alpha^3 c^2/3a_0) s\sqrt{|s|}$ for
small values of $s$, the strict inequality (b) is not satisfied at
$s = 0$. In other words, the scalar field equation
(\ref{eqPhiRAQUAL}) or (\ref{eqPhiRAQUAL2}) is no longer
hyperbolic on the hypersurfaces where $s$ vanishes, and the scalar
degree of freedom cannot cross them. Since $s \approx
(\partial_i\varphi)^2 \geq 0$ when considering the local physics
of clustered matter, but $s \approx -\dot\varphi^2 \leq 0$ when
considering the cosmological evolution of the Universe, there
always exist such singular surfaces around clusters. Therefore,
this model cannot describe consistently both cosmology and
galaxies, unless independent solutions are glued by hand on the
singular hypersurfaces $s=0$. However, a simple cure to this
discontinuity would be to consider, for small values of $s$, a
function
\begin{equation} f'(s) = \varepsilon + \sqrt{|\bar s|},
\label{Epsilon}
\end{equation}
where $\varepsilon$ is a small dimensionless positive number. Then
both conditions (a) and (b) would obviously be satisfied even at
$s=0$. In other words, the RAQUAL model (\ref{RAQUAL}) with
$V(\varphi) = 0$, $A(\varphi) = \exp(\alpha \varphi)$ and
\begin{equation}
f(s) = \varepsilon\, s + \frac{a_0^2\,\text{sgn}(s)}{\alpha^6 c^4}
\left[\sqrt{|\bar s|\left(1+|\bar s|\right)} -
\sinh^{-1}\left(\sqrt{|\bar s|}\right)\right], \label{GoodRAQUAL}
\end{equation}
where $\bar s \equiv \alpha^6 c^4 s/a_0^2$, does reproduce the
MOND dynamics and is free of mathematical inconsistencies. Of
course, there is no reason why this specific function, for
negative values of $s$, should reproduce the right cosmological
behavior. One may obviously look for other functions of $s<0$
connecting smoothly to (\ref{GoodRAQUAL}) for small values of
$|s|$, provided conditions (a) and (b) above remain always
satisfied. Therefore, Eq.~(\ref{GoodRAQUAL}) should just be
considered as an example of a mathematically consistent RAQUAL
model.

A second difficulty is that the above model is rather fine tuned,
since it needs the introduction of a small dimensionless constant
$\varepsilon$ besides Milgrom's MOND acceleration constant $a_0
\approx 1.2\times 10^{-10}\, \text{m}.\text{s}^{-2}$. Indeed, the
presence of $\varepsilon$ in
Eqs.~(\ref{Epsilon}),(\ref{GoodRAQUAL}) notably implies that
Newtonian gravity is recovered at very large distances, the MOND
regime manifesting only at intermediate ranges
\cite{Bruneton:2006gf}. When $s \rightarrow 0$, the derivative
$f'(s)$ entering Eq.~(\ref{eqPhiRAQUAL2}) tends to $\varepsilon$,
so that the scalar field reads $\varphi \approx -(\alpha/
\varepsilon)GM/rc^2$ faraway from a body of mass $M$. In this
regime, the total gravitational potential felt by a test particle
reads $-G(1+\alpha^2/\varepsilon)M/r$, and is therefore of the
Newtonian form with a renormalized gravitational constant
$G_{\infty} = G(1+\alpha^2/ \varepsilon)$, where the subscript
refers to the fact that the above form of the potential holds
exactly when $r \to \infty$ (remind that, if $r \to 0$, the
effective gravitational constant is given by
$G_{\textrm{eff}}=G(1+\alpha^2)$). Let us compute the range of
distances $r$ for which the MOND force dominates the Newtonian
contribution. One needs of course $f'(s)\propto \sqrt{s}$ and
thus $\varepsilon \ll \sqrt{\bar s}$. On the other hand, the MOND
force dominates the Newtonian one if $r \gg \sqrt{G M/a_0}$.
Using $s\approx (\partial_r\varphi)^2 \approx GMa_0/\alpha^2 c^4
r^2$, we thus find that the MOND force dominates within the
following range of distances $\sqrt{GM/a_0}\ll r \ll (\alpha^2/
\varepsilon) \sqrt{GM/a_0}$. Since solar system tests impose
$\alpha^2 < 10^{-5}$, Eq.~(\ref{alpha2}), and since rotation
curves of galaxies may be flat up to $r \sim 10 \sqrt{GM/a_0}$
\cite{Gentile:2006hv}, one needs therefore $\varepsilon \ll
10^{-6}$. This illustrates the fine tuning required to define a
consistent RAQUAL model even for $s\rightarrow 0$.

On the other hand, Eq.~(\ref{Epsilon}) has two great advantages,
besides the fact that it cures the singularity at $s=0$. First,
the fact that the field assumes a Newtonian form at large
distances, $r> (\alpha^2/ \varepsilon) \sqrt{GM/a_0}$, ensures
that a body of mass $M$ does not create a static (scalar) field
with infinite energy, contrary to the case where $f'(s) \sim
\sqrt{s}$ for small $s$ \cite{Bekenstein:1984tv}. Secondly, the
presence of $\varepsilon$ naturally leads to a theory where the
local constant of gravity $G_{\textrm{eff}}=G(1+\alpha^2)$ may
differ from $G_{\infty}$, i.e., the one in regions of spacetime
where the scalar field gradient is small. This may have some
interesting applications in cosmology. Indeed, provided that
$\sqrt{|\bar s|} \alt \varepsilon$, i.e., when $\dot{\varphi}^2
\alt \varepsilon a_0^2/ \alpha^6 c^2$ in a cosmological context,
a homogeneous Universe is described by Friedman
equations\footnote{Beware that the density and the pressure of the
scalar field must also be taken into account.} but the baryonic
matter contributes as $G_{\infty} \rho_{\textrm{baryons}}$. The
effective density of matter reads therefore
$\rho_{\textrm{baryons}} G_{\infty}/G_{\textrm{eff}} =
\rho_{\textrm{baryons}} (1+\alpha^2/\varepsilon)/(1+\alpha^2) \approx
\rho_{\textrm{baryons}} (1+\alpha^2/\varepsilon)$. In other words,
baryonic matter may contribute effectively as if there were true
(exotic) dark matter in that regime, provided that we tune the
free parameter $\varepsilon$ so that $\alpha^2/\varepsilon \approx
10$, which is consistent with galactic dynamics. Since the
cosmology of such models goes beyond the scope of the present
paper, we will not investigate further this phenomenology. Let us
however stress that there is here a potential threat for MOND-like
theories. Indeed, bounds may be found on the ratio
$G_{\infty}/G_{\textrm{eff}}$ from the analysis of primordial
nucleosynthesis, since the speed of the expansion of the Universe,
and therefore the relative proportion of hydrogen and helium in the
present Universe, depends on this ratio. Within the framework of
the so-called Einstein-aether theories
\cite{Jacobson:2000xp,Eling:2004dk}, it was shown in
Ref.\cite{Carroll:2004ai} that one must have
$|G_{\infty}/G_{\textrm{eff}}-1| \alt 1/8$ to be consistent with
the data. Of course, in the above RAQUAL model of MOND, such a
constraint only exists if the time derivative of the scalar field
is small at the epoch of nucleosynthesis. In that case, the above
bound would read $\alpha^2/\varepsilon < 1/8$, and the MOND regime
would therefore not exist at all (there is no more room for the
flattening of rotation curves, as shown in the above paragraph). A
solution would of course to impose a very small value of $\varepsilon$
such that $\dot{\varphi}^2 \gg \varepsilon a_0^2/ \alpha^6 c^2$ at the
nucleosynthesis epoch. However one should study the cosmology of
such models to investigate in more detail this effect and make
quantitative estimates; we leave this point for future
work.\footnote{Note that this phenomenon is very likely to occur
also in extended models we consider below (TeVeS-like theories),
since they notably involve a k-essence scalar field whose kinetic
term also needs to be cured at $s=0$; see Sec.~\ref{Disformal}.}

Finally, the third (and the most serious) problem of such RAQUAL
models of MOND, besides the one of light deflection discussed in
Sec.~\ref{Light} below, is that the MOND force $\propto
\sqrt{GMa_0}/r$ starts manifesting at quite small distances.
Indeed, the MOND potential $\propto \sqrt{GMa_0}\ln r$ appears at
values of field gradient $\bar s$ such that $f'(\bar s)\sim
\sqrt{\bar s}$. This typically happens at $\bar s \alt 1$ for
natural functions $f'$ interpolating between the MOND and the
Newtonian regimes, see Fig.~\ref{fig2}. Using $\bar s \alt 1$ we
find that the MOND potential appears at distance $r \agt
r_{\textrm{trans}}$ from a body of mass $M$, where
\begin{equation}
\label{RayonDeTransition} r_{\textrm{trans}} = \alpha^2
\sqrt{\frac{G M}{a_0}}.
\end{equation}
Since, in the present (conformally-coupled) scalar-tensor
framework, the matter-scalar coupling constant $\alpha$ is
severely constrained by solar-system tests, Eq.~(\ref{alpha2}),
we conclude that the extra MOND potential starts manifesting at a
radius $r_{\textrm{trans}} = \alpha^2 \sqrt{GM_\odot/a_0} < 0.1$
AU in the solar system, where $M_\odot$ denotes the mass of the
Sun. In other words, all planets (including Mercury, at 0.4 AU
from the Sun) should feel a total gravitational potential
$-GM_\odot/r+\sqrt{GM_\odot a_0}\ln r$, the first term coming
from the spin-2 field $g^*_{\mu\nu}$, and the second one from the
scalar field which is already in its MOND regime. This extra
potential $\sqrt{GM_\odot a_0}\ln r$ (or force in $1/r$) leads to
deviations from the Newtonian behavior and is tightly constrained
by tests of Kepler's third law; see \cite{Talmadge:1988qz} and
Sec.~\ref{Pioneer} below. Although this MONDian anomalous force
remains numerically small with respect to the Newtonian one
within the solar system, it is large with respect to
post-Newtonian corrections $\propto 1/c^2$, and its existence is
definitely ruled out experimentally. In conclusion, although the
RAQUAL model of Ref.~\cite{Bekenstein:1984tv} or its refinement
(\ref{GoodRAQUAL}) do predict a Newtonian behavior at distances
$r \rightarrow 0$ (i.e., $s \rightarrow\infty$), planets are far
enough to be already in the MOND regime! Note that this
surprizing problem comes from the experimental bound
(\ref{alpha2}), telling us that the scalar field must be very
weakly coupled to matter to reproduce GR within the solar system.
If one did not take it into account, by setting $\alpha = 1$ for
instance, one would easily reproduce Newtonian gravity up to
$\sqrt{GM_\odot/a_0}\approx 7000$ AU, and the MOND dynamics
beyond. A reasonable value $\varepsilon \sim 1/10$ would also
suffice for the refined model (\ref{GoodRAQUAL}). However, this
theory would be inconsistent by 5 orders of magnitude with
post-Newtonian tests in the solar system, because the scalar
field would be much too strongly coupled to matter.

A possible solution to the above problem would be to fine tune
the function $f(s)$ even further. In order to get the MOND regime
for $r\agt \sqrt{GM/a_0}$, as required by galaxy rotation curves,
one would need $f'(s) \approx \sqrt{\bar s}$ for $\bar s \alt
\alpha^4 < 10^{-10}$. On the other hand, in order to obtain the
Newtonian regime within the solar system, say for $r \alt
r_\text{max} \sim 20$ or $30$ AU, one would need $f'(s) \approx
1$ for $\bar s \agt (\alpha^4 GM_\odot/a_0 r_\text{max}^2)^2 \sim
10^{-10}$ (this second occurrence of $10^{-10}$ is a numerical
coincidence). Therefore, there would exist a brutal transition
between the MOND and Newtonian regimes around $\bar s \sim
10^{-10}$. Not only the introduction of such a small
dimensionless number would be quite unnatural, but this model
would also predict that the anomalous acceleration caused by the
scalar field remains approximately equal to the
constant\footnote{Note that the MOND constant $a_0$ is too small
by a factor $7$ to account for the anomalous acceleration of the
two Pioneer spacecrafts. Actually, we will also see in
Sec.~\ref{Pioneer} below that there exists a crucial difference
between the MOND dynamics and the Pioneer anomaly.} $a_0$ between
30 and 7000 AU.
\begin{figure}
\includegraphics[scale=0.75]{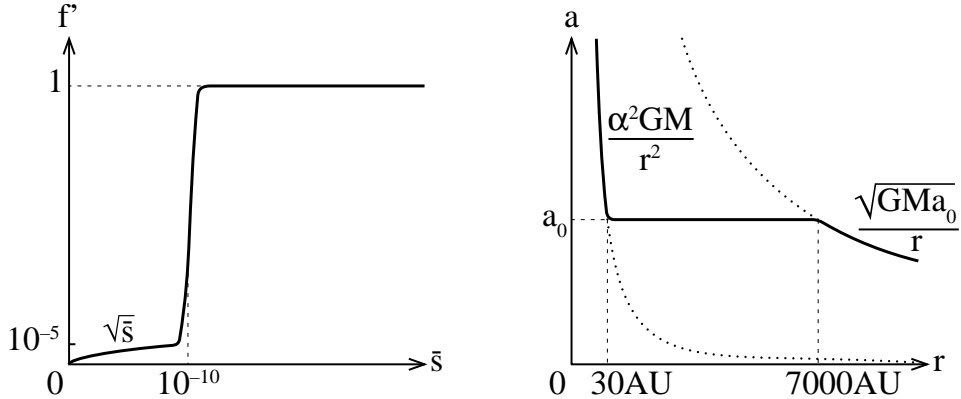}
\caption{Fine-tuned function $f'(s)$ such that Newtonian and
post-Newtonian predictions are not spoiled in the solar system,
although the MOND dynamics is predicted at large distances. The
right panel displays the quite unnatural contribution of the
scalar field to the acceleration of a test mass, as a function of
its distance with respect to the Sun.} \label{fig3}
\end{figure}
As illustrated in Fig.~\ref{fig3}, this would be a way to
reconcile the MOND acceleration $\sqrt{GMa_0}/r$ at large
distances with the experimentally small Newtonian contribution
$\alpha^2 GM/r^2$ of the scalar field at small distances. Although
this is not yet excluded experimentally, it would however suffice
to improve by one order of magnitude the post-Newtonian constraint
on $\alpha^2$ to rule out such a fine-tuned model (the planned
astrometric experiment GAIA \cite{GAIA} should
reach the $10^{-6}$ level for $\alpha^2$, and the proposed LATOR
mission \cite{Turyshev:2003wt,Turyshev:2004ga} should even reach the
$10^{-8}$ level). Therefore, one should not consider it seriously.

\begin{figure}
\includegraphics[scale=0.75]{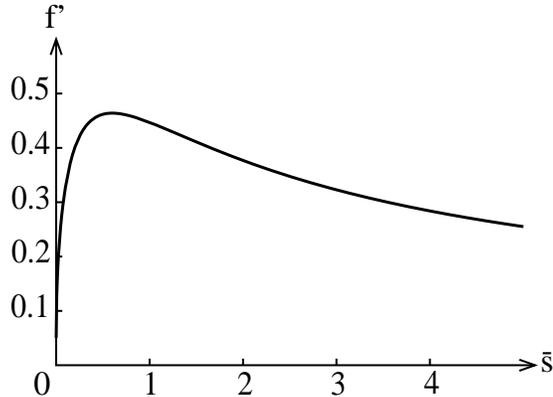}
\caption{Typical shape of a function $f'(s)$ such that
scalar-field deviations from GR are $\propto r^n$, $n > 0$,
in the large-$s$ limit.}
\label{fig4}
\end{figure}

An \textit{a priori} better solution to the above problem would be
to recall that the Newtonian limit $f'(s) \approx 1$ is actually
unnecessary. Since the metric $g^*_{\mu\nu}$ already generates a
Newtonian potential $-GM/r$, it suffices that the scalar field
contribution remain negligible (even at the post-Newtonian level)
in the solar system. One may try a function $f'(s)$ whose shape
looks like the one displayed in Fig.~\ref{fig4}, for instance
\begin{equation}
f'(s) = \varepsilon + \frac{\sqrt{|\bar s|}}{\left(1+|\bar
s|\right)^{1+1/n}},
\label{BadRAQUAL}
\end{equation}
where $n$ is a positive constant and as before $\bar s \equiv
\alpha^6 c^4 s/a_0^2$.
Then the transition occurs again at $\bar s \sim 1$, i.e., around
$r_\text{trans} = \alpha^2 \sqrt{GM/a_0}$, but Eq.~(\ref{eqPhiRAQUAL2})
shows that the force mediated by the scalar field reads $\alpha\,
\partial_r \varphi\, c^2 = (a_0/\alpha^2)\, (r/r_\text{trans})^n$ in
the large-$s$ limit [assuming that the $\varepsilon$ of
Eq.~(\ref{BadRAQUAL}) has a negligible influence]. Therefore, even if
$\alpha = 1$, so that the Newton-MOND transition occur at
$r_\text{trans} = \sqrt{GM/a_0}$ as expected, the anomalous scalar
force would be negligible with respect to post-Newtonian relativistic
effects for $r \ll r_\text{trans}$ and $n$ large enough. However, the
above scalar field $\partial_r\varphi \propto r^n$ happens
\textit{not} to be a solution of Eqs.~(\ref{eqPhiRAQUAL2}) and
(\ref{BadRAQUAL}) for $r < r_\text{trans}$, where $\varepsilon$
\textit{must} actually dominate. In such a case, the scalar force
takes the Newtonian form $\alpha \partial_r\varphi c^2 = \alpha^2
GM/r^2\varepsilon$ and is even increased by a factor $1/\varepsilon$
with respect to model (\ref{Epsilon}). This suffices to rule out the
class of models (\ref{BadRAQUAL}). One may also notice that for
$\varepsilon \rightarrow 0$, they do not satisfy the hyperbolicity
condition (b) in the large-$s$ limit unless $n = \infty$; but even for
this limiting case $n = \infty$, the scalar force is not negligible at
small distances.

In conclusion, the above discussion illustrates that RAQUAL models
are severely constrained, although they involve a free function
$f(s)$ defining the kinetic term of the scalar field. Contrary to
some fears in the literature, the possible superluminal
propagations do not threaten causality, and the two conditions (a)
and (b) are the only ones which must be imposed to guarantee the
field theory's consistency. For instance, monomials $f(s) = s^n$
are allowed if $n$ is positive and odd [except on the possible
hypersurfaces where $s$ vanishes, which would violate the strict
inequality (b)]. However, when taking into account simultaneously
these two consistency conditions \textit{and} experimental
constraints, it seems difficult to reproduce the MOND dynamics at
large distances without spoiling the Newtonian and post-Newtonian
limits in the solar system. It seems necessary to consider
unnatural functions $f(s)$ involving small dimensionless numbers.

In the next section, we will address the problem of light
deflection, and recall the solution which has been devised in the
literature. This solution will at the same time release the
experimental constraint on the matter-scalar coupling constant
$\alpha$ obtained from accurate measurement of the post-Newtonian
parameter $\gamma^{\textrm{PPN}}$ within the solar system, and
therefore cure the above fine-tuning problem. On the other hand, we
will show in Sec.~\ref{PSR} that the analysis of binary-pulsar data
also imposes a bound on $\alpha$, so that the problem of the
fine-tuning of the free function $f$ will anyway remain.

\section{The problem of light deflection}\label{Light}
\subsection{Conformally-coupled scalar field}\label{LightST}
\begin{figure}
\includegraphics[scale=0.75]{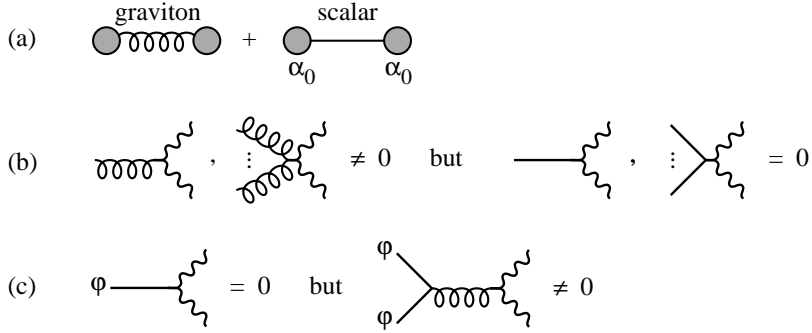}
\caption{Feynman diagrams in scalar-tensor theories, where
straight, curly and wavy lines represent respectively the scalar
field, gravitons and photons. Matter sources are represented by
blobs. (a)~Diagrammatic interpretation of the effective
gravitational constant $G_\text{\textrm{eff}} = G (1+\alpha_0^2)$,
where each vertex connecting matter to one scalar line involves a
factor $\alpha_0$. (b)~Photons are directly coupled to gravitons
but not to the scalar field. (c)~Photons feel nevertheless the
scalar field indirectly, via its influence on gravitons: The
energy-momentum tensor of the scalar field generates a curvature
of the Einstein metric $g^*_{\mu\nu}$ in which electromagnetic
waves propagate.} \label{fig5}
\end{figure}

Both scalar-tensor theories (\ref{TS1}) and RAQUAL models
(\ref{RAQUAL}) assume a conformal relation $\tilde g_{\mu\nu} =
A^2(\varphi) g^*_{\mu\nu}$ between the physical and Einstein
metrics. The inverse metrics are thus related by $\tilde
g^{\mu\nu} = A^{-2}(\varphi) g_*^{\mu\nu}$, and their
determinants read $\tilde g = A^{2d}(\varphi) g_*$ in $d$
spacetime dimensions. Since all matter fields, including gauge
bosons, are assumed to be minimally coupled to $\tilde
g_{\mu\nu}$, this leads to a simple prediction for the behavior
of light in such models. Indeed, the action of electromagnetism
is conformal invariant in $d=4$ dimensions:
\begin{eqnarray}
S_\text{EM} &=&
\int \frac{d^4x}{c}\,\frac{\sqrt{-\tilde g}}{4}
\tilde g^{\mu\rho}\tilde g^{\nu\sigma}
F_{\mu\nu}F_{\rho\sigma}
\nonumber\\
&=& \int \frac{d^4x}{c}\,\frac{A^4(\varphi)\sqrt{-g_*}}{4}
\left[A^{-2}(\varphi)g_*^{\mu\rho}\right]
\left[A^{-2}(\varphi)g_*^{\nu\sigma}\right]
F_{\mu\nu}F_{\rho\sigma}\nonumber\\
&=& \int \frac{d^4x}{c}\,\frac{\sqrt{-g_*}}{4}
g_*^{\mu\rho}g_*^{\nu\sigma}
F_{\mu\nu}F_{\rho\sigma}.
\label{EM}
\end{eqnarray}
Therefore, light is only coupled to the spin-2 field $g^*_{\mu\nu}$,
but does not feel at all the presence of the scalar field $\varphi$.
In terms of Feynman diagrams, there exist nonvanishing vertices
connecting one or several gravitons to two photon lines, but the
similar vertices connecting photons to scalar lines all vanish;
this is illustrated in line (b) of Fig.~\ref{fig5}.
It is then obvious that light behaves
strictly as in GR, in a geometry described by the Einstein metric
$g^*_{\mu\nu}$. In particular, the light deflection angle caused by
a spherical body of mass $M$ must be given by the same expression
as in GR (at lowest order)
\begin{equation}
\Delta\theta = \frac{4 GM}{b c^2},
\label{defl1}
\end{equation}
where $b$ denotes the impact parameter of the light ray.
An even simpler way to prove that light propagates in the
Einstein metric $g^*_{\mu\nu}$, without feeling the scalar
field, is to note that its geodesic equation reads $\tilde g_{\mu\nu}
dx^\mu dx^\nu = 0$ in the eikonal approximation. Dividing
by the nonvanishing factor $A^2(\varphi)$, this equation
implies $g^*_{\mu\nu} dx^\mu dx^\nu = 0$, giving thus the
standard geodesic equation for null rays in the metric
$g^*_{\mu\nu}$. [It is interesting to note that this second
reasoning would remain valid even if the action of electromagnetism
were not conformal invariant, for instance in dimension $d\neq4$,
or even if one multiplied it by an explicit scalar-photon
coupling function $B^2(\varphi)$, thereby violating the weak
equivalence principle. Then it is straightforward to prove
that the nonvanishing scalar-photon vertices would only
affect the amplitude of electromagnetic waves, in the
eikonal approximation, but not their polarization nor
their trajectory. One of us (G.E.F.) discussed this result
with B.~Bertotti several years ago, but it has been recently
rediscovered in \cite{Fujii:2006ic}.]

However, the crucial difference with GR is that massive matter
does feel the scalar field, via the matter-scalar coupling
function $A(\varphi)$. One may expand this function around the
background value $\varphi_0$ of the scalar field as
\begin{equation}
\ln A(\varphi) = \text{const.} + \alpha_0 (\varphi-\varphi_0)
+\frac{1}{2}\beta_0 (\varphi-\varphi_0)^2 + \cdots,
\label{lnA}
\end{equation}
where $\alpha_0$, $\beta_0$, \dots, are dimensionless constants.
In usual scalar-tensor theories (\ref{TS1}), as well as in the
Newtonian regime of RAQUAL models (\ref{RAQUAL}), the lowest-order
contribution of the scalar field to the gravitational potential
felt by a test mass reads then $-\alpha_0^2GM/r$
\cite{Willbook,DEF92}. This generalizes the results recalled in
Sec.~\ref{MONDRAQUAL} above in the particular case of an exponential
coupling function
$A(\varphi) = e^{\alpha \varphi}$. This spin-0 contribution adds
up to the standard Newtonian potential $-GM/r$ caused by the
spin-2 interaction (i.e., via the Einstein metric $g^*_{\mu\nu}$),
so that the total gravitational potential remains of the Newtonian
form but with an effective gravitational constant
$G_\text{\textrm{eff}} = G(1+\alpha_0^2)$; see line (a) of
Fig.~\ref{fig5}. In other words, when one determines the mass of
the Sun by analyzing the orbits of the planets, one actually
measures $G_\text{\textrm{eff}} M$ instead of the bare $GM$.
Therefore, if one expresses the light deflection angle
(\ref{defl1}) in terms of the observable quantity
$G_\text{\textrm{eff}} M$, one gets
\begin{equation}
\Delta\theta = \frac{4 G_\text{\textrm{eff}}M}{(1+\alpha_0^2)b
c^2} \leq \frac{4 G_\text{\textrm{eff}}M}{b c^2}. \label{defl2}
\end{equation}
For a given value of the observed constant $G_\text{\textrm{eff}}
M$, general relativity would have predicted $4
G_\text{\textrm{eff}}M/b c^2$. In conclusion, the light deflection
angle is actually \textit{smaller} in scalar-tensor theories than
in GR, although photons are strictly decoupled from the scalar
field! This paradoxical result is simply due to the fact that we
need to compare the measured angle $\Delta\theta$ with another
observable, $G_\text{\textrm{eff}} M$, which \textit{is}
influenced by the scalar field.

This reduction of light deflection in scalar-tensor theories has
been immediately recognized as a serious problem for MOND-like
field theories \cite{Bekenstein:1992pj,Bekenstein:1993fs}. Indeed,
if one interprets data within GR, the presence of dark matter
suggested by galaxy rotation curves is confirmed by weak lensing
observations~\cite{Fort:1994,Mellier:1998pk,Bartelmann:1999yn}.
Therefore, in a modified-gravity theory avoiding the assumption of
dark matter, one should predict simultaneously a larger
gravitational potential and a larger light deflection. However,
the interpretation of data is slightly more subtle than in the
solar system, because there exist at least three different notions
of mass which may be defined for a galaxy: (i)~its baryonic mass
$M_\text{b}$, assumed to be proportional to its luminous mass
$M_\text{L}$; (ii)~its total mass (baryonic plus ``dark'')
evaluated from its rotation curve, that we will denote
$M_\text{tot}^\text{rot}$; and (iii)~its total mass evaluated from
weak lensing, that we will denote $M_\text{tot}^\text{lens}$. The
only conclusion driven from the above result (\ref{defl2}) is that
the deflecting mass is smaller than the gravitational one in
scalar-tensor theories:
\begin{equation}
M_\text{tot}^\text{lens} \leq M_\text{tot}^\text{rot}.
\label{Mtot}
\end{equation}

Note that this conclusion is also valid in the
class of RAQUAL models reproducing the MOND dynamics
that we discussed in Sec.~\ref{MONDRAQUAL} above.
Indeed, the MOND force was caused by the exchange of
a scalar particle [cf. line (a) of Fig.~\ref{fig5}],
i.e., by the conformal factor $A^2(\varphi)$ relating
the physical metric $\tilde g_{\mu\nu}$ to the
Einstein one $g^*_{\mu\nu}$. Since light only feels the
latter, the total mass evaluated from weak lensing is
obviously smaller that the one evaluated from rotation
curves.

On the other hand, it should be underlined that Eq.~(\ref{defl2})
does \textit{not} imply $M_\text{tot}^\text{lens} \leq
M_\text{b}$, contrary to some erroneous or imprecise claims in the
literature (see, e.g., \cite{Bekenstein:1993fs}). Indeed, the
stress-energy tensor of the scalar field does contribute to the
curvature of the Einstein metric $g^*_{\mu\nu}$, so that light
deflection may be \textit{larger} than if only baryonic matter was
present. This is illustrated in terms of Feynman diagrams in line (c)
of Fig.~\ref{fig5}: Although the scalar-photon vertex vanishes, light
may nevertheless feel the scalar field indirectly via a graviton
exchange. The reason why this does not appear in
Eqs.~(\ref{defl1}) or (\ref{defl2}) above is because we derived
them at lowest order in powers of $1/c^2$, and in usual
scalar-tensor theories or in the Newtonian regime of RAQUAL
models. But when considering their MOND regime, one cannot any
longer perform a naive post-Newtonian expansion. Indeed, the
stress-energy of the scalar field may become non-negligible with
respect to that of baryonic matter, and even dominate it, although
this scalar field has been generated by its coupling to matter.
This mere sentence gives us a hint that such a model may be
unstable, but one should note that $M_\text{tot}^\text{lens} \gg
M_\text{b}$ is indeed possible in RAQUAL models.

We will provide an explicit example in Sec.~\ref{LightRAQUAL}
below, although it does not reproduce the right MOND dynamics.
Its aim is just to underline that the derivation of the
inequality $M_\text{tot}^\text{lens} \leq M_\text{b}$ proposed in
Ref.~\cite{Bekenstein:1993fs} is incorrect. It is instructive to
locate where its reasoning fails. It writes the deflection angle
in terms of the $T_r^r$ component of the total stress-energy
tensor, and decomposes it in two contributions, due to matter and
to the scalar field. It happens that the latter is negative. However,
what was identified as the matter contribution to $T_r^r$ happens
\textit{not} to vanish outside matter. Therefore, the decomposition
was flawed, and part of the \textit{positive} contribution of the
scalar field to light deflection was actually attributed to matter.
The simplest way to understand that a scalar field may indeed increase
light deflection is to consider it as a dark matter fluid, without any
coupling to baryonic matter and assuming GR as the correct theory of
gravity. If the scalar field is massive enough, it will obviously
cluster and contribute to the total gravitational potential deflecting
light.

Actually, a violation of inequality (\ref{Mtot}) is \textit{a priori}
also possible when the scalar's stress-energy contributes
significantly to the curvature of the Einstein metric $g^*_{\mu\nu}$.
Indeed, specific models could predict a negligible contribution to the
time component $g^*_{00}$ but a significant one to the spatial
components $g^*_{ij}$. In such a case, test particles would feel the
standard Newtonian potential $-\frac{1}{2}(1+g^*_{00})c^2 \approx
-GM/r$ [and the extra scalar force caused by the conformal factor
$A^2(\varphi)$, which may be tuned to be negligible], but light bending
would be directly affected by the scalar-induced corrections
entering $g^*_{ij}$. Therefore, $M_\text{tot}^\text{lens} >
M_\text{tot}^\text{rot}$ seems possible. However, such models may be
difficult to construct if one imposes the positivity of energy, as in
Ref.~\cite{Bekenstein:1993fs}. We will not attempt to study them here,
since there is anyway no experimental evidence for
$M_\text{tot}^\text{lens} > M_\text{tot}^\text{rot}$. Weak-lensing
observations only impose $M_\text{tot}^\text{lens} \sim
M_\text{tot}^\text{rot}$
\cite{Fort:1994,Mellier:1998pk,Bartelmann:1999yn}.

Let us end this section by recalling that the conclusion of
Ref.~\cite{Bekenstein:1993fs} remains correct in a limiting case
which is often implicitly assumed in MOND-like field theories
\cite{Soussa:2003vv,Soussa:2003sc}: If the stress-energy of the scalar
field is negligible with respect to that of baryonic matter, then the
curvature of the Einstein metric $g^*_{\mu\nu}$ is basically
generated by $M_\text{b}$ alone. Since light only feels
$g^*_{\mu\nu}$, weak lensing provides a measure of
$M_\text{tot}^\text{lens} \approx M_\text{b}$. In other words,
light is only deflected by baryonic matter, whereas experiment
tells us that the much greater amount of ``dark matter'' does
deflect it too. We will review in Secs.~\ref{Disformal} and
\ref{Stratified} how the same authors as
Ref.~\cite{Bekenstein:1993fs} devised a clever way to solve this
crucial difficulty of RAQUAL models.

\subsection{A RAQUAL example increasing light deflection}
\label{LightRAQUAL}

Let us illustrate that RAQUAL models may predict a much larger
light deflection than if there were baryonic matter alone in
GR, i.e., that $M_\text{tot}^\text{lens} \gg M_\text{b}$ is
possible. We consider the action
\begin{equation}
S = \frac{c^4}{16\pi G}\int\frac{d^4x}{c}\sqrt{-g_*}
\left\{
R^* - \ell_0^{2(n-1)} s^n\right\}
+S_\text{matter}\left[\psi; \tilde g_{\mu\nu} =
A^2(\varphi)g^*_{\mu\nu}\right],
\label{DEU}
\end{equation}
where as before $s\equiv
g_*^{\mu\nu}\partial_\mu\varphi\partial_\nu\varphi$, $\ell_0$ is a
length scale (e.g. $\ell_0 \propto c^2/a_0$ in MOND-like models),
and $n$ is a positive odd integer. As underlined at the end of
Sec.~\ref{Field}, such a model satisfies the two consistency
conditions (a) and (b), except on hypersurfaces where $s$ vanishes.
It suffices to add a small standard kinetic term $-\varepsilon s$
(with $\varepsilon > 0$) to the above action to cure the
singularity on such hypersurfaces. However, we will not take into
account this refinement because the above model will anyway not
reproduce the correct MOND dynamics.

The field equations deriving from action (\ref{DEU}) read
\begin{subequations}
\label{FieldEqRAQUAL}
\begin{eqnarray}
R^*_{\mu\nu}-\frac{1}{2}g^*_{\mu\nu}R^* &=&
\frac{8\pi
G}{c^4}\,\left(T_{\mu\nu}^{*\text{matter}} +
T_{\mu\nu}^{*\varphi}\right),
\label{EinsteinEqRAQUAL}\\
n\, \ell_0^{2(n-1)}
\nabla^*_\mu\left(s^{n-1}
\partial_*^\mu\varphi\right)&=&
-\frac{8\pi G}{c^4}\,\frac{d
\ln A(\varphi)}{d\varphi}\,T^*_\text{matter},
\label{ScalarEqRAQUAL}
\end{eqnarray}
\end{subequations}
where the stress-energy tensor of the scalar field
is given by
\begin{equation}
\frac{8\pi G}{c^4}\,T_{\mu\nu}^{*\varphi} =
\left(\ell_0^2\, s\right)^{n-1}
\left(n\, \partial_\mu\varphi \partial_\nu \varphi
-\frac{1}{2}\, s\, g^*_{\mu\nu}\right).
\label{TmunuScalar}
\end{equation}
Let us choose for instance an exponential (Brans-Dicke-like)
matter-scalar coupling function $A(\varphi) = e^{\alpha \varphi}$,
so that $d \ln A(\varphi)/d\varphi = \alpha$ is constant.
The solution of Eq.~(\ref{ScalarEqRAQUAL}), around a spherical
body of mass $M$, is then easy to write at lowest order
(i.e., neglecting the curvature of spacetime):
\begin{equation}
\varphi = \varphi_0 + \frac{2n-1}{2n-3}
\left(\frac{2\alpha GM\ell_0}{n r^2 c^2}\right)^{1/(2n-1)}
\frac{r}{\ell_0}.
\label{ScalarSolution}
\end{equation}
The stress-energy tensor of the scalar field reads thus
\begin{equation}
\frac{8\pi G}{c^4}\,T_{\mu\nu}^{*\varphi} =
\frac{\alpha GM}{\ell_0 r^2 c^2}
\left(\frac{2\alpha GM\ell_0}{n r^2 c^2}\right)^{1/(2n-1)}
\left(2\delta_\mu^r\delta_\nu^r -\frac{1}{n}g^*_{\mu\nu}\right),
\label{TmunuScalar2}
\end{equation}
and Einstein's equations (\ref{EinsteinEqRAQUAL}) can now
be solved straightforwardly, still to lowest order around
a flat Minkowski background. In particular, one may write
$(8\pi G/c^4)(T_{00}^{*\varphi} - \frac{1}{2}g^*_{00}T^*_\varphi)
= R^*_{00} \approx \frac{1}{2}\Delta \ln(-g^*_{00})$, and
this gives immediately
\begin{equation}
-g^*_{00} = 1 - \frac{2GM}{rc^2}
- \frac{(n-1)(2n-1)^2}{n(2n-3)}\,
\frac{\alpha GM}{\ell_0 c^2}
\left(\frac{2\alpha GM\ell_0}{n r^2 c^2}\right)^{1/(2n-1)}
+\mathcal{O}\left(\frac{1}{c^4}\right).
\label{g00}
\end{equation}
The $R^*_{rr}$ component of Einstein's equations can
now be used to solve for $g^*_{rr}$, but one must
specify the coordinate system to write the solution
explicitly. In order to compute easily the light deflection
angle below, it will be convenient to use the so-called
``Schwarzschild coordinates'' (which define an
``area radius'' $r$), such that the line element
takes the form
\begin{equation}
d s_*^2 =
-{\cal B}(r) c^2 dt^2 + {\cal A}(r) dr^2
+ r^2 (d\theta^2+\sin^2\theta\, d\phi^2).
\label{SchwarzschildForm}\end{equation}
One then finds straightforwardly
\begin{equation}
g^*_{rr} = {\cal A}(r) =
1 + \frac{2GM}{rc^2}
+\frac{\alpha GM}{n\ell_0 c^2}
\left(\frac{2\alpha GM\ell_0}{n r^2 c^2}\right)^{1/(2n-1)}
+\mathcal{O}\left(\frac{1}{c^4}\right).
\label{grr}
\end{equation}

The physical metric $\tilde g_{\mu\nu} = e^{2\alpha\varphi}
g^*_{\mu\nu}$ to which matter is coupled can now be deduced from
Eqs.~(\ref{ScalarSolution}), (\ref{g00}) and (\ref{grr}). In
particular, the gravitational potential felt by test masses reads
$V = -\frac{1}{2}(1+\tilde g_{00})c^2$, and its radial derivative
$\partial_r V$ gives their acceleration
\begin{equation}
a = \frac{GM}{r^2}
+ \left(\frac{2\alpha GM\ell_0}{n r^2 c^2}\right)^{1/(2n-1)}
\left[
\frac{\alpha c^2}{\ell_0}
+ \left(1-\frac{1}{n}\right) \left(1+\frac{2}{2n-3}\right)
\frac{\alpha GM}{\ell_0 r}
\right].
\label{accel0}
\end{equation}
For large enough values of the integer $n$, this model predicts
thus a MOND-like acceleration $\propto 1/r$, together with
the standard Newtonian one $GM/r^2$ and a constant contribution
$\alpha c^2/\ell_0$. This constant contribution comes from the
conformal factor $A^2(\varphi)$ entering the physical metric,
i.e., from a scalar exchange between the massive source $M$
and the test mass, as in line (a) of Fig.~\ref{fig5} above.
The standard Newtonian force comes from the curvature of the
Einstein metric $g^*_{\mu\nu}$ caused by the matter stress-energy
tensor. On the other hand, the $1/r$ force comes from the
curvature of $g^*_{\mu\nu}$ caused by the stress-energy
tensor of the scalar field, as illustrated in line
(c) of Fig.~\ref{fig5}. However, let us underline that
this model does \textit{not} reproduce the MOND dynamics,
first because the constant contribution happens to dominate
over the $1/r$ force, but above all because the coefficient
of this $1/r$ force is proportional to the baryonic mass $M$
instead of $\sqrt{M}$.

In order to compute the light deflection angle predicted by such a
model, let us use its exact integral expression \cite{Weinberg}
\begin{equation}
\Delta\theta = -\pi + 2\int_{r_0}^{\infty} \frac{dr}{r^2}
\left(\frac{\mathcal{A}(r)\mathcal{B}(r)}
{\mathcal{B}(r_0)/r_0^2 - \mathcal{B}(r)/r^2}\right)^{1/2},
\label{IntegralLight}\end{equation}
valid for any metric expressed in Schwarzschild
coordinates (\ref{SchwarzschildForm}).
In this integral (\ref{IntegralLight}), $r_0$ denotes
the smallest distance between the light ray
and the center of the deflecting body, which
may also be replaced by the impact parameter
$b = r_0 + \mathcal{O}(1/c^2)$ at lowest order.
Actually, since the coordinates we chose to write
Eq.~(\ref{grr})
are such that the Einstein metric $g^*_{\mu\nu}$ takes
the Schwarzschild form, this is no longer the case for
the physical metric $\tilde g_{\mu\nu} = A^2(\varphi)
g^*_{\mu\nu}$, because the radial dependence
of the scalar field (\ref{ScalarSolution}) spoils
the exact expressions $g^*_{\theta\theta} = r^2$
and $g^*_{\phi\phi} = r^2 \sin^2\theta$. However,
we saw in Sec.~\ref{LightST} that light is totally
insensitive to any global factor of the metric,
so that one may actually compute its deflection
directly from the Einstein metric $g^*_{\mu\nu}$.
The anomalous terms in Eqs.~(\ref{g00}) and
(\ref{grr}) are of the form
\begin{equation}
\mathcal{A}(r) = 1 + \frac{k_\mathcal{A}}{r^p}\ ,
\quad
\mathcal{B}(r) = 1 + \frac{k_\mathcal{B}}{r^p},
\label{g00grr}\end{equation}
where $p \equiv 2/(2n-1) = \mathcal{O}(1/n)$ is a small number,
and where $k_\mathcal{A} = \mathcal{O}(1/n)$ whereas
$k_\mathcal{B} = \mathcal{O}(n)$. For any value of these
coefficients, integral (\ref{IntegralLight}) gives
straightforwardly
\begin{equation}
\Delta\theta =
\left(\frac{k_\mathcal{A}}{p} - k_\mathcal{B}\right)
\frac{\sqrt{\pi}\,\Gamma\left(\frac{1+p}{2}\right)}{\Gamma(p/2)}\,
\frac{1}{b^p}
+ \mathcal{O}(k_\mathcal{A,B}^2),
\label{GeneralDeflection}
\end{equation}
and in our present case
\begin{subequations}
\label{LightDeflectionRAQUAL}
\begin{eqnarray}
\Delta\theta &=& \Delta\theta_\text{GR}
+ \frac{\pi}{2}\,
\frac{k_\mathcal{A} - p\,k_\mathcal{B}}{b^p}
+ \mathcal{O}(k_\mathcal{A,B}^2)
+ \mathcal{O}(p^2)
\label{LightDeflectionRAQUAL1}\\
&=& \frac{4 GM}{bc^2}
+\frac{\pi \alpha GM}{\ell_0 c^2}
\left(\frac{2\alpha GM\ell_0}{n b^2 c^2}\right)^{1/(2n-1)}
\left[1+\mathcal{O}\left(\frac{1}{n}\right)\right]
+\mathcal{O}\left(\frac{1}{c^4}\right).
\label{LightDeflectionRAQUAL2}
\end{eqnarray}
\end{subequations}
It is interesting to note that in Eq.~(\ref{LightDeflectionRAQUAL1}),
$p\,k_\mathcal{B} = \mathcal{O}(n^0)$ dominates over
$k_\mathcal{A} = \mathcal{O}(n^{-1})$, in spite of its
small factor $p$. For large values of $n$, the anomalous
contribution $\pi\alpha GM/\ell_0c^2$ to the
deflection angle (\ref{LightDeflectionRAQUAL2})
is precisely \textit{half} the result GR would have
predicted in presence of pressureless dark matter
reproducing the $1/r$ acceleration in (\ref{accel0})
[namely, $T^\text{dark}_{00} =
\alpha M c^2/4\pi \ell_0 r^2$
and $T^\text{dark}_{\mu i} = 0$].
This is due to the specific form of the scalar
stress-energy tensor (\ref{TmunuScalar2}), whose
spatial components $T^{*\varphi}_{ij}$ do not vanish.
But besides this subtle factor 2 discrepancy with
the standard dark matter prediction, let us also
recall that light is insensitive to the conformal
factor $A^2(\varphi)$, responsible for the dominant
anomalous contribution in (\ref{accel0}), namely the
constant one $\alpha c^2/\ell_0$. In conclusion,
using the notation introduced in Sec.~\ref{LightST},
the present model actually predicts
$M_\text{tot}^\text{lens} \ll M_\text{tot}^\text{rot}$,
which is even worse than the inequality (\ref{Mtot})
derived in standard scalar-tensor theories. On the
other hand, for a large enough impact parameter $b$, the light
deflection angle (\ref{LightDeflectionRAQUAL2}), $\Delta\theta
\approx \pi\alpha GM/\ell_0 c^2 = \text{const}$, is much larger
than the one predicted by GR in presence of $M$ alone, namely
$4GM/bc^2 \rightarrow 0$. Therefore, we have exhibited a RAQUAL
model predicting $M_\text{tot}^\text{lens} \gg M_\text{b}$.

One might try to refine the model (\ref{DEU}) by choosing
another matter-scalar coupling function, for instance
\begin{equation}
A(\varphi) = \text{exp}(\alpha\varphi^{-2n}),
\label{GoodAphi}\end{equation}
with the same integer $n$ as in the kinetic term of the
scalar field. A first useful consequence of such a
choice is that the extra constant acceleration we found
in Eq.~(\ref{accel0}) above becomes now negligible.
Indeed, it came from the conformal factor $A^2(\varphi)$
relating the physical and Einstein metrics, so that this extra
acceleration derives now from $\alpha\varphi^{-2n} c^2 =
\mathcal{O}(r^{-2n})$, and tends quickly towards 0
as $r$ grows. On the other hand, the $1/r$ acceleration
of Eq.~(\ref{accel0}) is still predicted, since
it comes from the curvature of the Einstein metric
caused by the scalar's stress-energy tensor $\propto 1/r^2$.
For large enough impact parameters, such a model predicts
thus
\begin{equation}
M_\text{tot}^\text{lens} \approx \frac{1}{2}
M_\text{tot}^\text{rot}\gg M_\text{b},
\label{PredictionDEU}\end{equation} the factor $\frac{1}{2}$
coming from the same calculation as
Eq.~(\ref{LightDeflectionRAQUAL}) above. The second consequence of
the choice (\ref{GoodAphi}) is that Eq.~(\ref{ScalarEqRAQUAL}) now
involves the non-constant coefficient $d\ln A(\varphi)/d\varphi =
-2n\alpha/\varphi^{2n+1}$. If one naively counts the powers of
$\partial_r\varphi$ and $\varphi$ entering this
Eq.~(\ref{ScalarEqRAQUAL}), one could then deduce that
$\varphi^{4n} \propto T^*_\text{matter} \propto M$. Using now
Eq.~(\ref{TmunuScalar}), this would imply $T_{\mu\nu}^{*\varphi}
\propto (\partial_r \varphi)^{2n} \propto \sqrt{M}$, which looks
like the dark matter MOND wishes to mimic. However, this would be
the case only if $\varphi \rightarrow 0$ as $r\rightarrow 0$,
within body $M$. Then one can prove that $\varphi \propto
\sqrt{r}$ for small radii, and $\int d^3x T_{\mu\nu}^{*\varphi}$
would diverge within the body. Therefore, such a solution is
unphysical. On the other hand, if $\varphi \rightarrow \varphi_0
\neq 0$ within the body, then one gets back an exterior solution
of the form (\ref{ScalarSolution}), where $\alpha$ is replaced
here by $-2n\alpha/\varphi_0^{2n+1}$, so that the anomalous force
is \textit{a priori} again proportional to $M$ instead of the
looked-for $\sqrt{M}$. There could still remain a possibility to
recover the MOND force with such a RAQUAL model, if one could prove
that $\varphi \rightarrow \varphi_0 \propto M^{1/4n}$ within the body.
This would be the case for instance if $\varphi \rightarrow 0$ at a
given radius, say at the surface of the body. However, this would
correspond to a diverging matter-scalar coupling strength at this
radius (singular and obviously forbidden by solar-system tests), and
we anyway do not see how to impose such a relation, nor the precise
value of $\varphi_0$. In standard scalar-tensor theories, the scalar
field tends towards a constant $\varphi_0$ when $r\rightarrow\infty$,
this constant being imposed by the cosmological evolution of the
Universe, and being constrained by post-Newtonian tests of gravity in
the solar system. Here, the scalar field $\varphi = \varphi_0 +
\text{const.}\, r^{1-2/(2n-1)}$ diverges at infinity, as soon as a
single particle exists in the Universe. This pathological behavior
is linked with the fact that the solution of
Eq.~(\ref{ScalarEqRAQUAL}) cannot be written unambiguously, the
integration constant $\varphi_0$ remaining undetermined. In the
model (\ref{ScalarSolution}), where $A(\varphi) =
e^{\alpha\varphi}$, this constant did not enter any physical
observable, but it does when the matter-scalar coupling function
is no longer a mere exponential, as in our present attempt
(\ref{GoodAphi}). Therefore, although this matter-scalar coupling
function looks promising to reproduce the MOND dynamics from a
simple RAQUAL model, it fails to define a fully predictive theory.
Its only solid prediction is that light bending would be
\textit{twice smaller} than the prediction of GR in presence of
dark matter, cf. Eq.~(\ref{PredictionDEU}). As far as we know,
this is still allowed by weak-lensing observations
\cite{Fort:1994,Mellier:1998pk,Bartelmann:1999yn}.

In conclusion, we have exhibited RAQUAL models
predicting a much larger light deflection than
GR in presence of baryonic matter alone, but this
framework does not seem to reproduce the MOND
dynamics consistently. Note that the class of RAQUAL theories
examined in the present subsection~\ref{LightRAQUAL}
are actually models of \textit{dark matter}.
Extra Newtonian force and light deflection are
caused by the stress-energy tensor of the scalar
field, which plays thus the role of dark matter.
The only difference with standard dark matter
models is that its distribution is strictly
imposed by the one of baryonic matter, whereas
they are usually only related via the dynamical
evolution of large-scale structures.

\subsection{Disformal coupling}\label{Disformal}

A nice trick exists to increase light deflection within
scalar-tensor theories or RAQUAL models. It is similar to Ni's
``stratified theory of gravity'' \cite{Ni,Willbook}, and was
introduced by Bekenstein in \cite{Bekenstein:1992pj}; see also
\cite{Bekenstein:1993fs}. Instead of assuming that matter is
coupled to a physical metric $\tilde g_{\mu\nu} = A^2(\varphi)
g^*_{\mu\nu}$, conformally related to the Einstein metric
$g^*_{\mu\nu}$, one may consider a ``disformal'' relation of the
form
\begin{equation}
\tilde g_{\mu\nu} = A^2(\varphi,s) g^*_{\mu\nu} + B(\varphi,s)
\partial_\mu \varphi
\partial_\nu \varphi,
\label{Disformal1}
\end{equation}
where as before $s \equiv g_*^{\mu\nu} \partial_\mu \varphi
\partial_\nu \varphi$. The crucial difference is that the $
\partial_\mu \varphi \partial_\nu \varphi$ contribution now
privileges a particular direction, namely the radial one in
spherically symmetrical systems. This extra contribution may remain
negligible in the equations of motion of test particles, but now
light can be directly sensitive to the presence of the scalar
field, contrary to the case of conformal coupling that we discussed
in Sec.~\ref{LightST}.

Let us emphasize that because of the disformal coupling above,
light (or matter) can travel faster or slower than gravitons
depending on the sign of $B$. In a similar way as in
Sec.~\ref{SecRAQUAL}, this does not imply a violation of
causality, provided that the matter metric is still of Lorentzian
signature. The matter metric Eq.~(\ref{Disformal1}) is causal if
\begin{equation}
A^{2}(\varphi,s) > 0\quad\text{and}\quad
A^{2}(\varphi,s)+ s B(\varphi,s)>0
\label{condDisformal}
\end{equation}
for all $\varphi$ and $s$. Notice that if $A$ does not depend on
$s$, $B$ must depend on $s$ for the latter inequality to hold.
Beware that these kind of models used in the literature as
realizations of varying speed of light scenarios
\cite{Clayton:1999zs,Bassett:2000wj} do not respect this last
inequality. Indeed, whereas $A$ was taken to be 1, $B$ was chosen
constant $B=-L^2$ with $L$ being some fixed length scale. Therefore
matter field equations are elliptic at local scales ($s>1/L^2$),
and do not have a well-posed Cauchy problem.

By analogy with the Schwarzschild metric, whose positive
contribution in the radial component $g_{rr} = 1+2
GM/rc^2+\mathcal{O}(1/c^4)$ contributes positively to light
deflection (together with the standard Newtonian potential involved
in $g_{00}$), one understands intuitively that positive values of
$B(s,\varphi)$ should increase light deflection. To prove it, let
us first recall again that light is insensitive to any global
factor of the metric, because of the conformal invariance of
electromagnetism in 4 dimensions, Eq.~(\ref{EM}). Therefore,
instead of the metric (\ref{Disformal1}), one may equivalently
consider $\bar g_{\mu\nu} \equiv g^*_{\mu\nu} + (B/A^2)
\partial_\mu \varphi \partial_\nu\varphi$. Let us then assume that
the coordinates have been chosen so that the Einstein metric
$g^*_{\mu\nu}$ takes the Schwarzschild form
(\ref{SchwarzschildForm}). Since, in a static and spherically
symmetric situation, the disformal contribution $(B/A^2)
\partial_\mu \varphi \partial_\nu\varphi$ only contributes to the
radial component $\bar g_{rr}$, the metric $\bar g_{\mu\nu}$ is
also of the Schwarzschild form, and one may thus use integral
(\ref{IntegralLight}) to compute the light deflection angle. The
only difference with GR is that the radial metric component now
reads $\mathcal{A} = \mathcal{A}_\text{GR} +
(B/A^2)(\partial_r\varphi)^2$. [Beware not to confuse the metric
components $\mathcal{A} \equiv \bar g_{rr}$ and $\mathcal{B} \equiv
-\bar g_{00}$ defined in (\ref{SchwarzschildForm}) with the
scalar-field functionals $A(s,\varphi)$ and $B(s,\varphi)$ entering
(\ref{Disformal1}).] At first post-Newtonian order, all
contributions coming from $g^*_{\mu\nu}$ reproduce the general
relativistic result, and the only extra contribution is thus the
one proportional to $(B/A^2)(\partial_r\varphi)^2$, assumed to be
of order $\mathcal{O}(1/c^2)$ too:
\begin{equation}
\Delta\theta = \Delta\theta_\text{GR}
+ \int_{r_0}^{\infty} \frac{dr/r}{\sqrt{r^2/r_0^2-1}}\,
\frac{B\,(\partial_r\varphi)^2}{A^2}
+ \mathcal{O}\left(\frac{1}{c^4}\right).
\label{IntegralLightDisformal}\end{equation}
Even without performing the explicit integration
in particular models, it is thus clear that a positive
value of $B(s,\varphi)$ always gives a positive
contribution to light bending, since the extra term
in (\ref{IntegralLightDisformal}) is the integral of
a positive function.

For instance, if one tunes the model such that
$(B/A^2)(\partial_r\varphi)^2 = 4\sqrt{GMa_0}/c^2$
is independent of $r$ (as we will do in Sec.~\ref{Nonmin}
below), then
\begin{equation}
\Delta\theta = \Delta\theta_\text{GR} +
\frac{2\pi\sqrt{GMa_0}}{c^2} +
\mathcal{O}\left(\frac{1}{c^4}\right),
\label{LightDeflectionDM}
\end{equation}
and the second term reproduces exactly the deflection GR would
have predicted in presence of dark matter (which is independent
of the impact parameter,\footnote{Note that although the
deflection angle is constant, the \textit{direction} of
deflection is radial with respect to the deflecting body, so that
distant objects are deformed and weak lensing observations can be
performed. A constant deflection angle in a constant direction
would have just shifted globally the image of the distant
objects, without any deformation.} as if there were a deficit
angle in a flat conical space). Indeed, it is straightforward to
prove that a dark matter density $T_{00}/c^2 = \sqrt{GMa_0}/4\pi
Gr^2$ generates a logarithmic MOND-like potential in the time
component of the metric, $-\tilde g_{00} = 1 - 2 GM/rc^2 +
2\sqrt{GMa_0}\ln r/c^2 +\mathcal{O}(1/c^4)$, and that the
corresponding radial metric reads $\tilde g_{rr} = 1 + 2GM/rc^2 +
2\sqrt{GMa_0}/c^2 +\mathcal{O}(1/c^4)$ in Schwarzschild
coordinates. A similar calculation as above then yields the same
light deflection angle (\ref{LightDeflectionDM}) [see
Eq.~(\ref{IntegralLightDisformal2}) below for an explicit
derivation]. Note in passing that $\tilde g_{rr} \neq -1/\tilde
g_{00}$ in such coordinates, contrary to the standard
Schwarzschild solution in vacuum. On the other hand, if one
rewrites this same metric in isotropic coordinates, $d\tilde s^2
= -\tilde g_{00} c^2 dt^2 + \tilde g_{\rho\rho}
\left(d\rho^2+\rho^2 d\theta^2 + \rho^2\sin^2\theta
d\phi^2\right)$, one indeed recovers the usual relation $\tilde
g_{\rho\rho} = -1/\tilde g_{00} +\mathcal{O}(1/c^4) = 1 +
2GM/\rho c^2 - 2\sqrt{GMa_0}\ln\rho/c^2 +\mathcal{O}(1/c^4)$.
This remark reminds us that a metric is coordinate-dependent, and
that observable quantities cannot be read naively in its
coefficients. Even the signs of the predicted effects are not
obvious. For instance, if one had naively tried to impose
$A^2(s,\varphi) = (1 + 2\sqrt{GMa_0}\ln r/c^2)$ but $\tilde
g_{rr}= -1/\tilde g_{00}$ in Eq.~(\ref{Disformal1}), then one
would have needed a negative $B(s,\varphi)$, and the predicted
deflection angle would have been negative (and diverging),
$\Delta\theta = \Delta\theta_\text{GR} -2\pi\sqrt{GMa_0}\ln
2b/c^2$.

Equation (\ref{IntegralLightDisformal}) proves that a positive
value of $B(s,\varphi)$ always increases light deflection.
However, light rays satisfy then $\tilde g_{\mu\nu} dx^\mu dx^\nu
= 0 \Leftrightarrow g^*_{\mu\nu} dx^\mu dx^\nu = - B/A^2
(\partial_\mu \varphi dx^\mu)^2 \leq 0$, and photons are thus
timelike with respect to the Einstein metric $g^*_{\mu\nu}$. This
has been considered as a serious flaw by the authors of
Refs.~\cite{Bekenstein:1992pj,Bekenstein:1993fs}. Indeed, this
means that gravitons (i.e., perturbations of the Einstein metric
$g^*_{\mu\nu}$) are superluminal, and it was believed that this
implied causality violations. Reference \cite{Bekenstein:1993fs}
even presents the reasoning the other way: If one imposes from the
beginning that no field (even gravity) can propagate faster than
light, then a disformal coupling (\ref{Disformal1}) needs $B \leq
0$, therefore light deflection is even \textit{smaller} that the
one predicted in standard scalar-tensor or RAQUAL models, see
Sec.~\ref{LightST}. This sufficed for the authors of these
references to discard such models. Actually, as already discussed
in Sec.~\ref{SecRAQUAL}, such a superluminal propagation does not
threaten causality. This is even clearer in the present case,
since one is used to write the field equations in terms of the
Einstein metric $g^*_{\mu\nu}$ (i.e., in what is usually called
the ``Einstein frame''), in order to diagonalize the kinetic terms
and prove that the Cauchy problem is well posed. To guarantee
causality, like in GR, one needs to assume that spacetime does not
involve any closed timelike curve with respect to $g^*_{\mu\nu}$.
Since the causal cone of matter, defined by the physical metric
$\tilde g_{\mu\nu}$, is always interior to the one defined by
$g^*_{\mu\nu}$ if $B>0$, there cannot exist any CTC for any matter
field. In conclusion, the nice trick of disformal coupling,
Eq.~(\ref{Disformal1}), is consistent from the point of view of
causality.

It remains however to study its consistency within matter. Indeed,
the physical metric $\tilde g_{\mu\nu}$ is involved in many places
of the matter action, to contract indices, to define covariant
derivatives, and in the volume element $\sqrt{-\tilde{g}}\, d^4x$.
One should thus take into account the matter contribution $\delta
S_\text{matter}[\psi; \tilde g_{\mu\nu}]/\delta \varphi$ in the
Euler-Lagrange equation (\ref{eqPhiRAQUAL}) for $\varphi$, and
study simultaneously the matter field equations $\delta
S_\text{matter}[\psi; \tilde g_{\mu\nu}]/\delta \psi = 0$ (where
as before $\psi$ denotes any matter field). One should then
identify any second derivative of $\varphi$ which appears in them,
and prove that the full set of field equations, for matter, the
scalar field and the Einstein metric $g^*_{\mu\nu}$, has a
well-posed Cauchy problem. This is usually done by diagonalizing
the kinetic terms, i.e., by separating all second derivatives in
the left-hand side of the field equations, and by proving that
their differential operator is hyperbolic. In the general case of
a theory involving any kind of matter, this is quite involved and
we postpone this crucial study to a future work. Indeed, if one
considers an action of the form (\ref{RAQUAL}) with a physical
metric (\ref{Disformal1}), the scalar field equation reads
\begin{eqnarray}
&&\sqrt{-g^*}\, \nabla_{\mu}^* \left[ \frac{\partial f}{\partial s}\,
\nabla^{\mu}_* \varphi
\right]\nonumber\\
&&- \frac{4 \pi G}{c^4}\,
\partial_{\mu}\biggl\{\sqrt{-\tilde{g}}\biggl[
B\, \tilde{T}^{\mu\nu} + 2\,\frac{\partial\ln A}{\partial s}\,
\tilde{T} g^{\mu\nu}_*
+\left(\frac{\partial B}{\partial s}- 2B\, \frac{\partial
\ln A}{\partial s} \right)\tilde{T}^{\rho\sigma} \partial_{\rho}
\varphi \partial_{\sigma} \varphi g^{\mu\nu}_*
\biggl] \partial_{\nu} \varphi \biggr\}\nonumber\\
&&=
\sqrt{-g^*} \left(\frac{1}{2}\,\frac{\partial f}{\partial
\varphi} + \frac{\partial V}{\partial\varphi}\right)
-\frac{4 \pi G}{c^4}
\sqrt{-\tilde{g}}\left[\frac{\partial \ln A}{\partial
\varphi}\, \tilde{T} +
\left(\frac{1}{2}\,\frac{\partial B}{\partial \varphi}
- B \frac{\partial
\ln A}{\partial \varphi} \right)
\tilde{T}^{\rho\sigma} \partial_{\rho} \varphi
\partial_{\sigma} \varphi\right],\nonumber\\
\label{EqPhidanslamatiere}
\end{eqnarray}
where $\tilde{T}^{\mu\nu} \equiv
\left(2c/\sqrt{-\tilde{g}}\right) \delta S_\text{matter}/\delta
\tilde{g}_{\mu\nu}$ is the physical stress-energy tensor, $\tilde
T\equiv \tilde{T}^{\mu\nu}\tilde{g}_{\mu\nu}$, and $\tilde{g} =
A^8(1+sB/A^2)g_*$ from definition (\ref{Disformal1}). Note
that Eq.~(\ref{EqPhidanslamatiere}) reduces to
Eq.~(\ref{eqPhiRAQUAL2}) for $A = e^{\alpha\varphi}$ and $B=0$. The
right hand side of the above equation does not involve any second
derivative and is merely a source term [if $\partial A/ \partial
\varphi$ were vanishing, $\partial_{\mu} \varphi = 0$ would always be a
solution]. The left hand side however contains second derivatives
of the scalar field \textit{and} of matter fields, through terms
like $\partial \tilde{T}$. The equation is not diagonal and,
moreover, the effective metric multiplying the second derivative
of $\varphi$ has a rather complicated expression that notably
involves terms like $\tilde{T}^{\mu\nu}$, $\tilde{T}^{\rho\sigma}
\partial_{\rho} \varphi \partial_{\sigma} \varphi$. The
Lorentzian signature of the latter is therefore a quite involved
question, even if we expect that it may depend on generic energy
conditions in the matter sector. Note that the literature on such
MOND-like field theories did \textit{not} discuss the consistency
of the field equations within matter. We will not examine either
the fully general case in the present paper, but we wish to
underline that this \textit{should} be done. We will however
analyze below the particular case where matter is described by a
pressureless and isentropic perfect fluid. Although much simpler
than the general case, it is relevant for most matter in the
universe, and exhibits the subtle conditions which must be
imposed on the coupling constants $A$ and $B$ for the scalar
field equation to remain always consistent. As far as we are
aware, this is the first time such a discussion is presented.

The action defining a pressureless and isentropic perfect fluid
reads $S_\text{matter} = -\int d^4 x \sqrt{-\tilde g}\,
\tilde\rho c$, where $\tilde\rho$ is the conserved matter density
as observed in the Jordan frame. If $\tilde u^\lambda \equiv d
x^\lambda / \sqrt{-\tilde g_{\mu\nu} dx^\mu dx^\nu}$ denotes the
matter unit 4-velocity, one has thus $\tilde
\nabla_\lambda(\tilde \rho \tilde u^\lambda) = 0$, which can be
equivalently written in terms of a partial derivative as
$\partial_\lambda(\sqrt{-\tilde g}\, \tilde\rho \tilde u^\lambda)
= 0$. This equation implies that the matter current density
$\sqrt{-\tilde g}\, \tilde\rho \tilde u^\lambda$ is determined
everywhere once initial conditions are given at a point of the
flowlines, and it is therefore unchanged if the metric $\tilde
g_{\mu\nu}$ is varied: $\delta(\sqrt{-\tilde g}\, \tilde\rho
\tilde u^\lambda) /\delta \tilde g_{\mu\nu} = 0$. In particular,
the coordinate-conserved density\footnote{The
coordinate-conserved density $\bar\rho$ is often denoted as
$\rho^*$ in the general relativistic literature, but this star
index would be confusing in our present context where
Einstein-frame quantities already bear such an index.} $\bar\rho
\equiv \sqrt{-\tilde g}\, \tilde\rho \tilde u^0$ is independent
of the Jordan metric $\tilde g_{\mu\nu}$, and thereby independent
of both the Einstein metric $g^*_{\mu\nu}$ and the scalar field
$\varphi$ (as well as its derivatives $\partial_\mu\varphi$). The
pressureless perfect fluid action may thus be written as
$S_\text{matter} = -\int d^4 x\, \bar\rho c/\tilde u^0 = -\int
d^3 x\, \bar\rho c\sqrt{-\tilde g_{\mu\nu} dx^\mu dx^\nu}$, which
does depend explicitly on the scalar field via the Jordan metric
(\ref{Disformal1}), and therefore contributes to the scalar-field
dynamics within matter. Introducing the Einstein-frame unit
4-velocity $u_*^\lambda \equiv d x^\lambda / \sqrt{-g^*_{\mu\nu}
dx^\mu dx^\nu}$ (independent of $\varphi$ and its derivatives),
one may thus write the action of the scalar field as
\begin{equation}
S_\varphi = -\frac{c^4}{8\pi G}
\int\frac{d^4 x}{c}\sqrt{-g_*}\bigl[f(s,
\varphi)+2V(\varphi)\bigr] - \int \frac{d^4 x}{c} \frac{\bar\rho
c^2}{u_*^0}
\sqrt{A^2(\varphi,s) - (u_*^\mu\partial_\mu\varphi)^2 B(\varphi,s)},
\label{Sphimatter}
\end{equation}
where all the scalar-field dependences are now explicitly exhibited
(neither $\sqrt{-g_*}$ nor $\bar\rho c^2 / u_*^0$ depend on it).
In conclusion, the dynamics of the scalar field within matter is
described by an action generalizing the RAQUAL form (\ref{RAQUAL}),
in which the function $f(\varphi, s)$ is replaced by
\begin{equation}
\tilde f(\varphi, s, u_*^\mu\partial_\mu\varphi) =
f(\varphi,s) +\frac{8\pi G\bar\rho}{\sqrt{-g_*}\,u_*^0c^2} A
\sqrt{1-(u_*^\mu\partial_\mu\varphi)^2 B/A^2},
\label{ftilde}
\end{equation}
where $A(\varphi, s)$ and $B(\varphi, s)$ are themselves
functions of the scalar field and $s \equiv g_*^{\mu\nu}
\partial_\mu \varphi \partial_\nu \varphi$. Since the derivatives
of the scalar field do not enter only via $s$ but also via
$u_*^\mu\partial_\mu\varphi$, the conditions that such a function
$\tilde f$ must satisfy to define a consistent scalar-field
equation (hyperbolicity and Hamiltonian bounded by below) are
significantly more complicated than (a) and (b) of
Sec.~\ref{SecRAQUAL} above [that $f$ itself must still satisfy to
define a consistent scalar-field equation in vacuum]. Using as before
a prime to denote partial derivation with respect to $s$, i.e.,
$\tilde f' \equiv \partial \tilde f(\varphi, s,
u_*^\mu\partial_\mu\varphi) / \partial s$, and introducing the
shorthand notation $\hat\rho \equiv 2\pi
G\bar\rho/\sqrt{-g_*}\,u_*^0c^2$, $C \equiv
B/\sqrt{A^2-(u_*^\mu\partial_\mu\varphi)^2 B}$, and $D \equiv C +
(u_*^\mu\partial_\mu\varphi)^2 C^3/B$, the hyperbolicity conditions
for the scalar field equation within matter may be written as
\begin{enumerate}
\item[(a1)]$\tilde f' > 0$,
\item[(b1)]$s \tilde f'' + \tilde f'
+ \hat\rho D - 4 \hat\rho (u_*^\mu\partial_\mu\varphi)^2 C'$
\item[]$\pm \left\{
\left(s\tilde f'' - \hat\rho D\right)^2
- 4\hat\rho\, (u_*^\mu\partial_\mu\varphi)^2
\left(\tilde f'' + 2\hat\rho C'\right)
\bigl(2 s C' + D \bigr) \right\}^{1/2} > 0$.
\end{enumerate}
Note that in condition (b1), the inequality must be satisfied for
both signs of the square root, and it suffices thus to check its
lowest value, involving a minus sign.
At the lowest post-Newtonian level, in the limit
$(u_*^\mu\partial_\mu\varphi)^2 \ll |s|$, one may simplify these
conditions as
\begin{enumerate}
\item[(a2)]$\tilde f' > 0$,
\item[(b2)]$2 s \tilde f'' + \tilde f' > 0$,
\item[(c2)]$\tilde f' + 2\hat\rho C > 0$.
\end{enumerate}
If $B$ is chosen positive in Eq.~(\ref{Disformal1}), in order to
increase light deflection in the MOND regime, then $\hat\rho C >
0$ and condition (c2) is implied by (a2). On the other hand, in
the ultrarelativistic regime, or in the cosmological one where
$|\partial_0\varphi| \gg |\partial_i\varphi|$, conditions (a1)
and (b1) reduce to
\begin{enumerate}
\item[(a3)]$\tilde f' > 0$,
\item[(b3)]$2 s \tilde f'' + \tilde f'
+ 2\hat \rho\left(4 s C'+ D\right) > 0$.
\end{enumerate}
Note however that conditions (a1) and (b1) should in principle be
satisfied in any situation (within matter), in order for the model
to always remain a well-defined field theory.

In addition to (a1) and (b1), the contribution of the scalar
field to the Hamiltonian should also be bounded by below. In the
rest frame of the perfect fluid, such that $u_*^\mu = (1,0,0,0)$,
and simultaneously with coordinates which diagonalize locally the
Einstein metric $g^*_{\mu\nu} = \text{diag}(-1,1,1,1)$, this
contribution is proportional to $\tilde f +
2(\partial_0\varphi)^2(\tilde f' + 2\hat\rho C)$. The general
conditions needed to ensure that it is bounded by below are more
involved. It suffices to check that this is indeed the case for
each particular model that one may consider.

Let us finally mention that no extra condition is necessary to
ensure that the argument of the square root is positive in
Eqs.~(\ref{Sphimatter})-(\ref{ftilde}). Indeed, since matter is
assumed to be metrically coupled to the Jordan metric $\tilde
g_{\mu\nu}$, we \textit{know} that $\tilde g_{\mu\nu} \tilde
u^\mu \tilde u^\nu = -1$, i.e., that the matter 4-velocity must
be timelike with respect to $\tilde g_{\mu\nu}$. This implies
$1-(u_*^\mu\partial_\mu\varphi)^2 B/A^2 > 0$, which should be
understood as a condition (automatically) satisfied by the
Einstein-frame 4-velocity $u_*^\mu$, but not as a constraint on
the functions $A(\varphi, s)$ and $B(\varphi, s)$.

Besides this important question of the consistency of the scalar
field equation within matter, it also remains to study the
precise predictions of this class of models. Specific examples
have been analyzed in
Refs.~\cite{Bekenstein:1992pj,Bekenstein:1993fs,Clayton:1999zs,Bassett:2000wj}.
We will examine a very different one in Sec.~\ref{NonminST}
below, but still within this class of disformally coupled
scalar-tensor theories. We will notably show that the predicted
light deflection can be consistent both with weak-lensing
determinations of dark matter and with post-Newtonian tests in
the solar system.

\subsection{Stratified theories}\label{Stratified}
Because the above superluminal propagation of gravity was
considered as deadly, Sanders proposed in
\cite{Sanders:1996wk,Sanders:2005vd} another kind of disformal
coupling, where the physical metric to which matter is universally
coupled takes the form
\begin{equation}
\tilde g_{\mu\nu} = A^2(\varphi) g^*_{\mu\nu}
+ B(\varphi) U_\mu U_\nu.
\label{Disformal2}
\end{equation}
This is thus strictly the same form as Eq.~(\ref{Disformal1})
above, but the gradient $\partial_\mu \varphi$ has been replaced
by a vector field $U_\mu$ (not to be confused with the matter unitary
velocities $\tilde u^\mu$ and $u_*^\mu$ introduced in
Sec.~\ref{Disformal} above). Bekenstein's theory TeVeS
\cite{Bekenstein:2004ne,Bekenstein:2004ca} also assumes the above
disformal relation. The dynamics of the vector field is defined by a
specific kinetic term of the Proca form
$(\partial_{[\mu} U_{\nu]})^2$, and its norm is imposed to be
$g_*^{\mu\nu} U_\mu U_\nu = -1$ thanks to a Lagrange parameter.
The reason why it is necessary to fix its norm is because the
magnitude of scalar-field effects would otherwise depend on the
background norm of the vector field. Since backgrounds where
$U_\mu = 0$ would be possible, one would not predict any extra
light deflection caused by the scalar field in such situations.
Moreover the Lagrange multiplier ensures that the vector field
equation is consistent with any type of matter one considers.
Indeed, without the Lagrange multiplier, the equation of motion
reads
\begin{equation}
\label{VecteurSansLambda} \nabla_{\mu}^*(\nabla_*^{[\mu} U_*^{\nu]})
\propto \frac{\sqrt{-\tilde{g}}}{\sqrt{-g_*}} \tilde{T}^{\rho \nu}
U_{\rho},
\end{equation}
where $U_*^\nu \equiv g_*^{\mu\nu} U_\mu$, and its divergence imposes
that the stress-energy tensor of matter must satisfy the non-trivial
property $\nabla_{\nu}^*(\tilde{T}^{\rho \nu}
U_{\rho}\sqrt{-\tilde{g}}/\sqrt{-g_*})=0$, that may be satisfied
only by very particular type of matter. On the other hand, a term like
$\lambda (U_{\mu} U_*^{\mu}+1)$ in the action, where $\lambda$ is a
Lagrange multiplier, gives rise to a term proportional to $\lambda
U_*^{\nu}$ in the left hand side of Eq.~(\ref{VecteurSansLambda}),
whose divergence equals the \textit{a priori} non vanishing divergence
of the right hand side, thus giving its value to $\lambda$. Note
finally that the consistency of the field equations requires that
the above matter metric (\ref{Disformal2}) has a Lorentzian
signature. This condition reads $A^{2}>0$ and $A^{2}>B$ for all
$\varphi$, where we used $U_{\mu} U_*^{\mu}=-1$.

One could think that the vector field is chosen timelike (negative
norm), contrary to the gradient $\partial_\mu \varphi$ in the
vicinity of clustered matter, in order to preserve isotropy of
space which is very well tested experimentally. However, this is
not the case, since a timelike vector does define a preferred
frame (i.e., an ``ether''), namely the one where it takes the
canonical form $U_\mu = (\pm 1,0,0,0)$. We will briefly discuss
the related experimental difficulties in Sec.~\ref{Preferredframe}
below.

The bonus added by this vector field $U_\mu$, with respect to the
purely scalar case (\ref{Disformal1}), is threefold: (i)~it
allows us to increase light deflection without necessitating
superluminal propagation of gravity; (ii)~it can also be used to
tune a specific kinetic term for the scalar field, in order to
avoid its own superluminal propagation identified in
Sec.~\ref{MONDRAQUAL} above; (iii)~the dynamics of the various
fields is easier to analyze even within matter, and there does
not seem to exist any generic difficulty with the Cauchy problem.
Of course, the third point is actually the only useful one, since
we know that superluminal propagation does not threaten causality
provided there always exists a nonvanishing exterior to the union
of all causal cones, as discussed in Sec.~\ref{SecRAQUAL} above
and in Refs.~\cite{Bruneton:2006gf,Babichev:2007dw}. However, the real
reason why this vector was considered in
Refs.~\cite{Sanders:1996wk,Bekenstein:2004ne,Bekenstein:2004ca,
Sanders:2005vd} is indeed to avoid any superluminal propagation.
Point (ii) is easy to understand: If the kinetic term of the
scalar field is defined as $(g_*^{\mu\nu} - U_*^\mu U_*^\nu)
\partial_\mu \varphi \partial_\nu\varphi$, instead of the
standard $s = g_*^{\mu\nu} \partial_\mu \varphi
\partial_\nu\varphi$, then a violation of condition (c) of
Sec.~\ref{SecRAQUAL} only implies that the scalar causal cone is
exterior to the one defined by the inverse of $g_*^{\mu\nu} -
U_*^\mu U_*^\nu$, namely $g^*_{\mu\nu} +\frac{1}{2} U_\mu U_\nu=
\text{diag}(-\frac{1}{2},1,1,1)$ in the preferred frame and in
locally inertial coordinates. It becomes thus possible to
reproduce the MOND dynamics while avoiding a propagation of
$\varphi$ outside the causal cone defined by $g^*_{\mu\nu} =
\text{diag}(-1,1,1,1)$. Since point (i) is subtler, we discuss it
now in more detail.

It is actually quite simple to understand why a coupling of
matter (including electromagnetism) to the disformal metric
(\ref{Disformal2}) suffices to increase light deflection. The
basic idea is similar to the case of the previous disformal
coupling (\ref{Disformal1}): Photons are now directly coupled to
the scalar field, whereas they do not feel any global factor of
the metric. More precisely, in the preferred frame where $U_\mu =
(\pm 1,0,0,0)$, and in locally inertial coordinates such that
$g^*_{\mu\nu} = \text{diag}(-1,1,1,1)$ at a given spacetime
point, one notices that $-U_\mu U_\nu = \text{diag}(-1,0,0,0)$
behaves like the $g^*_{00}$ component of the metric, and
conversely that $g^*_{\mu\nu}+U_\mu U_\nu = \text{diag}(0,1,1,1)$
behaves like its spatial components $g^*_{ij}$. The vector field
is thus a mere tool to separate by hand the space and time
components of a tensor (implying obviously a violation of Lorentz
invariance, namely the existence of an ether). Its interest is
that one may now multiply $g^*_{00}$ and $g^*_{ij}$ by
\textit{different} functions of the scalar field. Inverse factors
$e^{2\alpha\varphi}$ and $e^{-2\alpha\varphi}$ have been chosen
in
Refs.~\cite{Bekenstein:2004ne,Bekenstein:2004ca,Sanders:1996wk,
Sanders:2005vd}, in order to mimic GR in isotropic coordinates.
One could consider more generally inverse factors $A^2(\varphi)$
and $A^{-2}(\varphi)$, but an exponential (Brans-Dicke-like)
coupling $A(\varphi) = e^{\alpha\varphi}$ suffices to predict the
right phenomenology. The physical metric is thus chosen as
\begin{equation}
\tilde g_{00} = e^{2\alpha\varphi} g^*_{00}\ ;\quad
\tilde g_{ij} = e^{-2\alpha\varphi} g^*_{ij}.
\label{Disformal3}
\end{equation}
This can be rewritten covariantly thanks to the
vector field $U_\mu$ defining the preferred frame:
\begin{eqnarray}
\label{Disformal4}
\tilde g_{\mu\nu} &=& -e^{2\alpha\varphi}
U_\mu U_\nu + e^{-2\alpha\varphi} (g^*_{\mu\nu}+U_\mu
U_\nu)
\nonumber \\
&=& e^{-2\alpha\varphi} g^*_{\mu\nu} -2\, U_\mu
U_\nu\sinh(2\alpha\varphi).
\end{eqnarray}
This matter metric has a Lorentzian signature, as can be easily
checked. Beware that our notation differs slightly from that of
Refs.~\cite{Bekenstein:2004ne,Bekenstein:2004ca}. In these
references, the matter-scalar coupling constant $\alpha$ reads
$\sqrt{k/4 \pi}$ where $k$ is a dimensionless parameter.

In order to reproduce the MOND dynamics at large distances, the
model must be tuned so that $e^{2\alpha\varphi} = 1 + 2
\sqrt{GMa_0}\ln r/c^2$ in this regime. Moreover, if one insists on
the fact that photons should propagate faster than gravitons (even
if it is actually not necessary), one must have $\alpha\varphi
\geq 0$ at every spacetime point. The logarithm entering this MOND
potential should therefore be understood as $\ln (r/\ell)$, where
$\ell$ is a length scale significantly smaller than the size of the
considered galaxy (either imposed dynamically by the model, or
simply a universal constant much smaller than the size of any
galaxy). Indeed, $d\tilde s^2 = 0 \Leftrightarrow
ds_*^2 = 2 e^{2\alpha\varphi} \sinh(2\alpha\varphi) (U_\mu
dx^\mu)^2 \geq 0$ because $\alpha\varphi \geq 0$, and the cone
defined by $\tilde{g}_{\mu\nu}$ is thus wider than the one defined
by $g_{\mu\nu}^*$.

Note that the character spacelike of $\partial_\mu\varphi$ or
timelike of $U_\mu$ does not change the sign of the square
$(\partial_\mu\varphi dx^\mu)^2$ or $(U_\mu dx^\mu)^2$. The
crucial difference is that we now consider a \textit{negative}
factor for the term $U_\mu U_\nu$ entering the disformal
metric (\ref{Disformal4}), whereas we needed a \textit{positive}
value of $B(s,\varphi)$ to increase light bending with metric
(\ref{Disformal1}). In fact, it may seem paradoxical to increase
light bending with the negative term entering
Eq.~(\ref{Disformal4}), since photons now propagate faster than in
GR (i.e., faster than gravitons), if $\alpha \varphi >0$.
Intuitively, the trajectories of fast particles are almost
straight lines, whereas slower ones are significantly curved by
the Newtonian potential. This reasoning just happens to be
erroneous,\footnote{As an illustration of this point, let us recall
that the scalar field surrounding a body of mass $M$ is not only
given by $\varphi_{\textrm{local}}=-\alpha G M /r c^2$ (or
$\sqrt{GMa_0}\ln r/\alpha c^2$, depending on the regime), but
more generally by $\varphi_0 + \varphi_{\textrm{local}}$ when one
takes into account the asymptotic value of the scalar field
$\varphi_0$ which evolves with cosmic time. Since the ``speed'' of
photons (or, more precisely, their character timelike, null or
spacelike with respect to the Einstein metric $g^*_{\mu\nu}$) depends
on the sign of $\varphi_0 + \varphi_{\textrm{local}}$, one may expect
that light bending would be affected by $\varphi_0$. Actually
this is not the case, as shown by the first line of
Eq.~(\ref{IntegralLightDisformal2}).} because light bending
depends very differently on the temporal ($\tilde g_{00}$) and
spatial ($\tilde g_{ij}$) components of the metric, as
illustrated by Eq.~(\ref{IntegralLight}) in Schwarzschild
coordinates. At linear order, Ref.~\cite{Bekenstein:1993fs}
actually proves that this integral involves the radial derivative
$\partial_r \tilde g_{00}$ of the time component, thanks to a
partial integration, whereas the radial component $\tilde g_{rr}$
enters it without any derivation.\footnote{This also explains why
the time component $k_\mathcal{B}$ enters multiplied by a factor
$-p$ in Eq.~(\ref{LightDeflectionRAQUAL1}), contrary to the
radial component $k_\mathcal{A}$.} In other words, one needs an
additional positive term in $\tilde g_{rr}$ to increase light
bending, as proven in Eq.~(\ref{IntegralLightDisformal}) above,
but one needs a term whose radial derivative is negative if it
enters $\tilde g_{00}$, as in Eq.~(\ref{Disformal4}) above. This
is the reason why the Newtonian and MOND potentials enter so
differently in the metric predicted by GR in presence of dark
matter $\tilde g_{00} = -1 + 2GM/rc^2 - 2\sqrt{GMa_0}\ln r/c^2$
and $\tilde g_{rr} = 1 + 2GM/rc^2 + 2\sqrt{GMa_0}/c^2$ [see below
Eq.~(\ref{LightDeflectionDM})]. In spite of their relatives
signs, all the above terms contribute positively to light
bending, because those entering $\tilde g_{rr}$ are positive, and
because the derivatives of those entering $\tilde g_{00}$ are
negative.

The above discussion illustrates once again that observable
quantities, or even their signs, can be subtle to identify in the
components of a metric tensor. The only way to be sure of an
observable prediction is to compute it explicitly. We already
mentioned, below Eq.~(\ref{LightDeflectionDM}), that the GR
metric in presence of dark matter takes precisely the form
(\ref{Disformal3}) in isotropic coordinates. Let us however
derive now the light deflection angle it predicts. First of all,
since light is insensitive to any global factor, let us multiply
it by $e^{2\alpha\varphi}$, i.e., consider instead the metric
$\bar g_{00} \equiv e^{4\alpha\varphi} g^*_{00}$ and $\bar g_{ij}
\equiv g^*_{ij}$. If the coordinates have been chosen so that the
Einstein metric $g^*_{\mu\nu}$ takes the Schwarzschild form
(\ref{SchwarzschildForm}), this is also the case for $\bar
g_{\mu\nu}$ and one may thus use the integral expression
(\ref{IntegralLight}). At first post-Newtonian order, all
contributions coming from $g^*_{\mu\nu}$ reproduce the general
relativistic result,\footnote{Actually, there is a subtle
modification of the Einstein metric $g^*_{\mu\nu}$ itself in
TeVeS, because of the contribution of the energy-momentum tensor
of the unit vector $U_\mu$; see Eq.~(58) of
\cite{Bekenstein:2004ne} and Eq.~(6.2) of
\cite{Bekenstein:2004ca}. However, this correction is chosen to
be negligible in TeVeS.} and the only extra contribution is
proportional to the scalar field, namely $\alpha\varphi = \alpha
\varphi_0 + \sqrt{GMa_0}\ln r/c^2 +\mathcal{O}(1/c^4)$ in the
MOND regime. One thus gets straightforwardly
\begin{eqnarray}
\Delta\theta &=& \Delta\theta_\text{GR} + 4 \alpha
\int_{r_0}^{\infty}
\frac{\varphi(r)-\varphi(r_0)}{(1-r_0^2/r^2)^{3/2}}\, \frac{r_0
dr}{r^2} + \mathcal{O}\left(\frac{1}{c^4}\right)
\nonumber\\
&=& \Delta\theta_\text{GR} + \frac{4\sqrt{GMa_0}}{c^2}
\int_{r_0}^{\infty} \frac{\ln(r/r_0)}{(1-r_0^2/r^2)^{3/2}}\,
\frac{r_0 dr}{r^2} + \mathcal{O}\left(\frac{1}{c^4}\right)
\nonumber\\
&=&\Delta\theta_\text{GR}
+ \frac{4\sqrt{GMa_0}}{c^2}
\left[\arcsin\left(\frac{r_0}{r}\right)
-\frac{(r_0/r)\ln(r_0/r)}{\sqrt{1-r_0^2/r^2}}
\right]_{r_0}^{\infty}
+ \mathcal{O}\left(\frac{1}{c^4}\right)
\nonumber\\
&=&\Delta\theta_\text{GR}
+ \frac{2\pi\sqrt{GMa_0}}{c^2}
+ \mathcal{O}\left(\frac{1}{c^4}\right),
\label{IntegralLightDisformal2}
\end{eqnarray}
which confirms that the light deflection angle is positive
although photons are faster than in GR (with respect to the
Einstein metric $g^*_{\mu\nu}$). This angle is exactly the one
predicted by GR, when a spherical halo of dark matter is assumed
to generate a Newtonian potential $\sqrt{GMa_0}\ln r$. Moreover,
the first line makes it clear that the background value of
$\varphi$ does not contribute to the integral. It must be
stressed that the sign of $\varphi$ is not crucial in itself; what
is important is that its \textit{derivative} is positive.
It is therefore possible to have a cosmological background such
that $\alpha \varphi < 0$ locally whereas the amount of light
deflection remains the same. This remark will be of great
importance in Sec.~\ref{Cherenkov}, since we will show that an
empirical fact strongly supports that gravitons should propagate at
a speed greater (or equal) than the speed of photons (i.e., that
$\alpha \varphi$ should be negative).

Although the above construction seems natural (apart from the
existence of a preferred frame), let us however underline that it
is actually quite \textit{ad hoc}. Indeed, the relation
(\ref{Disformal3}) has been imposed by hand precisely to mimic
GR's metric, but any other relation between $g^*_{00}$ and
$g^*_{ij}$ would have predicted another light deflection angle.
Even without considering the general case (\ref{Disformal2}),
where $A(\varphi)$ and $B(\varphi)$ are fully independent
functions of the scalar field, it is instructive to analyze a
disformal coupling slightly generalizing Eq.~(\ref{Disformal4}),
namely
\begin{equation}
\tilde g_{\mu\nu} =
-e^{2\alpha\varphi} U_\mu U_\nu
+ e^{-2\kappa\alpha\varphi} (g^*_{\mu\nu}+U_\mu U_\nu),
\label{Disformal5}
\end{equation}
where $\kappa$ and the matter-scalar coupling constant $\alpha$
are dimensionless numbers to be constrained by experiments. The
tightest experimental constraints come from solar-system tests,
i.e., in the Newtonian regime of such theories, where one must
have $\varphi =\varphi_0-\alpha GM/rc^2$. A nonvanishing
background value $\varphi_0$ of the scalar field is necessary to
respect the inequality $\alpha\varphi \geq 0$, imposed in
Ref.~\cite{Bekenstein:2004ne} to avoid superluminal gravitons.
Since $GM/Rc^2 \leq \frac{1}{2}$ for any body of mass $M$ and
radius $R$, it suffices to impose $\varphi_0/\alpha \geq
\frac{1}{2}$. (Let us however recall that superluminal gravitons
are actually \textit{not} a problem, as discussed in
Sec.~\ref{SecRAQUAL} above and in
Refs.~\cite{Bruneton:2006gf,Babichev:2007dw}.) In this Newtonian
regime, metric (\ref{Disformal5}) reads thus
$\tilde g_{00} = \exp(2\alpha\varphi_0) \left[-1 +
2(1+\alpha^2)GM/rc^2\right] +\mathcal{O}(1/c^4)$ and $\tilde
g_{ij} = \exp(-2\kappa\alpha\varphi_0)\delta_{ij} [1 +
2(1+\kappa\alpha^2)GM/rc^2]+\mathcal{O}(1/c^4)$, where the
constant exponential factors may be eliminated by a redefinition
of (physical) time and length units. The Eddington post-Newtonian
parameter $\gamma^\text{PPN}$ is thus given by
\begin{equation}
\gamma^\text{PPN} = \frac{1+\kappa\alpha^2}{1+\alpha^2}
= 1 + \frac{(\kappa-1)\alpha^2}{1+\alpha^2}.
\label{gammaDisformal}
\end{equation}
The Cassini constraint \cite{Bertotti:cassini} on this
parameter imposes therefore $|\kappa -1|\alpha^2 /(1+\alpha^2)<
2\times 10^{-5}$. In conclusion, experiment forces us to fine
tune $\kappa \approx 1$, otherwise the matter-scalar coupling
constant $\alpha$ needs to be small and we get back the serious
problem pointed out in Sec.~\ref{MONDRAQUAL} above: The MOND
logarithmic potential starts manifesting at too small distances,
and one needs unnatural RAQUAL kinetic terms like the one of
Fig.~\ref{fig3} to pass solar-system tests. [Note that $\kappa =
1\pm 0.1$ suffices for the Newton-MOND transition to be much
smoother that in this Figure, allowing a Newtonian regime up to
distances $\sim 100$ AU instead of $30$ AU, but this remains
nevertheless a quite fine-tuned behavior.] On the other hand, if
$\kappa = 1$ strictly, as chosen in
Refs.~\cite{Bekenstein:2004ne,Bekenstein:2004ca,Sanders:1996wk,
Sanders:2005vd}, there is no longer any post-Newtonian
solar-system constraint on the matter-scalar coupling constant
$\alpha$, and one may thus \textit{a priori} choose $\alpha = 1$
and a RAQUAL kinetic term of the natural form of Fig.~\ref{fig2}.
[The fact that $\kappa=1$ implies $\gamma^{\textrm{PPN}}=1$ has
been noticed in Ref.~\cite{Bekenstein:2004ne}. However, its
author did not choose $\alpha=1$ (i.e., $k = 4\pi$ in the TeVeS
notation) but a smaller value $\alpha \approx 5 \times 10^{-2}$
(i.e., $k \approx 0.03$). The reason was that he wished to
neglect the energy-momentum tensor of the scalar field $\propto
\alpha^2$, but it \textit{is} actually negligible even when
$\alpha = 1$ since it is of second post-Newtonian order
$\mathcal{O}(1/c^4)$.]

This discussion illustrates that several experimental constraints
have already been used to construct the model proposed in
Refs.~\cite{Bekenstein:2004ne,Bekenstein:2004ca,Sanders:1996wk,
Sanders:2005vd}. Moreover, one cannot claim that it
\textit{predicts} the right light deflection, but rather that it
has been \textit{tuned} in order to do so. A vivid way to
illustrate the fine tuning hidden in the choice
(\ref{Disformal3})-(\ref{Disformal4}) is to exhibit different
metrics of the form (\ref{Disformal2}), reproducing GR's light
deflection in the Newtonian regime of the solar system but a
different one in the MOND regime. For instance, if one writes
this physical metric as\footnote{To simplify the discussion, we
assume here that the asymptotic value $\varphi_0$ vanishes, or,
stated differently, we merely denote here $\varphi - \varphi_0$
as $\varphi$.}
\begin{equation}
\tilde g_{\mu\nu} =
-f_0(\varphi) U_\mu U_\nu
+ f_r(\varphi) (g^*_{\mu\nu}+U_\mu U_\nu),
\label{Disformal6}
\end{equation}
then the choice $f_0(\varphi) = e^{2\varphi}$ and
$f_r(\varphi) = e^{-2\varphi/(1+\varepsilon \varphi)}$
reproduces the above model when $\varphi \ll
1/\varepsilon$, but predicts a smaller light
deflection when $\varphi > 1/\varepsilon$, notably
when $\varphi \propto \ln r \rightarrow \infty$ in
the MOND regime. On the contrary, the choice
$f_0(\varphi) = e^{2\varphi}$ and $f_r(\varphi) =
e^{-2\varphi(1+\varepsilon \varphi)}$ would also mimic
the above model for $\varphi \ll 1/\varepsilon$, but
predict a much larger light deflection when
$\varphi > 1/\varepsilon$, in the MOND regime.
Therefore the specific choice
(\ref{Disformal3})-(\ref{Disformal4}),
mimicking GR, may almost be considered as a fit
of some experimental data. It is nevertheless
rather natural, and the fact that it allows to
reproduce GR's predictions in presence of dark
matter is anyway a great achievement, since
light deflection was a crucial difficulty
of previous constructions.

Further analysis of such stratified theories however
undermines this conclusion, and we will see that they cannot
survive other phenomenological constraints, unless they are much
more fine-tuned than they already are.

\section{\label{Section4}Difficulties with stratified theories}

\subsection{\label{Stabilite}Stability issues}

One of the great advantages of the stratified TeVeS metric
(\ref{Disformal2}), over the disformally coupled scalar field
(\ref{Disformal1}), is the simplicity of the field equations
within matter. However, as we have seen, the vector field needs
to have a constant norm unless consistency problems occur in its
field equation, and this has two serious drawbacks. First, this
leads to a mixing of the vector and the spin-2 degrees of freedom
and thereby to a nondiagonal system of differential equations.
This means that the actual degrees of freedom have to be computed
carefully (as done in \cite{Eling:2004dk} in the specific case
$\tilde{g}_{\mu\nu}=g^{*}_{\mu\nu}$). Second, the fact that the
vector field must be of constant norm spoils its stability.
Indeed, as was shown in Ref.~\cite{Clayton:2001vy}, the
Hamiltonian of the vector field is not bounded by below, and the
theory is unstable. The dynamics of the vector field derives from
the action
\begin{equation}
\label{ActionVecteur} S= \int \mathcal{L} d^4 x = - \frac{K
c^3}{32 \pi G} \int \sqrt{-g^*} d^4x
\bigl[g_*^{\mu\rho}g_*^{\nu\sigma}F_{\mu\nu}F_{\rho\sigma} - 2
\lambda (g_*^{\mu\nu} U_{\mu} U_{\nu} +1) \bigr],
\end{equation}
where the stress $F_{\mu\nu}$ is given by $\partial_{\mu}
U_{\nu}-\partial_{\nu} U_{\mu}$, $\mathcal{L}$ is the Lagrangian
density, and $K$ is a positive dimensionless number. The
Hamiltonian can be computed in the standard way, noticing that
both $\lambda$ and $U_{0}$ must be treated as Lagrange multipliers
since the action does not involve their time derivatives. The momenta
read then
\begin{equation}
\pi^i = -\frac{K c^3}{8 \pi G} F_*^{0i},
\end{equation}
where latin indices run from 1 to 3, and indices are raised using
the inverse Einstein metric $g_*^{\mu\nu}$. In the limit of decoupling
gravity ($g^*_{\mu\nu} \to \eta_{\mu\nu}$), the Hamiltonian reads
\begin{eqnarray}
H&=&\int d^3x \left( \pi^i \partial_{0} U_{i} - \mathcal{L}
\right) \nonumber \\ &=&\frac{K c^3}{8 \pi G}\int d^3x \left(
\frac{F_{\mu\nu}F_*^{\mu\nu}}{4} -F_*^{0i}\partial_{0} U_{i} \right),
\end{eqnarray}
where we used that $U_{\mu} U_*^{\mu}=-1$ which derives from the
equation $\delta S/ \delta \lambda =0$. One then writes
$\partial_{0} U_{i}$ as $F_{0i}+\partial_i U_0$ and integrates by
parts the last term. The Hamiltonian, as a functional of the
independent field configurations $U_i$ and $\pi^i$, then reads
\begin{equation}
H=\int d^3x \left[ \frac{4 \pi G}{K c^3} \boldsymbol{\pi}^2 +
\frac{K c^3}{4 \pi G} (\boldsymbol{\nabla} \times \mathbf{U})^2 +
\sqrt{1+\mathbf{U}^2}\,\boldsymbol{\nabla} \boldsymbol{\pi} \right],
\label{VectorHamiltonian}
\end{equation}
where bold-faced symbol denotes three-vectors. The first two terms
are just proportional to $E^2+ B^2$ where $E$ and $B$ are the
usual electric and magnetic fields associated to $U_{\mu}$.
Contrary to the Maxwell case where $\boldsymbol{\nabla}
\boldsymbol{\pi} \propto \boldsymbol{\nabla} \mathbf{E}=0$ in
vacuum, the last term does not vanish here (even in vacuum), and
can take arbitrarily large and negative values. This can be made more
explicit, for instance, by considering the special case where the
vector field derives from a static potential $\phi$, i.e., $U_i
=\partial_i \phi(x)$; see Ref.~\cite{Clayton:2001vy}. The
Hamiltonian is thus unbounded by below, and the theory is
unstable, at least as it is defined in action (\ref{ActionVecteur}).
Note that empty space ($U_i = 0$) is not even a locally stable
vacuum, because when $U_i =\partial_i \phi(x) \rightarrow 0$, the
dominant term is the last one in Hamiltonian
(\ref{VectorHamiltonian}). Of course, there always remains the
possibility that higher-order terms could stabilize the model (around
a \textit{different} vacuum), but one should exhibit such corrections
explicitly to prove the consistency of the model.

We stressed in Sec.~\ref{ST} that a competing model (phase
coupling gravity) also reproduces the MOND dynamics. The theory is
however unstable because of the presence of a negative potential.
Recently, an improved version of this theory (called BSTV) was given in
Ref.~\cite{Sanders:2005vd}. It involves, besides two scalar
fields, a dynamical unit vector field in order to reproduce
the right amount of light deflection using a disformal metric of
type (\ref{Disformal4}).
This improved theory is however still unstable. Three different
terms of the full Hamiltonian can take it to arbitrary large and
negative values. To begin with, there is the instability driven by
the vector field that we described above, if its kinetic term if of
the Proca form (like in TeVeS). The dynamics of the two scalar fields
$q$ and $\varphi$ derives from the action\footnote{Beware that
Ref.~\cite{Sanders:2005vd} uses the opposite (mostly $-$) signature.
We have translated its action in our mostly~$+$ convention.}
\begin{equation}
S=-\int \sqrt{-g^*} d^4x \left( \frac{1}{2} (\partial_\mu q)^2 +
\bigl[h(q) g^{\mu\nu}_* +
\bigl(f(q)-h(q)\bigr)U^{\mu}_*U^{\nu}_*\bigr]
\partial_{\mu} \varphi \partial_{\nu} \varphi + V(q)\right).
\end{equation}
Because the author of Ref.~\cite{Sanders:2005vd} wrote a
quadratic potential $V(q)=-A q^2$ with $A>0$, the field $q$ is
tachyonic and thus unstable. Note that this potential is not
crucial to obtain the MOND dynamics, but is merely added to
produce oscillations of the scalar field that may generate
the cosmic microwave background (CMB) spectrum predicted
in cold dark matter theories. Curing the instability
by changing the sign of $A$ may however spoil this interesting
phenomenology. The computation of the full Hamiltonian reveals
another instability. The momenta read $p_q =\partial_0 q$ and
$p_{\varphi} = 2 h \partial_0 \varphi - 2 (U_{\mu} \partial^{\mu}
\varphi) (h-f) U_0$. This last equality can be inverted to find
$\partial_0 \varphi$ as a function of (notably) $p_{\varphi}$.
Since the resulting Hamiltonian has a quite complicated
expression, we only report its value on the particular field
configuration such that $U_i = 0$. Up to the contribution of the
vector field, we have:
\begin{equation}
H=\int d^3 x \left(\frac{p_q^2 + (\partial_i q)^2}{2} -A q^2 +
\frac{p_{\varphi}^2}{4[2h(q)-f(q)]} + h(q) (\partial_i \varphi)^2
\right).
\end{equation}
Changing the sign of $A$ would not be sufficient to guarantee the
stability of the theory. Indeed, the functions $h$ and $f$ must
behave as $h(q)=q^2$ and $f(q)=q^6$ (at least in some regime)
in order to obtain MOND dynamics. Therefore the term
$p_{\varphi}^2/4(2h-f)$ can be made arbitrarily large and
negative.

\subsection{\label{Preferredframe}Preferred-frame effects}

Ni's stratified theory \cite{Ni} is a particular case of disformal
coupling (\ref{Disformal4}) where the Einstein metric is no longer
dynamical, and assumed to be flat: $g^*_{\mu\nu} = \eta_{\mu\nu}$.
Although it was \textit{built} to reproduce the Schwarzschild solution
at linear order, by imposing $e^{2\alpha\varphi} = 1-2GM/rc^2$, it is
nevertheless known to be inconsistent with preferred-frame
tests. Indeed, it predicts a post-Newtonian parameter $\alpha_1
\approx -8$, whereas the present experimental constraint is
$|\alpha_1| < 10^{-4}$ \cite{Willbook,Willreview}. One may
thus wonder if the more general models of Sec.~\ref{Stratified}
present the same difficulty.

Their experimental constraints have been discussed in
Ref.~\cite{Sanders:1996wk}, as well in Ref.~\cite{Foster:2005dk}
in a more general context. In the case where the preferred frame
defined by $U_\mu$ is assumed to be the CMB rest frame (as is usually
done when analyzing the weak-field predictions of vector-tensor
theories), Ref.~\cite{Sanders:1996wk} has proven that the
post-Newtonian parameter $\alpha_1 \approx -16 \alpha^2$ (where
$\alpha$ without any index denotes as before the matter-scalar
coupling constant, related to the parameter $\eta$ of
Ref.~\cite{Sanders:1996wk} by
$\eta = 4\alpha^2$). The other PPN parameters characterizing
preferred-frame effects vanish identically, like in general
relativity ($\alpha_2 = \alpha_3 = 0$). The present experimental
bound on $\alpha_1$ would thus impose $\alpha^2 < 6\times
10^{-6}$, even tighter than the light-deflection constraint
$\alpha^2 < 10^{-5}$ on conformally-coupled scalar-tensor
theories. The fine-tuning problem discussed in
Sec.~\ref{MONDRAQUAL} above would thus be even more serious in
the present disformal framework.

However, the vector field $U_\mu$ is dynamical, and
Refs.~\cite{Bekenstein:2004ne,Bekenstein:2004ca} claim that its
field equation drives it to be parallel to the matter proper
time, i.e., to read $U_\mu = (\pm 1,0,0,0)$ in the
\textit{matter} rest frame. In such a case, there would be no
preferred-frame effect due to the motion at $370 \text{
km}.\text{s}^{-1}$ of the solar-system with respect to the CMB,
but there could still exist observable effects due to the
planets' velocities around the Sun. The Earth's velocity at about
$30 \text{ km}.\text{s}^{-1}$ would \textit{a priori} induce
effects constrained by the tracking of artificial
satellites~\cite{DEF94a}, and although the bound on $\alpha^2$
might be increased by a factor 10, the fine-tuning described in
Sec.~\ref{MONDRAQUAL} would still be necessary. It remains
possible that the vector field is \textit{also} driven to be
parallel to the Earth's proper time in its vicinity, but there
would always exist a transition region between the barycentric
and geocentric frames in which test particle should undergo
observable effects due to the variation of $U_\mu$.

Anyway, the derivation proposed in
\cite{Bekenstein:2004ne,Bekenstein:2004ca} actually only proves
that $U_\mu = (\pm 1,0,0,0)$ is \textit{a} solution, but not
\textit{the} solution. In other words, it just tells us that a
situation where matter is at rest with respect to the preferred
frame is \textit{possible}. This obviously does not prove that
the model is consistent with high-precision tests of local
Lorentz invariance of gravity. There remains to perform a careful
analysis of the specific models proposed in
\cite{Bekenstein:2004ne,Bekenstein:2004ca,Sanders:2005vd} in
order to draw definite conclusions about this problem of
preferred-frame effects. We will not discuss it any longer in the
present paper, because stratified theories anyway present several
other serious difficulties.

\subsection{\label{Cherenkov}Cherenkov radiation and high-energy
cosmic rays}

Throughout this paper, we insisted on the fact that superluminal
fields do not necessarily threaten causality, provided they
propagate along cones in spacetime, i.e., if their equations of
motion remain hyperbolic. We even announced in
Sec.~\ref{MONDRAQUAL} that experiment strongly suggests that no
field should propagate slower than light. Indeed highly
relativistic matter coupled to a subluminal field emits Cherenkov
radiation and thereby loses energy. In particular, the various
gravitational fields one may consider should not propagate slower
than light, otherwise ultra high energy cosmic rays would emit
``gravi-Cherenkov'' radiation (following the term used in
\cite{Moore:2001bv,Elliott:2005va}), and those that we detect
should thus have been produced near the Earth. However, no source
has been identified within a range of the order of the
kiloparsec, and this yields a tight bound on how much photons
(and hence relativistic matter particles) are allowed to travel
faster than gravity.

Cherenkov radiation occurs in electrodynamics when charged
particles propagate faster than light in a medium with refractive
index $n > 1$. The emission of a real photon through the vertex
of QED is kinematically allowed in that case because electromagnetic
waves do not follow the usual dispersion law $k_{\mu}k^{\mu}=0$, but
satisfy the law $|\mathbf{k}|=n |k_{0}|$ in a refractive
medium and have thus a larger wave vector than angular frequency if
$n>1$; see \cite{Moore:2001bv,Elliott:2005va}.

If light propagates faster than gravitational waves, as is the case
in the above stratified theory with $\alpha \varphi >0$, a
similar phenomenon occurs: (ultra relativistic) matter particles
propagating faster than gravity emit gravitational waves. The
same reasoning as above shows that this tree process is
kinematically allowed. We may however stress a subtlety here.
Since spacetime is endowed with two different metrics of
Lorentzian signature, one can define two different sets of
locally inertial coordinates (for which one of the two metrics
$g_{\mu\nu}^*$ or $\tilde{g}_{\mu\nu}$, but not both, is written
in its canonical form). There exists therefore two (equivalent)
formulations of conservation laws. Gravitational waves follow the
null geodesic of the gravitational metric\footnote{Beware that
the spectrum of gravitational waves may be affected by the
presence of the vector field. However, when the kinetic term of
the vector field is of the Proca type, there is no such
modification; see Ref.~\cite{Jacobson:2004ts}.} and satisfy thus the
equation $g_{*}^{\mu\nu}k_{\mu}k_{\nu}=0$, where $k_{\mu}$ denotes
their wave four-vector. Their dispersion relation in the matter
(Jordan) frame reads thus $\tilde{g}^{\mu\nu}k_{\mu}k_{\nu}=
(U_*^{\mu}k_{\mu})^2B/[A^2(B-A^2)]$, and takes the form
\begin{equation}
|\mathbf{k}|= |k_0|\sqrt{1+2 \sinh (2 \alpha \varphi)}
\end{equation}
in coordinates such that $\tilde g_{\mu\nu} = \eta_{\mu\nu}$ and
simultaneously $U_*^{\mu} = (\pm 1,0,0,0)$ locally (which can
always be imposed by a local Lorentz transformation keeping
$\tilde g_{\mu\nu} = \eta_{\mu\nu}$). Using $\alpha \varphi \ll
1$, we get $|\mathbf{k}|= |k_0|(1 + 2 \alpha \varphi)$. In
other words, everything behaves as if gravitons were travelling within
a refractive medium of index $n \approx 1 + 2\alpha \varphi$.
The emission of real gravitational waves by ultra relativistic matter
particles is thus allowed once $\alpha \varphi>0$. The authors of
Ref.~\cite{Moore:2001bv} calculated the rate of energy loss by
this process and found:
\begin{equation}
\frac{dE}{dt} \approx \frac{G p^4 (n-1)^2}{\hbar^2
c_{\textrm{light}}}
\end{equation}
for a particle of momentum $p$. Here we approximate pure numbers
by $1$, which is correct in view of the exact results derived in
\cite{Moore:2001bv}. [This result is actually obtained for a
scalar particle of matter; the case of a proton is more involved
but leads to similar formulae; see Ref.~\cite{Elliott:2005va}].
This equation is easily integrated for an ultra-relativistic
particle $E \approx p c_{\textrm{light}}$. Assuming that the
initial value of the momentum $p_i$ is much larger than the final
one $p_f$ and that $n$ is approximatively constant along the
particle's path, one finds that it can travel only over a distance
of order
\begin{equation}
L_{\textrm{max}} \sim \frac{\hbar^2 c_{\textrm{light}}^3}{G p_f^3
(n-1)^2},
\end{equation}
whose value reads
\begin{equation}
L_{\textrm{max}} \sim \frac{2 \times 10^{-11}}{(n-1)^2} \textrm{m}
\end{equation}
for $p_f \sim 10^{20}$ eV. Therefore, cosmic rays detected on
Earth with such a high momenta as $10^{20}$ eV cannot have been
produced farther than $L_{\textrm{max}}$. Assuming that these
cosmic rays have at least a galactic origin, we find
$L_{\textrm{max}} \agt 10^{21}$ m and therefore $n-1 \alt
10^{-16}$. Note of course that gravi-Cherenkov radiation can only
be emitted by matter particles propagating at a speed greater than
$c_{\textrm{gravitons}} = c_{\textrm{light}}/n$, where $n-1 \alt
10^{-16}$. Therefore only ultra-relativistic particles with a
Lorentz-Fitzgerald factor of order $\gamma \geq 1/\sqrt{2 (n-1)}
\sim 10^8$ can emit gravitational waves in this way. Assuming that
the observed high-energy cosmic rays with momenta $p \sim 10^{20}$
eV are protons, we find that their $\gamma$ reads $10^{11}$, so
that our analysis is self-consistent.

Since the gravitational field produced by the Milky Way is
approximatively MONDian in the vicinity of the Sun, we have $n-1
\approx 2 \alpha \varphi \approx 2 \sqrt{G M_{\textrm{Milky Way}} a_0}
\ln (r/\ell) /c^2$, where $\ell$ is a fixed length scale and where,
numerically, $\sqrt{G M_{\textrm{Milky Way}} a_0}/c^2 \sim 10^{-7}$.
If one insists on the fact that $\alpha \varphi$ should be
positive (so that gravity propagates slower than photons), the only way
inequality $n-1 \alt 10^{-16}$ could be satisfied would be to fine tune
the length scale $\ell$ to the galactocentric distance of the Sun up to
nine decimals! In other words, such a theory predicts that high energy
cosmic rays should not be seen unless they are produced very close to
the Earth, or if the Sun were very specifically located within our
galaxy. This is clearly unacceptable, and $\alpha\varphi$ would
anyway become negative for slightly smaller galactocentric distances.
In other words, gravity \textit{must} propagate faster than light at
least in some regions.

The only reasonable way to avoid the above fine tuning problem is
simply to have $\alpha \varphi < 0$. In such a case, light propagates
slower than gravitons and there is no gravi-Cherenkov radiation at
all.\footnote{Note that one may also consider Cherenkov radiation made
of waves of the vector field. Indeed, depending on the sign of the
parameters, such ``Einstein-aether'' waves \cite{Jacobson:2004ts} may
propagate slower than light. In any case, however, it would be
possible to require that $\alpha \varphi$ is negative enough so that
light (and hence matter) propagates slowly enough, to ensure that
matter particles do not emit such Cherenkov radiation at all.}
Let us emphasize that it is however in great disagreement with the
whole spirit of the TeVeS theory. Indeed the vector field was
precisely introduced to produce the right amount of light deflection
while avoiding any superluminal field. Actually, the above discussion
shows that subluminal fields coupled to the matter sector are
generically ruled out by the existence of ultra high-energy cosmic
rays. Therefore MOND-like stratified theories reproducing the right
light deflection must either be based on the disformal relation
(\ref{Disformal1}) (where the vector field assumes the form $U_{\mu}
=\partial_{\mu} \varphi$) with $B>0$, or on the disformal metric
Eq.~(\ref{Disformal4}) with $\alpha \varphi < 0$. In conclusion, light
should in any case travel slower than gravitons. Let us recall that it
does not imply a violation of causality, provided all field equations
remain hyperbolic.

\subsection{\label{PSR}Binary-pulsar constraints}

The fourth and probably most serious difficulty of stratified
theories is their consistency with binary-pulsar observations.
Indeed, even if the choice $\kappa=1$, in
Eq.~(\ref{gammaDisformal}), releases the constraint on $\alpha$
arising from precise measurement of the Eddington parameter
$\gamma^\text{PPN}$ in the solar system, binary-pulsar data \textit{a
priori} impose a tight constraint on it too. The literature on
binary-pulsar tests has studied wide classes of alternative theories
of gravity
\cite{Willbook,DEF92,DEF96b,DEF98,GefMGX,Damour:2007uf,DEF07} but not
the disformal coupling (\ref{Disformal2}). A specific analysis should
thus be performed, but it goes beyond the scope of the present
article. Let us just mention here that a strong matter-scalar coupling
constant $\alpha \approx 1$ means that the spin-2
($g^*_{\mu\nu}-\eta_{\mu\nu}$) and spin-0 ($\varphi-\varphi_0$) fields
are of the same order of magnitude in the vicinity of massive bodies,
i.e., in the Newtonian regime of the model. Near a binary system,
these fields cannot be static because they are driven by the orbital
motion. This causes the emission of gravitational and scalar waves
(and even waves of the vector field $U_\mu$, in the present
TeVeS model), which extract energy from the system and make the orbit
shrink. Because of the specific form of the kinetic terms (i.e., of the
helicities of the degrees of freedom), the energy loss via spin-2
waves start at the quadrupolar order $\mathcal{O}(1/c^5)$, but it is
of the much larger order $\mathcal{O}(1/c^3)$ for the dipolar spin-0
waves [monopolar scalar waves also exist but cause an energy loss of
order $\mathcal{O}(1/c^5)$ instead of $\mathcal{O}(1/c)$ if the two
orbiting bodies are at equilibrium]. The orbit should thus shrink much
more quickly in the present model than in GR, and this is known to be
inconsistent with several binary-pulsar observations
\cite{Eardley75,Will-Eardley77,DEF92,DEF96b,DEF98,GefMGX,Damour:2007uf,DEF07}.
Since this reasoning only depends on the \textit{local} dynamics
of the scalar field, its subtle MOND-like behavior at larger
distances should not change the conclusion. The scalar waves will
obviously behave in a non-standard way at distances $r \agt
\sqrt{GM/a_0}$, but they have anyway extracted too much energy
from the binary system in the near zone. Note also that the
peculiar matter-scalar coupling
(\ref{Disformal3})-(\ref{Disformal4}) should not change either
the above conclusion. It tells us that $\varphi$ does not feel
the matter current and pressure $T^{*\text{matter}}_{ij}$ as in
standard scalar-tensor theories, but the dominant source remains
$\alpha T^{*\text{matter}}_{00} \propto \alpha M c^2$ and
generates a scalar field $\varphi-\varphi_0 = \mathcal{O}(\alpha
GM/rc^2)$. Therefore, we expect present binary-pulsar data to
impose $\alpha^2 < 4\times 10^{-4}$, as in Brans-Dicke theory
\cite{GefMGX,Damour:2007uf,DEF07}. Although this is less
constraining than the $10^{-5}$ bound imposed by solar system
\cite{Bertotti:cassini} on conformally-coupled scalar fields
(\ref{TS1})-(\ref{RAQUAL}), this binary-pulsar constraint on the
matter-scalar coupling constant is thus \textit{a priori} also
valid for the disformally coupled ones
(\ref{Disformal1})-(\ref{Disformal2}).\footnote{This discussion
is reminiscent of a particular tensor-bi-scalar theory
constructed in \cite{DEF92} as a contrasting alternative to GR.
It depends on two free parameters characterizing how the two
scalar fields are coupled to matter. Even for strong couplings,
the model was fine tuned to pass anyway solar-system tests. The
combination of various binary-pulsar tests however severely
constrains both of the parameters \cite{DEF92,DEF07}.}

Note that this constraint is one order of magnitude smaller than
the value $\alpha^2 \approx 2\times 10^{-3}$ chosen in TeVeS
\cite{Bekenstein:2004ca} (where $\alpha^2 = k/4\pi$ in
Bekenstein's notation). This reference needed to choose $\alpha^2
> 1.4\times 10^{-3}$ in order to predict negligible MONDian
effects at distances smaller than Saturn's orbit, assuming a
natural RAQUAL kinetic term like the one displayed in
Fig.~\ref{fig2} above. Since this is inconsistent with
binary-pulsar tests, we recover again the problem discussed in
Sec.~\ref{MONDRAQUAL}: $\alpha$ needs to be small and we need an
unnatural kinetic term $f(s)$. In particular, the acceleration
cannot be purely Newtonian beyond $r \sim 150 \text{ AU}$ (if
$\alpha^2 = 4\times 10^{-4}$), and an approximately constant
extra contribution $a_0$ must exist between this distance and
$\sqrt{GM/a_0} \approx 7000 \text{ AU}$, where it becomes of the
$1/r$ MOND form (see Fig.~\ref{fig3}). Although this is still
allowed experimentally, this behavior seems anyway too
fine-tuned.

\subsection{Discussion}
As Secs.~\ref{Stratified} and \ref{Section4} above underline, the
TeVeS model
\cite{Bekenstein:2004ne,Bekenstein:2004ca,Sanders:2005vd} is
certainly the best field theory reproducing the MOND dynamics in
the literature, but it has still some drawbacks. One of them is
probably unavoidable: It is fine tuned to obtain the wanted
phenomenology. A second one seems generic but hopefully curable
in some specific models: It presents some instabilities. Other
theoretical problems should be solved, and have already received
some answers: It involves discontinuities, too many different
fields, and the various arbitrary functions it contains look
quite complicated. But the most serious difficulties seem to be
\textit{experimental}, as discussed in the present section. In
particular, it very probably does not pass binary-pulsar tests in
its present form (or needs to be even more fine-tuned in a quite
unnatural way). We will also mention in Sec.~\ref{Concl} other
generic problems that any MOND-like field theory should address.
Therefore, in spite of the great achievements of
Refs.~\cite{Bekenstein:2004ne,Bekenstein:2004ca,Sanders:2005vd},
it remains interesting to look for other theoretical frameworks
able to reproduce the MOND dynamics in a consistent way. We will
examine a rather new one in the following section, although it is
still in the class of RAQUAL and disformal models. Its original
feature is that the theory will be particularly simple in vacuum
(outside matter sources), namely pure general relativity in a
first model, and pure Brans-Dicke theory in a second one.
However, we will see that the consistency of the field equations
\textit{within} matter exhibits subtle but crucial difficulties.

\section{Nonminimal metric couplings}\label{Nonmin}
\subsection{Changing the matter--spin~2 coupling}\label{MatterSpin2}

In the present section, we adopt a mixing of the
modified-inertia and modified-gravity viewpoints discussed
in Sec.~\ref{GR} above. Like in the modified-inertia
framework, the dynamics of gravity in vacuum will be
defined by the Einstein-Hilbert action (\ref{EH}).
On the other hand, like in modified gravity, the matter
action will be assumed to depend only on the matter fields
and their first derivatives, or, in the case of test particles,
on their positions and velocities only. Moreover, in order
to preserve the weak equivalence principle, we will assume
that all matter fields are coupled to a single second
rank symmetric tensor $\tilde g_{\mu\nu}$. In other words,
the action of the theory reads
\begin{equation}
S = \frac{c^4}{16 \pi G}
\int\frac{d^4 x}{c}\sqrt{-g_*}\, R^*
+S_\text{matter}[\psi; \tilde g_{\mu\nu}],
\label{NonminAction}
\end{equation}
where as before $\psi$ denotes globally any matter field.
Therefore, we do not seem to have changed anything with respect
to GR, and this looks like a model in which \textit{neither}
inertia \textit{nor} gravity are modified! However, the crucial
difference with GR is that we will assume the physical metric
$\tilde g_{\mu\nu}$ to depend not only on the Einstein (spin-2)
metric $g^*_{\mu\nu}$, but also on the curvature tensors one may
construct from it and their covariant derivatives. Several
definitions of $\tilde g_{\mu\nu}$ yield the right MOND
phenomenology at leading order, and differ only at the
post-Newtonian level, i.e., for relativistic corrections of order
$\mathcal{O}(1/c^2)$ with respect to the Newtonian and MOND
accelerations. Let us choose a particular example:
\begin{equation}
\tilde g_{\mu\nu} \equiv g^*_{\mu\nu} + \frac{\sqrt{a_0/3}}{4\,c}\,
\frac{\left(\partial_\lambda \text{GB}\right)^2 h(X)\,g^*_{\mu\nu}
+ 2\, \partial_\mu\text{GB}\,\partial_\nu\text{GB}}
{(\Box^*\text{GB}/10)^{7/4}},
\label{gtilde}
\end{equation}
where $\text{GB} \equiv R^{*2}_{\mu\nu\rho\sigma} - 4
R_{\mu\nu}^{*2} + R_*^2$ is the Gauss-Bonnet topological invariant
(which may be replaced by $R_{\mu\nu\rho\sigma}^{*2}$ or the square
of the Weyl tensor $C_{\mu\nu\rho\sigma}^{*2}$ without changing the
results at first order in $\sqrt{a_0}$), where $\Box^*$ denotes the
d'Alembertian operator (covariant with respect to the Einstein
metric $g^*_{\mu\nu}$), and where
\begin{equation}
X \equiv \frac{1}{\ell_0}
\sqrt{\frac{30\,\text{GB}}{\Box^*\text{GB}}}\,.
\label{X1}
\end{equation}
The fixed length scale $\ell_0$ is introduced to make this
quantity dimensionless. The numerical factors entering
Eqs.~(\ref{gtilde})-(\ref{X1}) have been chosen to simplify
the expression of the function $h(X)$, which will be specified
below. Note that $\tilde g_{\mu\nu}$ is indeed a second-rank
symmetric \textit{tensor}, since it is constructed only from tensors,
their covariant derivatives and their contractions. The matter action
$S_\text{matter}[\psi; \tilde g_{\mu\nu}]$
thereby satisfies exactly Lorentz invariance. Moreover, this is
still a local action since the physical metric (\ref{gtilde})
involves a finite number of derivatives of $g^*_{\mu\nu}$. It
should be noted that Eq.~(\ref{gtilde}) does not make sense in flat
spacetime, because it would involve ratios of vanishing
expressions, but our Universe is never strictly flat since massive
bodies do exist.

Since in Eq.~(\ref{NonminAction}), the physical metric
$\tilde g_{\mu\nu}$ only enters the matter action,
the curvature tensors it involves do not contribute
to the dynamics of gravity in vacuum. This implies
notably that the Birkhoff theorem is satisfied, like
in GR. The metric generated by a spherically symmetric
distribution of matter is thus of the Schwarzschild type,
i.e., reads in Schwarzschild coordinates
\begin{equation}
g^*_{\mu\nu} dx^\mu dx^\nu = -\left(1-\frac{2 G M}{rc^2}\right)
c^2 dt^2 + \left(1-\frac{2 G M}{rc^2}\right)^{-1} dr^2 + r^2
\left(d\theta^2 +
\sin^2\theta\, d\phi^2\right),
\label{Schw}
\end{equation}
where $M$ is a constant, very close to the one that GR would
have defined (but slightly modified because of the nonminimal
coupling of matter to curvature).
It is then straightforward to compute the gravitational
potential $V = -\frac{1}{2}(1+\tilde g_{00})c^2$ felt by
a test particle slowly moving in such a spacetime. The
Schwarzschild solution (\ref{Schw}) implies
\begin{eqnarray}
\text{GB} &=& R_{\mu\nu\rho\sigma}^{*2} = C_{\mu\nu\rho\sigma}^{*2} =
48\left(\frac{GM}{r^3 c^2}\right)^2,
\label{GB}\\
\Box^*\text{GB} &=& 1440\left(\frac{GM}{r^4 c^2}\right)^2
\left(1-\frac{12}{5}\,\frac{GM}{rc^2}\right),
\label{BoxGB}
\end{eqnarray}
so that the time-time component of the physical metric reads
\begin{equation}
\tilde g_{00} = -1+\frac{2 GM}{rc^2}
- 2\frac{\sqrt{GM a_0}}{c^2}\,h\left(\frac{r}{\ell_0}\right)
+\mathcal{O}\left(\frac{1}{c^4}\right).
\label{gtilde00}
\end{equation}
One thus recovers the MOND potential if one chooses $h(X) = \ln
X$, and the length scale $\ell_0$ is then arbitrary since it does
not enter any observable prediction. However, we know that the
MOND potential should not be felt at small distances, because it
would be inconsistent with solar-system tests of Kepler's third
law \cite{Talmadge:1988qz}. One should thus choose a more fine-tuned
function $h(X)$, such that its derivative gives a $1/r$ force at
large distances $r \gg \ell_0$ but a vanishing one when $r \ll
\ell_0$. One may choose for instance $h'(X) = X/(1+X)^2$, i.e.,
after integration,
\begin{equation}
h(X) = (1+X)^{-1} +\ln(1+X).
\label{fX}
\end{equation}
Now the constant $\ell_0$ does enter the physical
predictions, and in order to obtain the correct
phenomenology, it should be chosen much larger than
the solar-system size but small with respect to
the radius of any galaxy. This underlines that the
present model is also quite fine tuned, in a
similar way as the RAQUAL models discussed in
Secs.~\ref{MONDRAQUAL} and \ref{Stratified} above.
They needed specific aquadratic kinetic terms
for the scalar field, and the transition between
the Newtonian and MOND regimes then occurred
at a radius $\sim \alpha^2\sqrt{GM/a_0}$,
generically much too small to be consistent
with solar-system tests. Here the transition scale
$\ell_0$ is fixed by hand in Eq.~(\ref{X1}),
which may be considered as a bonus or a drawback.
Note anyway that it would be easy to impose
$\ell_0 = \sqrt{GM/a_0}$, for instance, merely
by using again Eqs.~(\ref{GB}) and (\ref{BoxGB})
to define this quantity and replace it in (\ref{X1}).
However, the corresponding definition of $X\approx
(\Box^*\text{GB})^{5/4}\sqrt{a_0\text{GB}}/
(\partial_\lambda\text{GB})^2c$ would be even
more unnatural!

The light deflection can also be derived straightforwardly from
the physical metric (\ref{gtilde}) expanded at first order in
$1/c^2$. Since photons do not feel any global factor of the
metric, we may actually discard the term proportional to
$\sqrt{a_0}\, g^*_{\mu\nu}$ in Eq.~(\ref{gtilde}), modulo higher
corrections $\mathcal{O}(a_0/c^4)$. Let us denote as $\bar
g_{\mu\nu}$ the resulting metric, which suffices to derive the
light path. The only remaining modification with respect to
general relativity is the final term of Eq.~(\ref{gtilde}),
proportional to $\partial_\mu\text{GB}\,\partial_\nu\text{GB}$.
In the case of the Schwarzschild solution (\ref{Schw}), this term
only contributes to the radial-radial component of the metric
\begin{equation}
\bar g_{rr} = 1+\frac{2 GM}{rc^2}
+ \frac{4\sqrt{GM a_0}}{c^2}
+\mathcal{O}\left(\frac{1}{c^4}\right).
\label{gbarrr}
\end{equation}
We thus recover the disformal metric constructed in
Sec.~\ref{Disformal} above, and the corresponding light
deflection, Eq.~(\ref{LightDeflectionDM}), reproduces exactly
(by construction) the result predicted by GR in presence
of dark matter. Note that the extra (constant) deflection
angle $2\pi\sqrt{GM a_0}/c^2$ is \textit{also} predicted
in the solar system. It would be possible to suppress
it by introducing another fine-tuned function $b(X)$
multiplying the
$\partial_\mu\text{GB}\,\partial_\nu\text{GB}$
term in Eq.~(\ref{gtilde}). However, this is not
necessary, since this extra deflection angle is
$\approx 2\times 10^{-6} \text{ arcsec}$ for
$M = M_\odot$, i.e., negligible with respect to
present and foreseen experimental accuracy.

Therefore, the above model reproduces both the MOND
dynamics and the correct light deflection in a
particularly simple way. It does not introduce any
superfluous degree of freedom in the vacuum, outside
matter sources: It has the same spectrum as GR,
namely a massless spin-2 field carrying
positive energy, and does not contain any ghost
nor tachyon. Moreover, it satisfies general covariance,
Lorentz invariance (no preferred frame), locality, and
the weak equivalence principle. Binary-pulsar tests
are also obviously passed, since the metric
$g^*_{\mu\nu}$ generated by a binary system takes
exactly the same form as in GR. The only difference
is that the masses slightly differ from those GR
would have predicted for a given amount of matter.
However, since these masses are not directly measured,
but determined from the combination of various
orbital observables, the analysis of binary-pulsar
data is actually \textit{strictly} the same as in GR.

It should also be
underlined that nonminimal couplings to curvature,
such as those introduced in (\ref{gtilde}), do appear
in scalar-tensor theories and in GR itself when
describing the motion of extended
bodies~\cite{Nordtvedt:1994,DEF98,Goldberger:2004jt}.
Indeed, when tidal effects can be neglected (i.e., up to order
$1/c^9$ included, in GR), compact bodies are characterized
only by their mass besides their position and velocity,
and their action can thus be written
$S_\text{pp} = -\int mc \sqrt{-g^*_{\mu\nu} dx^\mu dx^\nu}$.
As soon as tidal deformations start
influencing their motion (from order $1/c^{10}$ in GR),
they must be described by more ``form factors'' characterizing how
they are coupled to derivatives of the metric. Phenomenological
couplings to $R_{\mu\nu\rho\sigma}^{*2}$ and higher derivatives
are thus expected to occur even within GR. Therefore, those
entering Eq.~(\ref{gtilde}) are rather natural, and define
a consistent dynamics for massive point particles.

However, the above model also presents several difficulties.
First of all, it has been fine-tuned to yield the right
phenomenology, but other functions of the curvature tensors
would obviously give a fully different physics. For instance,
it suffices to change the numerical coefficient of the
disformal contribution
$\partial_\mu\text{GB}\,\partial_\nu\text{GB}$ in
Eq.~(\ref{gtilde}) to change the prediction for light
bending by any amount. Therefore, this model is
\textit{unpredictive}, and just manages to fit some
experimental data. Actually, this is probably one of
its main interests: It illustrates, in a very simple
framework, all the hypotheses which are \textit{also}
needed in other models of the literature. As
discussed at the end of Sec.~\ref{Stratified}
above, the best present MOND-like field theory
\cite{Bekenstein:2004ne,Bekenstein:2004ca,Sanders:2005vd}
could also predict a fully different light bending
than GR if one had chosen different functions of
the scalar field in Eq.~(\ref{Disformal3}).

A related problem is that there exist many different realizations
of the physical metric $\tilde g_{\mu\nu}$ yielding the same
phenomenology at lowest order. Indeed, Eq.~(\ref{gtilde}) is
actually a mere rewriting of the needed metric
(\ref{gtilde00})-(\ref{gbarrr}) in a covariant way. Any other
combination of the Riemann curvature tensor and its covariant
derivatives, reducing to the same metric at order
$\mathcal{O}(1/c^2)$, is thus equally valid. For instance, the
crucial factor $(2\sqrt{GMa_0}/c^2) g^*_{\mu\nu}$ may
equivalently be written in terms of $(\Box^*\text{GB})^{-3/4}
R^*_{\mu\rho\sigma\tau} R_{\nu}^{*\rho\sigma\tau}$. Any other
nonvanishing contraction, like $(\nabla^*_\mu
R^*_{\nu\rho\sigma\tau})^2$ or
$R^{\mu\nu}_{*\hphantom{\nu}\rho\sigma}
R^{\rho\sigma}_{*\hphantom{\sigma}\kappa\lambda}
R^{\kappa\lambda}_{*\hphantom{\lambda}\mu\nu}$, may also be used
to build the appropriate MOND potential. [On the other hand,
neither the Ricci tensor $R^*_{\mu\nu}$, the scalar curvature
$R^*$, nor any of their covariant derivatives, can be used to
construct the physical metric $\tilde g_{\mu\nu}$, since they
vanish in vacuum and thereby cannot be felt by a test particle
far from the gravitational source.] The physical metric
(\ref{gtilde}) that we chose above is convenient to write because
it involves very few explicit indices, but it has no special
property which distinguishes it from other possible choices. This
freedom actually teaches us another important lesson: There is no
reason, for a model reproducing the right phenomenology at lowest
order, to remain valid at higher orders. In the present case,
there is an infinity of ways to complete it at higher
post-Newtonian orders, by adding arbitrary combinations of the
curvature tensor and its covariant derivatives. In the case of
the TeVeS model
\cite{Bekenstein:2004ne,Bekenstein:2004ca,Sanders:2005vd}, there
is no reason either to trust its higher post-Newtonian
predictions.

Another related problem is that the variable $X$ defined
in Eq.~(\ref{X1}) may become imaginary in particular
backgrounds where $\Box^*\text{GB} < 0$. One may of
course take its absolute value in order to always
define $X$ consistently, but it would anyway diverge
on the hypersurfaces where $\Box^*\text{GB} = 0$.
The full metric (\ref{gtilde}) is actually ill-defined
on such hypersurfaces, since our ``correcting terms''
$\mathcal{O}(\sqrt{a_0})$ may blow up. This illustrates
once again that the specific choice (\ref{gtilde}) is
not justified by any symmetry principle nor any
fundamental theory. This is just a possible
phenomenological realization of the MOND dynamics
at lowest order, but not a complete theory remaining
consistent in all physical situations. It is always
possible to refine the model to avoid such pathological
situations, as we did in Sec.~\ref{MONDRAQUAL} to avoid
the singularity of RAQUAL models on the hypersurfaces
where $s \equiv g_*^{\mu\nu} \partial_\mu\varphi
\partial_\mu\varphi$ vanishes. However, such
refinements do not make the models more fundamental,
and there is still no reason to believe their
predictions in situations very different from
clustered matter (i.e., solar system and galaxies).

\begin{figure}
\includegraphics[scale=0.75]{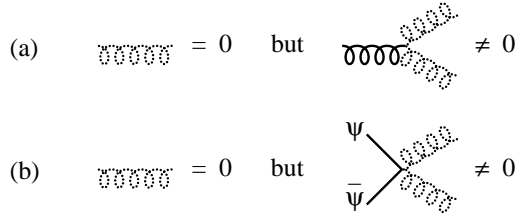}
\caption{Diagrammatic illustration of the negative-energy modes
spoiling the stability of (a)~Lagrangians of the form
$R+\gamma\text{GB}+\alpha(R_{\mu\nu\rho\sigma}^2)^n$, $n \geq 2$,
and (b)~the present nonminimal metric coupling
(\ref{NonminAction})-(\ref{gtilde}). As before, curly lines represent
spin-2 degrees of freedom, namely the usual massless and
positive-energy graviton in plain line, and the massive ghost in
dotted lines. Contrary to Fig.\ref{fig5}, straight lines represent
here matter fields $\psi$ instead of the scalar field $\varphi$. In
both cases (a) and (b), no ghost degree of freedom can be identified
around a flat or empty background. However, a ghost mode does exist in
curved backgrounds in case (a), and within matter in case (b). They
use the background gravitons or matter fields as ``catalyzers'' in
order to propagate and develop instabilities.} \label{fig6}
\end{figure}

But the most serious difficulty of the nonminimally-coupled model
(\ref{NonminAction})-(\ref{gtilde}) is its consistency
\textit{within} matter. First of all, the physical metric $\tilde
g_{\mu\nu}$ should always remain hyperbolic, in order to define a
well-posed Cauchy problem for matter fields. This is clearly the
case if the correcting terms $\mathcal{O}(\sqrt{a_0})$ entering
(\ref{gtilde}) remain small with respect to the flat background
metric $\eta_{\mu\nu} = \text{diag}(-1,1,1,1)$. In more general
situations where they become of order 1, higher-order corrections
are anyway expected to grow too, and the above phenomenological
model just stops being predictive. Once again, the lowest-order
MOND dynamics does not suffice to fully specify the physics in all
regimes. The deadly problem arises in the field equations for
\textit{gravity} within matter. Indeed, since higher derivatives
of $g^*_{\mu\nu}$ enter the matter action, the full Hamiltonian is
not bounded by below, because of Ostrogradski's theorem discussed in
Sec.~\ref{R2} above. Therefore this model is unstable, in the same
way as higher-order gravity. Although one cannot identify any ghost
degree of freedom around a flat empty background, a Lagrangian of
the form $R+\gamma\text{GB}+\alpha(R_{\mu\nu\rho\sigma}^2)^n$
($n \geq 2$, and probably even $R+f(\text{GB})$ as in
Refs.~\cite{Navarro:2005da,Cognola:2006eg,DeFelice:2006pg})
does contain negative-energy modes around curved backgrounds,
and the present model (\ref{NonminAction})-(\ref{gtilde}) involves
negative-energy modes within matter. This is illustrated in terms
of Feynman diagrams in Fig.~\ref{fig6}.

As a simple toy model of such unstable theories, one may
consider for instance the Lagrangian $\mathcal{L} =
-(\partial_\mu\psi)^2 \left[1+\lambda(\partial_\nu\varphi)^2\right]$
in flat spacetime, where $\lambda$ is a nonzero constant. The scalar
$\psi$ is an analogue of the matter fields in our present model
(\ref{NonminAction})-(\ref{gtilde}), and of the usual massless
graviton in Lagrangians of the form
$R+\gamma\text{GB}+\alpha(R_{\mu\nu\rho\sigma}^2)^n$. One
cannot identify any kinetic term for the second scalar
$\varphi$ around a background $\psi = \text{const}$, but it is
obvious that it defines a ghost degree of freedom in backgrounds
where $\lambda(\partial_\mu\psi)^2 < 0$. More generally, one can
easily compute the Hamiltonian $\mathcal{H} =
\bigl[\dot\psi^2 +(\partial_i\psi)^2\bigr]\bigl[1 + \lambda
(\partial_j\varphi)^2\bigr] +\lambda \dot\varphi^2
\bigl[(\partial_i\psi)^2 - 3\dot\psi^2\bigr]$, and prove that
there always exist initial conditions such that
$\mathcal{H}\rightarrow -\infty$. If $\lambda > 0$, it suffices
for instance to choose $\partial_i\psi = \partial_i\varphi = 0$,
$\dot\varphi^2 = 1/\lambda$, $\dot\psi^2 \rightarrow \infty$,
and if $\lambda < 0$, to choose $\partial_i\varphi = 0$,
$\dot \psi = 0$, $(\partial_i\psi)^2\dot\varphi^2 \rightarrow
\infty$. Both cases can be expressed the same way in terms
of the conjugate momenta $\pi_\psi \equiv \partial
\mathcal{L}/\partial\dot\psi$ and $\pi_\varphi \equiv
\partial \mathcal{L}/\partial\dot\varphi$: Whatever the sign
of $\lambda \neq 0$, one gets $\mathcal{H}\rightarrow -\infty$
for $\partial_i\varphi = 0$, $\pi_\psi = 0$, $\partial_i\psi
\rightarrow 0$ and $\pi_\varphi^2 \rightarrow \infty$.

In the present model (\ref{NonminAction})-(\ref{gtilde}), the
simplest way to understand the existence of negative-energy
modes, confined within matter, is to consider a homogeneous
density (either filling the Universe, as in standard cosmological
models, or in the interior of a massive body). Then the matter
action $S_\text{matter}[\psi; \tilde g_{\mu\nu}]$ actually
defines a gravity theory of the form $f(R^*_{\mu\nu\rho\sigma},
R^*_{\mu\nu}, \nabla^*_\lambda R^*_{\mu\nu\rho\sigma}, \dots)$,
which is known to contain a massive spin-2 ghost degree of
freedom~\cite{Stelle:1976gc,Hindawi:1995cu,Tomboulis:1996cy}, as
recalled in Sec.~\ref{R2}. Of course, one may invoke again
higher-order corrections to the phenomenological model
(\ref{gtilde}), which may be able to stabilize it. However, it
remains that it \textit{is} unstable without considering such
higher-order corrections, and therefore that its lowest-order
truncation (\ref{gtilde}) cannot describe consistently massive
bodies. At the classical level, one should analyze carefully this
instability to determine its characteristic timescale. Naively,
one could expect it to be related to the MOND acceleration scale
$a_0$, i.e., of order $c/a_0 \approx 6/H_0 \approx 8\times10^{10}
\text{ yr}$. Since this is several times larger than the age of
the Universe, this would obviously not be a problem for massive
stars. However, other dimensionful constants do exist in the
model. The length scale $\ell_0$ entering (\ref{X1}) does not
change any physical prediction at small enough distances, but
within matter, all masses or coupling constants entering the
matter action $S_\text{matter}$ may contribute to the instability
timescale. Therefore, the analysis of this instability is highly
nontrivial. It may actually happen that it does \textit{not}
develop at the classical level. Indeed, it is quite difficult to
exhibit toy models in which the presence of ghost degrees of
freedom (negative kinetic energy) imply classical instabilities,
whereas they are obvious in presence of tachyons (maximum of a
potential). On the other hand, ghost degrees of freedom are known
to be deadly at the quantum level, because the vacuum then
disintegrates instantaneously~\cite{Woodard:2006nt}.

In conclusion, the nonminimally-coupled model
(\ref{NonminAction})-(\ref{gtilde}) has many interesting features,
notably the fact that it is rather simple and that its predictions
can be easily computed from the general relativistic metric. But
it is unstable and therefore one cannot trust its predictions. It
may still be useful as a phenomenological framework at lowest
order in powers of $\sqrt{a_0}$, since it goes beyond the original
MOND formulation \cite{Milgrom:1983ca}. For instance, when a test
particle is located between two massive bodies, it may happen that
its total acceleration is smaller than $a_0$ although each individual
force happens to be in the standard Newtonian regime ($f > m a_0$). In
such a case, one may wonder what MOND should predict, whereas the
present nonminimally-coupled model unambiguously defines the dynamics
from the local curvature tensor and its covariant derivatives. This
model also allows us to compute consistent MOND effects for
non-spherically symmetric sources. But its main interest is to show
that the MOND dynamics is consistent with most of GR's symmetries. It
also illustrates in a clear way many generic problems of other
MOND-like field theories of the literature, notably their
unpredictiveness (because they are just tuned to reproduce the needed
physics) and their almost systematic instability.

\subsection{A nonminimal scalar-tensor model}\label{NonminST}

The most serious problem of the above nonminimally-coupled
model is its instability because of the presence of
higher derivatives of the metric in the matter action.
By introducing back a single scalar field, we will now show
that this instability may be avoided (but that a more subtle
consistency problem anyway remains, as discussed at the end of the
present section). It will not even be necessary to give it an
aquadratic kinetic term as in Sec.~\ref{SecRAQUAL}. A standard
scalar-tensor theory of the form (\ref{TS1}) will suffice for our
purpose, but with a more complex physical metric $\tilde g_{\mu\nu}$.

The idea is very similar to the previous section~\ref{MatterSpin2}.
We showed in Eqs.~(\ref{GB})-(\ref{BoxGB}) that a combination of
curvature tensors and their covariant derivatives gave us a
\textit{local} access to the baryonic mass $M$ and to the distance
$r$, independently from each other. It was thus possible to
construct from them the needed MOND potential $\sqrt{GMa_0}\ln r$.
In standard scalar-tensor theories, with an exponential
matter-scalar coupling function $A(\varphi) = e^{\alpha\varphi}$,
we know that the scalar field takes the form $\varphi = -\alpha
GM/rc^2 + \mathcal{O}(1/c^4)$ in the vicinity of a spherical body
of mass $M$, if the background value $\varphi_0$ vanishes. Then it
is obvious that combinations of $\varphi$ and its derivative
$\partial_r \varphi = \alpha GM/r^2c^2 + \mathcal{O}(1/c^4)$ also
give us access to $M$ and $r$ independently. One may thus imitate
the previous model (\ref{NonminAction})-(\ref{gtilde}) by defining
\begin{equation}
S = \frac{c^4}{4\pi G} \int\frac{d^4 x}{c}
\sqrt{-g_*}\left\{\frac{R^*}{4}-\frac{s}{2}
- \frac{1}{2}\left(\frac{mc}{\hbar}\right)^2\varphi^2\right\}
+S_\text{matter}\left[\psi; \tilde g_{\mu\nu}\right],
\label{actionNonminST}
\end{equation}
where $s \equiv g_*^{\mu\nu} \partial_\mu \varphi
\partial_\nu \varphi$ denotes as before the standard
kinetic term of the scalar field, and where
$\tilde g_{\mu\nu} \equiv A^2 g^*_{\mu\nu} + B \partial_\mu \varphi
\partial_\nu \varphi$ takes the disformal form (\ref{Disformal1})
with
\begin{subequations}\label{NonminSTDef}
\begin{eqnarray}
A(\varphi, s) &\equiv& e^{\alpha\varphi}
-\frac{\varphi X}{\alpha} h(X),
\label{NonminSTDefA}\\
B(\varphi, s) &\equiv& -
4\frac{\varphi X}{\alpha}\,\frac{1}{s},\\
X &\equiv& \frac{\sqrt{\alpha a_0}}{c}\, s^{-1/4}.
\label{X2}
\end{eqnarray}
\end{subequations}
The function $h(X) = (1+X)^{-1} + \ln(1+X)$ was
chosen in Eq.~(\ref{fX}) to reproduce the MOND
$1/r$ force at large distances but a negligible
one at small distances. The potential
$V(\varphi) = \frac{1}{2}(mc/\hbar)^2\varphi^2$
introduced in Eq.~(\ref{actionNonminST}) has
a negligible influence on local physics
(solar system, binary pulsars, galaxies
and clusters) if the mass it involves is
small enough, say $mc^2/\hbar \alt H$ where
$H$ denotes the Hubble constant. However,
it plays a role in the cosmological evolution
of the scalar field, driving it to the
minimum $\varphi_0 = 0$. This term ensures
thus that the background value of the
scalar field vanishes. Note however that
the above model must be refined
to remain consistent in the cosmological
regime ($s < 0$), on the hypersurfaces
$s = 0$ surrounding any galaxy or cluster
(see Sec.~\ref{MONDRAQUAL} above), and
near the center of any massive body where
$s \rightarrow 0$. It suffices, for instance,
to replace $s$ by $\sqrt{s^2+\varepsilon^2}$
in Eqs.~(\ref{NonminSTDef}), where
$\varepsilon \sim (mc/\hbar)^2 \alt H^2/c^2$
is small enough to have negligible influence
both on Newtonian and MONDian physics. As
already underlined below
Eq.~(\ref{GoodRAQUAL}) and in Sec.~\ref{MatterSpin2},
such a refinement just \textit{illustrates} how
to define a consistent field theory, but there
is no reason to trust its predictions in
the cosmological regime $s \leq 0$. Indeed,
many other functions of $s$ could connect
smoothly to those entering (\ref{NonminSTDef})
for small values of $s$, and they are
\textit{not} imposed by the MOND phenomenology.

Since the above model defines a standard scalar-tensor theory
outside matter (in vacuum), we know that the scalar field takes
the form $\varphi \propto - GM/rc^2 + \mathcal{O}(1/c^4)$ in the
vicinity of a spherical body of mass $M$. The multiplicative
factor is just $\alpha$ in scalar-tensor theories, but it may now
be modified by the MOND-like corrections we added in the physical
metric $\tilde g_{\mu\nu}$. One expects it to remain close to
$\alpha$ if the matter sources are compact enough to be in the
Newtonian regime of the model. This is not only the case for
solar-system bodies and binary pulsars, but also for almost all
the baryonic mass of galaxies and clusters. We will thus merely
assume that $\varphi \approx -\alpha GM/rc^2$ in the present
discussion. We will come back below to the (crucial) problem of
the scalar field equation within matter, and thereby to the
actual coupling constant $\alpha$ entering this expression. The
value $\varphi \approx -\alpha GM/rc^2$ implies that $X =
r/\sqrt{GM/a_0}$, which not only gives us a direct access to the
radial distance $r$, but also involves the MOND characteristic
radius $\ell_0 = \sqrt{GM/a_0}$. This is thus already a bonus
with respect to the purely metric model
(\ref{NonminAction})-(\ref{gtilde}), since one needed to impose
by hand that the arbitrary scale $\ell_0$ was much larger than
the solar-system size but smaller than the radius of any galaxy.
Here, this is obtained automatically from a rather simple
definition of $X$, Eq.~(\ref{X2}). One also finds that $\varphi
X/\alpha = -\sqrt{GMa_0}/c^2$, i.e., precisely (by construction)
the needed factor of the MOND logarithmic
potential.\footnote{This negative value ensures that the matter
field equations are well posed, since inequalities
(\ref{condDisformal}) are satisfied: $A^2+s B = A^2 +
4\sqrt{GMa_0}/c^2 > A^2 > 0$.} Therefore, the physical metric
admits the following post-Newtonian expansion:
\begin{equation}
\tilde g_{\mu\nu} =
\left[1- \frac{2\alpha^2GM}{rc^2}
+\frac{2\sqrt{GMa_0}}{c^2}\,
h\left(\frac{r}{\sqrt{GM/a_0}}\right)\right] g^*_{\mu\nu} +
\frac{4\sqrt{GMa_0}}{c^2} \delta_\mu^r \delta_\nu^r
+ \mathcal{O}\left(\frac{1}{c^4}\right).
\label{gtilde3}
\end{equation}
For distances $r \gg \sqrt{GM/a_0}$, the function
$h(X)$ has been chosen to reproduce the MOND
logarithmic potential, and the disformal contribution
$\propto \delta_\mu^r \delta_\nu^r$ in (\ref{gtilde3})
is precisely the one needed to reproduce the same
light deflection as GR in presence of dark matter,
Eq.~(\ref{LightDeflectionDM}), as already noticed
below Eq.~(\ref{gbarrr}) for the previous nonminimal
metric model. On the other hand, for $r \ll
\sqrt{GM/a_0}$, the function $h(X)$ tends to 1
and does not contribute to any physical
observable, whereas the disformal contribution
$\propto \delta_\mu^r \delta_\nu^r$ still
gives the constant anomalous light deflection
(\ref{LightDeflectionDM}), but too small to be
of observational significance in the solar system.
Therefore, this model does reproduce the right
phenomenology. Solar-system tests just impose
the tight bound $\alpha^2 < 10^{-5}$, as in
standard scalar-tensor theories, because the
$-2\alpha^2GM/rc^2$ term contributes to our
\textit{interpretation} of light deflection,
as explained in Eq.~(\ref{defl2}) above.

Since this model has been devised to reproduce
the standard Brans-Dicke predictions for distances
$r \ll \sqrt{GM/a_0}$, we can also conclude that
it passes binary-pulsar tests, which impose the
constraint $\alpha^2 < 4\times 10^{-4}$
\cite{GefMGX,Damour:2007uf,DEF07}, weaker than the above solar-system
one. Contrary to our discussion of the TeVeS model in Sec.~\ref{PSR},
the (spin-2 and spin-0) gravitational waves are
now strictly the same as in Brans-Dicke theory,
even at large distances where MOND effects can
be observed on material test bodies. Indeed, this
model is a \textit{pure} scalar-tensor theory in
vacuum, so that gravitational waves themselves
do not feel any MONDian effect. This is an
interesting feature of this model, and also
its main difference with the previous ones proposed
in the literature. In previous models, the helicity-0
(scalar) waves undergo MONDian effects at large
distances from massive sources, whereas they follow
the same geodesics (of the Einstein metric
$g^*_{\mu\nu}$) as the helicity-2 waves in the
present model. Unfortunately, this different
prediction cannot be used to experimentally
discriminate between these models, because
the solar-system constraint $\alpha^2 < 10^{-5}$
on the matter-scalar coupling constant implies
that gravitational-wave detectors (made of matter)
are almost insensitive to scalar waves. On the
other hand, let us underline that they could
in principle discriminate between such MOND-like
field theories and the standard dark matter
hypothesis. Indeed, all RAQUAL and disformal
models, including TeVeS and the present one,
predict that the helicity-2 waves are not directly
coupled to the scalar field, and thereby are not
deflected by what we usually interpret as dark
matter haloes, contrary to light (coupled to the
physical metric $\tilde g_{\mu\nu}$ describing
MONDian effects). Therefore, when the LISA
interferometer will detect gravitational waves
emitted from optically known sources, their
direction should be slightly different from
the optical direction if there exists at least
one intermediate gravitational deflector.
Unfortunately again, the LISA angular resolution
will be about the angular size of the Moon (30
arcmin), much too large to measure such a
deflection angle between electromagnetic and
gravitational waves. But the \textit{time-delay}
between gravitational waves and optical or
neutrino pulses from supernovae would allow
us to discriminate between MOND and the dark
matter hypothesis, as recently shown
in \cite{Kahya:2007zy}.

The deadly problem of the previous model
(\ref{NonminAction})-(\ref{gtilde}) was its Ostrogradskian
instability. One may thus wonder if the present scalar-tensor
variant avoids it. Indeed, since the matter action \textit{a
priori} involves covariant derivatives of the matter fields, it
depends on the first derivatives of the physical metric $\tilde
g_{\mu\nu}$, and thereby of the \textit{second} derivatives of
$\varphi$, see Eqs.~(\ref{NonminSTDef}). The same instability as
before, caused by second and higher derivatives, seems thus to
spoil the present model too. However, let us recall that the
standard model of particle physics only involves gauge bosons
described by 1-forms, and fermions described by Dirac spinors. If
is well known that all covariant derivatives may be replaced by
ordinary ones in forms, because their antisymmetry cancels all
the Christoffel symbols $\tilde\Gamma^\lambda_{\mu\nu}$.
Therefore, there is actually no derivative of $\tilde g_{\mu\nu}$
involved in the action describing the dynamics of gauge bosons,
and only $\varphi$ and its \textit{first} gradient
$\partial_\mu\varphi$ enter it. On the other hand, the action of
a spinor in curved spacetime does depend of the first derivative
of the metric (via the derivative of a tetrad, cf. Sec.~12.5 of
\cite{Weinberg}), but only \textit{linearly}. Schematically, it
reads $\bar\psi \tilde g^m/\!\!\!\partial\psi + \bar\psi \tilde
g^n(/\!\!\!\partial \tilde g)\psi$. Therefore,
$\nabla^*_\mu\partial^{\vphantom{*}}_\nu\varphi$ enters this
action, but if one tries to define a conjugate momentum $p_2
\equiv \partial\mathcal{L}/\partial\ddot\varphi$, then its
expression does not depend on any second derivative of $\varphi$.
Therefore, one cannot express $\ddot\varphi$ as a function of
$\varphi$, $\dot\varphi$ and $p_2$ as in Eq.~(\ref{ddotq}), and
we are thus precisely in the case of a degenerate Lagrangian for
which the Ostrogradskian instability does not occur. More
explicitly, all second time derivatives of the scalar field,
$\ddot\varphi$, may be eliminated by partial integration. In
conclusion, the action of the standard model of particle physics
actually does \textit{not} depend on $\ddot\varphi$, and the
above model is thus free of Ostrogradskian instability.

However, this absence of generic instability does
not suffice to prove that the model is indeed stable.
One must also check that the scalar field equation
remains consistent \textit{within} matter. In the
most general case of arbitrary matter fields and
arbitrary initial conditions, this problem is quite
involved because the matter action depends on the
derivative of the scalar field. As discussed in
Sec.~\ref{Disformal}, one should write all the
field equations, including for matter, and prove
that the second derivatives of the scalar field
define a hyperbolic operator. The simpler case of
a pressureless perfect fluid will however suffice
to exhibit the subtle difficulties appearing in
this class of models. The generalized RAQUAL kinetic
term
\begin{equation}
\tilde f(\varphi, s, \partial_0\varphi) =
s +\frac{8\pi G\bar\rho}{c^2} A
\sqrt{1-(\partial_0\varphi)^2 B/A^2},
\label{NonminSTftilde}
\end{equation}
written here in the rest frame of the perfect fluid
and in coordinates locally diagonalizing the Einstein
metric $g^*_{\mu\nu} = \text{diag}(-1,1,1,1)$, should
satisfy inequalities (a1) and (b1) quoted below
Eq.~(\ref{ftilde}). The contribution of the scalar field
to the Hamiltonian should also be bounded by below.
Since the model has been constructed to reproduce
the MOND dynamics at the lowest (Newtonian) order,
without imposing its behavior at higher post-Newtonian
orders nor in the cosmological regime, let us focus
on the conditions (a2), (b2) and (c2) that $\tilde f$
should at least satisfy at the Newtonian level.
As mentioned in Sec.~\ref{Disformal}, condition (c2)
is actually implied by (a2), because we imposed
a positive value of the coupling function $B$
in order to increase light deflection in the MOND
regime. Moreover, one can check easily that the terms
proportional to $(\partial_0\varphi)^2$ may be neglected
in (a2) and (b2). On the other hand, note that
$A'/c^2$ and $sA''/c^2$ are actually of Newtonian
order $\mathcal{O}(1/c^0)$, in spite of the explicit
factor $1/c^2$. At this Newtonian order, it suffices
thus to check conditions
\begin{enumerate}
\item[(a2)]$\tilde f' > 0$,
\item[(b2)]$2 s \tilde f'' + \tilde f' > 0$,
\end{enumerate}
where $\tilde f$ may be approximated as
$\tilde f \approx s + (8\pi G\bar \rho/c^2) A$,
as if we were considering a RAQUAL model in vacuum;
see conditions (a) and (b) of Sec.~\ref{SecRAQUAL}.
In such a case, these hyperbolicity conditions
suffice to ensure that the scalar-field contribution
to the Hamiltonian is bounded by below. [The coupling
function $B$ plays an even more negligible role in
the Hamiltonian than in the hyperbolicity conditions,
namely of order $\mathcal{O}(1/c^4)$ smaller than the
lowest term involving the matter density $\bar \rho$.]

In the Newtonian regime, i.e., when the MONDian correction
involving $\varphi X/\alpha$ in Eqs.~(\ref{NonminSTDef}) are
negligible, it is obvious that the hyperbolicity conditions (a2)
and (b2) are satisfied. The model reduces indeed to the standard
Brans-Dicke theory, whose consistency is well known. If all
matter in the Universe was clustered as compact enough bodies,
like stars and planets, then this Newtonian regime of the model
would always be reached when considering the scalar field
equation \textit{within} matter, and its Cauchy problem would
always be well posed. As mentioned below
Eqs.~(\ref{NonminSTDef}), there would still remain a difficulty
near the center of any body, where the spatial gradient of the
scalar field tends to 0, and therefore where MOND corrections may
become large (before they are saturated by our \textit{ad hoc}
replacement of $s$ by $\sqrt{s^2+\varepsilon^2}$). However, it is
not even necessary to study such centers of compact bodies to
point out an extremely serious difficulty of this model. Indeed,
there does exist dilute gas in the outer regions of galaxies
where the MOND phenomenology applies. One should therefore check
whether conditions (a2) and (b2) hold when the $\varphi X/\alpha$
corrections start dominating even over the Newtonian force.
Actually, for the coupling functions (\ref{NonminSTDef}), one
finds that condition (b2) is easily satisfied, but it happens
that (a2) is \textit{not}. Indeed, $\tilde f' \approx 1 + (8\pi
G\bar \rho/c^2) A'$, where $A' = (h+X\, dh/dX)\, \varphi
X/4\alpha s$. The coefficient $(h+X\, dh/dX)$ varies from 1 to
$3.3$ in the range of distances relevant to the MOND regime,
namely $X = r/\sqrt{GM/a_0} < 10$. On the other hand, $\varphi
X/\alpha s = -(\sqrt{GMa_0}/c^2)(\alpha GM/r^2c^2)^{-2}$ is a
huge negative number, because the kinetic term $s$ of the scalar
field appears in its denominator. Even multiplied by the small
gas density $G\bar \rho/c^2$ existing in the outer regions of
galaxies, this negative contribution to $\tilde f'$ dominates by
several orders of magnitude over the 1 coming from the standard
kinetic term in vacuum. In other words, $\tilde f' < 0$ within
such a gas, and the scalar perturbations are thus ill defined. In
spite of its nice features and its relative simplicity, the model
(\ref{NonminSTDef}) is thus inconsistent, because the scalar
field equation is not always hyperbolic \textit{within} matter.
Such a difficulty was not discussed in previous works on
MOND-like field theories, although it seems generic for
disformally-coupled scalar-tensor models (\ref{Disformal1}).

Of course, one may try to tune the coupling function $A$,
Eq.~(\ref{NonminSTDefA}), to cure this inconsistency. For
instance, since $\varphi X/\alpha$ takes the constant value
$-\sqrt{GMa_0}/c^2$ outside matter, one may add to $A$ any
function of $\varphi X/\alpha$ without changing its spatial
gradient, i.e., the force felt by a test mass. However, it is
straightforward to show that there is not much freedom for such a
function. In order to ensure condition (a2) without spoiling
(b2), it must be close to $k (\varphi X/\alpha)^{-2}$, where $k$
is a constant. This precise expression has the peculiarity not to
contribute to condition (b2), whereas it adds a positive
contribution to $\tilde f'$. The crucial difficulty is that the
proportionality constant $k$ must take a fixed value, imposed in
the action of the model, and should not be chosen independently
for each galaxy or cluster (cf. our discussion in Sec.~\ref{Lit}
above). To guarantee condition (a2) in any situation, $k$ should
thus be chosen large enough, and one finds that $k > 2
(GM_\text{cluster} a_0/c^4)^{3/2}$ would suffice, where
$M_\text{cluster}$ is the largest cluster mass existing in the
Universe. Obviously, this appearance of another dimensionful
constant in the model underlines that it would be even more fine
tuned than before. [One might be tempted to simply set $k = 1$ to
make the model more natural, but the coupling constant $A$ would
then take large values instead of admitting a post-Newtonian
expansion $A = 1 +\mathcal{O}(1/c^2)$, and the predicted MONDian
effects would then be reduced by the inverse of this large
value.] But it happens that such a tuning of the coupling
function $A$ actually does \textit{not} work. Indeed, although a
term $k (\varphi X/\alpha)^{-2}$ would not contribute to the
force felt by a test mass, it does contribute to the scalar field
equation $2\nabla^*_\mu(\tilde f' \nabla_*^\mu\varphi) =
\partial\tilde f/\partial\varphi$. In the MOND regime, it even
dominates over the other contributions in both sides of this
field equation, in spite of its coefficient $(8\pi G \bar
\rho/c^2)$ involving the small gas density $\bar \rho$. In a
region where $\bar\rho$ is almost constant, one may thus divide
each side by it, and one gets a scalar field equation fully
different from the one satisfied in vacuum,\footnote{This problem
actually already arises with the initial coupling function $A$,
Eq.~(\ref{NonminSTDefA}), whose MONDian terms dominate within the
gas in outer regions of galaxies. However, one can check that the
scalar field and its derivative keep the same orders of magnitude
as the vacuum (Brans-Dicke) behavior $\varphi \approx -\alpha
GM/rc^2$ we assumed, so that the phenomenological predictions
remain correct up to factors of order unity. On the other hand,
the scalar field behavior is fully changed within this outer gas
when the correcting term $k (\varphi X/\alpha)^{-2}$ is
included.} although it does not depend any longer on the precise
value of $\bar\rho$. Its exact solution is difficult to find, but
the signs involved anyway show that $\varphi$ would quickly tend
to $0$ within matter (in the MOND regime), instead of keeping the
form $\varphi \approx -\alpha GM/rc^2$ that we assumed from the
beginning of this section to construct the present model. Even
without invoking the signs on the different terms in this field
equation, it is anyway obvious to check that $\varphi \approx
-\alpha GM/rc^2$ cannot be valid within the gas in outer regions
of a galaxy. In other words, the extra force predicted by such a
model, if any, would \textit{not} take the MOND form that we
wished. In conclusion, a tuning of the coupling function $A$ to
ensure the hyperbolicity of the scalar field equation within
matter suffices to ruin the predictions of the model.

The above hyperbolicity problem underlines again how difficult it
is to construct a MOND-like field theory, even fine tuned. In
spite of the presence of several free functions in the model,
which could \textit{a priori} fit any phenomenology one wishes,
the consistency of the field equations within matter is a strong
enough condition to rule it out. One should however underline
that this conclusion cannot be considered as a no-go theorem. The
same framework as Eqs.~(\ref{actionNonminST})-(\ref{NonminSTDef})
might be able to reproduce consistently the MOND dynamics with
different coupling functions $A$ and $B$. If the scalar field
does not take the form $-\alpha GM/rc^2$ in the MOND regime, then
specific combinations of $\varphi$ and its kinetic term $s$ might
be able to give us access independently to the crucial factor
$\sqrt{GMa_0}/c^2$ and to the radius $r/\sqrt{GM/a_0}$. The MOND
metric (\ref{gtilde3}) might thus be obtained again, and our
discussion above shows that the Ostrogradskian instability would
still be avoided. We expect the hyperbolicity conditions (a2) and
(b2) to still cause serious problems within matter in the MOND
regime, but we do not have any proof. Our investigations indicate
anyway that the coupling functions $A$ and $B$ would very
probably involve unnaturally complicated expressions (like large
powers of $s$ or transcendental functions).

\subsection{The Pioneer anomaly}\label{Pioneer}
As mentioned in Sec.~\ref{Intro}, the two Pioneer spacecrafts
exhibited an anomalous extra acceleration $\delta a \approx
8.5\times 10^{-10}\, \text{m}.\text{s}^{-2}$ towards the Sun,
between 30 and 70 AU
\cite{Anderson:2001sg,Nieto:2003rq,Turyshev:2005ej}. From a
theoretical point of view, this anomaly is \textit{a priori} much
easier to explain than dark matter by a fine-tuned model modifying
Newtonian gravity. Indeed, it has been observed in our single solar
system, mainly characterized by the solar mass $M_\odot$, whereas
galaxy and cluster rotation curves have confirmed the relation
$M_\text{dark} \propto \sqrt{M_\text{baryon}}$ for many different
values of the masses. A possible explanation (or, rather,
description) of the Pioneer acceleration would be to couple
matter to a scalar field, whose potential is merely fitted to
reproduce observed data. However, in order not to spoil the
precision tests of Kepler's third law in the solar system, such a
potential should not manifest at distances smaller than Saturn's
orbit ($\approx 10$ AU), and cause an almost constant extra force
beyond $30$ AU. Besides the fact that this would be quite
unnatural, one may also wonder if this would correspond to a
stable theory. Indeed, the simplest stable potential for a scalar
field is a mass term $\propto \varphi^2$, and it causes a force
which \textit{decreases} (exponentially) as the distance grows.
Even within the \textit{a priori} simpler case of the Pioneer
anomaly, the construction of a stable model does not seem
obvious, thus.

A first, important, step has been understood in the series of
papers \cite{Jaekel:2005qe,Jaekel:2005qz,Jaekel:2006me} by
Jaekel and Reynaud. From a phenomenological viewpoint, it is much
more natural to describe the Pioneer anomaly by a modification of
the spatial metric $g_{ij}$ rather than the Newtonian potential
involved in $g_{00}$. Indeed, if the metric is written in the
Schwarzschild form (\ref{SchwarzschildForm}), then the geodesics
equation for a test mass $m_0$ reads
\begin{equation}
\mathcal{E}^2/c^2 - \left(m_0^2 c^2
+ \mathcal{J}^2 u^2\right)\mathcal{B} =
\mathcal{J}^2 \mathcal{A}\mathcal{B}\, u'^2,
\label{geodSchw}
\end{equation}
where $\mathcal{E}$ is the particle conserved energy (including
rest mass $m_0 c^2$), $\mathcal{J}$ its angular momentum, $u
\equiv 1/r$, and a prime denotes here derivation with respect to
the angle in polar coordinates. For a circular orbit, $u' = 0$ so
that the spatial metric component $g_{rr} = \mathcal{A}$ does not
enter this equation.\footnote{Beware that in the Schwarzschild
coordinates we are using, the post-Newtonian term in the time
component $g_{00}$ of the metric is proportional to the
combination $\beta - \gamma$ of the Eddington parameters, whereas
it is proportional to $\gamma$ in the radial component. It would
thus be misleading to claim that circular orbits are merely
independent of $\gamma$. The correct statement should underline
that this is only true for a fixed value of $\beta - \gamma$.}
More generally, its right-hand side is proportional to the square
of the orbit's eccentricity, $e^2$. In conclusion, planets on
nearly circular orbits are almost insensitive to a small
modification of $\mathcal{A}$, whereas they would feel any extra
potential entering the time component $\mathcal{B}$. On the other
hand, hyperbolic trajectories,\footnote{Recall that the virial
theorem implies $v^2/c^2 = \mathcal{O}(GM/rc^2)$ for bound
orbits, whereas hyperbolic ones reach a nonzero asymptotic
velocity $v^2/c^2 \gg GM/rc^2$ for large distances $r$. As
compared to the time-time contribution $g_{00} c^2 dt^2$ to the
line element $ds^2$, the spatial one $g_{ij} dx^i dx^j = g_{ij}
v^i v^j dt^2$ is thus of post-Newtonian order for bound orbits,
but much larger for hyperbolic ones (and actually of the same
order as the time-time contribution for light rays, yielding the
famous $(1+\gamma)$ factor of the light deflection formula).}
like those of light rays and the Pioneer spacecrafts, are
directly sensitive to a modification of $\mathcal{A}$. This is
actually the crucial difference between galaxy rotation curves
(which exhibit a discrepancy with Newton's law for the
\textit{circular} orbits of outer stars) with the Pioneer anomaly
(occurring for hyperbolic trajectories). While it was necessary
to modify the time-time metric $g_{00}$ to reproduce the MOND
phenomenology (together with the spatial one $g_{ij}$ to predict
the correct light deflection), it is possible to account for the
Pioneer anomaly by a modification of $g_{rr}$ alone.

However, post-Newtonian tests in the solar system, notably those
of light deflection and Shapiro time delay, severely constrain
the magnitude of the deviations from general relativity in the
spatial component $\mathcal{A}$. In particular, it is impossible
to account for the Pioneer anomaly within the parametrized
post-Newtonian formalism
\cite{Eddington,Schiff,Baierlein,Nordtvedt:1968qs,Will71,
Will-Nordtvedt72,Willbook}, which describes all metrically-coupled
theories of gravity at order $\mathcal{O}(1/c^2)$ with respect to
the Newtonian force, under the hypothesis that the gravitational
field does not involve any characteristic length scale. This
hypothesis implies that all post-Newtonian parameters, including
the famous Eddington parameters $\beta$ and $\gamma$, must be
independent of the radial distance $r$. Within this wide class of
theories, the Pioneer anomaly is merely inconsistent with other
precision tests in the solar system.

The lesson of
Refs.~\cite{Jaekel:2005qe,Jaekel:2005qz,Jaekel:2006me} is that a
spatial dependence of the Eddington parameter $\gamma$, going
thus beyond the PPN formalism, suffices to account for the
Pioneer anomaly without spoiling the classic tests. This
parameter should of course remain very close to the general
relativistic value $\gamma = 1$ at small radii, so that
light-deflection and time-delay predictions are consistent with
observation, but it may change at distances $\agt 30$ AU to fit
Pioneer data. In their first analysis,
Refs.~\cite{Jaekel:2005qe,Jaekel:2005qz} showed that an
expression $\gamma = 1 + \kappa r^2$ would \textit{a priori}
suffice, for an appropriate constant $\kappa$ tuned to reproduce
the Pioneer extra acceleration. In other words, the radial
component of the metric should take the form $g_{rr} = 1 +
2\gamma GM_\odot/rc^2 + \mathcal{O}(1/c^4) = 1 +
(2GM_\odot/c^2)(1/r + \kappa r) + \mathcal{O}(1/c^4)$, in which a
small term linear in $r$ corrects the general relativistic $1/r$
potential. The refined discussion of Ref.~\cite{Jaekel:2006me}
actually leads to a more complex radial dependence of the
parameter $\gamma$, cubic in $r$, but the qualitative conclusion
remains the same.

Let us just summarize here the simpler result of
\cite{Jaekel:2005qe,Jaekel:2005qz}. The extra acceleration
$\delta \mathbf{a}$ caused by a term $k_n r^n$ added to $g_{rr} =
\mathcal{A}$ is a straightforward consequence of the geodesics
equation (\ref{geodSchw}). One finds
\begin{equation}
\delta \mathbf{a} = - \frac{1}{2}\,k_n\,n\, r^{n-1}\, v^2\,
\mathbf{n},
\label{deltaa}
\end{equation}
where $v$ denotes the particle's velocity and $\mathbf{n} \equiv
\mathbf{x}/r$ the radial unit vector (pointing away from the
Sun). Since the velocities of the Pioneer spacecrafts were almost
constant on the range of distances where their anomalous
acceleration was observed, one thus needs $n = 1$ and a positive
value of $k_1$ to obtain a constant $\delta \mathbf{a}$ directed
towards the Sun. Actually, one does not directly observe the
acceleration of the spacecrafts, but deduce it from Doppler
tracking. One should thus also take into account the fact that
electromagnetic waves travel in the same metric, and feel the
anomalous potential too. References
\cite{Jaekel:2005qe,Jaekel:2005qz} show that this leads to a
factor 2 difference in the prediction of the recorded anomalous
acceleration, i.e., $\delta a = k_1\, v^2$. Numerically, for the
Pioneer spacecrafts, $v \approx 1.2\times 10^4
\text{m}.\text{s}^{-1}$ and $\delta a \approx 8.5\times 10^{-10}
\text{m}.\text{s}^{-2}$, therefore
\begin{equation}
k_1 \approx \frac{0.5\,
\text{m}.\text{s}^{-2}}{c^2} \approx (10^6 \text{AU})^{-1}.
\label{k1}
\end{equation}
If interpreted as a varying Eddington parameter $\gamma = 1 +
\kappa r^2$, with $2\kappa G M_\odot/c^2 = k_1$, this gives
$\kappa \approx 45\, \text{AU}^{-2}$. The refined analysis of
Ref.~\cite{Jaekel:2006me} yields a quadratic extra potential $k_2
r^2$ with $k_2 \approx 8\times 10^{-8}\, \text{AU}^{-2}$, i.e., a
varying Eddington parameter $\gamma \approx 1 + 4
(r/\text{AU})^3$. It further tunes this extra potential in the
form $k_2 (r-r_\oplus)^2 + k'_2 (r-r_\oplus)$, where $r_\oplus$
is the radius of the Earth's orbit and $k'_2$ another constant
characterizing the anomalous acceleration at $r = r_\oplus$. We
shall not take into account such refinements in the following,
but just mention that they would not change our conclusion.

The last difficulty is to exhibit a consistent field theory
reproducing such a radial dependence of $\gamma$, both stable
and admitting a well-posed Cauchy problem. Because of the
generic difficulties with modified gravity models recalled
in Sec.~\ref{Field}, and notably the stability issues of
higher-order gravity discussed in Sec.~\ref{R2}, it is
\textit{a priori} not obvious that such a theory exists,
and Refs.~\cite{Jaekel:2005qe,Jaekel:2005qz} actually write
nonlocal field equations without discussing these crucial
mathematical requirements. But it happens that the scalar-tensor
framework of Sec.~\ref{NonminST} above suffices to reproduce
consistently a varying Eddington parameter $\gamma$. Indeed,
the disformal contribution $B \partial_\mu \varphi \partial_\nu
\varphi$ to the physical metric (\ref{Disformal1}) is precisely
what is needed to change only the radial component $g_{rr}$, i.e.,
what is denoted as $\mathcal{A}$ in the geodesics equation
(\ref{geodSchw}). As in Sec.~\ref{NonminST}, one may thus consider a
scalar field weakly coupled to matter via a Brans-Dicke conformal
factor $A^2(\varphi) = \text{exp}(2\alpha\varphi)$, and build the
appropriate correction\footnote{A correction to the time-time metric
$\tilde g_{00}$ would also be easy to define by using $\varphi$ and
$\partial_r\varphi$. This might be necessary to avoid too large
modifications of the planets' perihelion shifts, which depend on both
of the PPN parameters $\beta$ and $\gamma$.} to the radial metric by
using $\varphi
\approx -\alpha GM_\odot/rc^2$ and $\sqrt{s} = \partial_r\varphi
\approx \alpha GM_\odot/r^2 c^2$. Since we do not have any
experimental evidence that the Pioneer anomalous acceleration is
related to the solar mass (nor its square root nor any function of
it), several expressions can actually be used to build an extra
gravitational potential linear in $r$, for instance $1/\varphi$ or
$s^{-1/4}$, or any function like $\varphi^{-1}f(\varphi^4/s)$. The
simplest choice would be to impose $B = -\lambda_5/\varphi^5$,
where $\lambda_5$ is a positive constant, i.e., a physical metric
of the form
\begin{equation}
\tilde g_{\mu\nu} = e^{2\alpha\varphi} g^*_{\mu\nu} -
\frac{\lambda_5}{\varphi^5}\, \partial_\mu \varphi
\partial_\nu \varphi,
\label{DisformalPioneer}
\end{equation}
so that $\tilde g_{rr} = e^{2\alpha\varphi} g^*_{rr} -
(\lambda_5/\varphi)(s/\varphi^4)$ involves a positive correction
proportional to $r$. [The more complex, quadratic, potential
derived in Ref.~\cite{Jaekel:2006me} can be obtained in the
same way, by choosing for instance $B = \mu_6/\varphi^6
+\mu_5/\varphi^5 +\mu_4/\varphi^4$, where $\mu_{4,5,6}$ are three
constants such that $B > 0$, in order to predict an extra
acceleration directed towards the Sun.] Of course, expression
(\ref{DisformalPioneer}) is valid only in the vicinity of the Sun,
where $s > 0$ and $\varphi < 0$. It should be refined to remain
valid in other regimes, but the Pioneer data do \textit{not} tell
us how. Since the condition $A^2 + s B > 0$ must be satisfied to
ensure the hyperbolicity of the matter field equations, see
Eq.~(\ref{condDisformal}), it might be better to consider coupling
functions $B = \lambda_9 s/ |\varphi|^9$ or $B =
\lambda_1/(|\varphi| s)$. However, it would also suffice to
multiply the above $B = -\lambda_5/\varphi^5$ by a function $h(s)$,
such that $h(s) = 0$ for $s < 0$ and $h(s) = 1$ for $s > (\alpha
GM_\odot/r_\text{max}^2c^2)^2$, where $r_\text{max} \approx
70\,\text{AU}$ is the largest distance for recorded Pioneer data.
The above metric (\ref{DisformalPioneer}) should also be refined to
avoid divergences when $\varphi$ passes through zero with a
nonvanishing derivative. A possible solution would be to replace
$-\varphi$ by $\sqrt{\varphi^2+\varepsilon^2}$, where $\varepsilon
\ll \alpha GM_\odot/r_\text{max}c^2$ is a tiny dimensionless
number, negligible for the Pioneer phenomenology but eliminating
our anomalous potential at larger distances. Numerically, one would
need $\varepsilon \ll 10^{-10} \alpha < 3\times 10^{-13}$, if one
takes into account the solar-system constraints on the
matter-scalar coupling constant $\alpha < 3\times 10^{-3}$. On the
other hand, to reproduce the observed Pioneer anomaly, the constant
$\lambda_5$ entering Eq.~(\ref{DisformalPioneer}) should take the
numerical value
\begin{eqnarray}
\lambda_5 \approx \alpha^3 (10^{-4}\text{m})^2 < (2\times
10^{-8}\text{m})^2,
\label{lambda5}
\end{eqnarray}
the upper limit corresponding to the largest
allowed value for $\alpha$. [The other possible expressions for $B$
quoted above would need factors
$\mu_6 \approx \alpha^4 (4\times 10^{-9}\text{m})^2 < (4\times
10^{-14}\text{m})^2$,
$\lambda_9 \approx \alpha^5
(0.4\,\text{m})^4 < (3\times 10^{-4}\text{m})^4$ and
$\lambda_1 \approx 10^{-14} \alpha < 3\times 10^{-17}$ respectively.]

Such numerical constants underline that we are merely fine-tuning a
model to account for experimental data. However, the crucial point
that we wish to stress is that the model (\ref{actionNonminST})
with a physical metric (\ref{DisformalPioneer}) is stable and
admits a well-posed Cauchy problem. The hyperbolicity conditions
(a1) and (b1) of Sec.~\ref{Disformal} are indeed trivially
satisfied, because both coupling functions $A = \text{exp}(\alpha
\varphi)$ and $B = -\lambda_5/\varphi^5$ are independent of the
kinetic term $s$, and because $B > 0$ in the regime where this
model is defined. Actually, for such a simple case, one may even
check the hyperbolicity of the fully general scalar field equation
(\ref{EqPhidanslamatiere}) within any kind of matter. The fact that
the Hamiltonian is bounded by below is also an obvious consequence
of the positivity of $B$. [Note that all these consistency
conditions are also obvious for the more complex expression $B =
\mu_6/\varphi^6 +\mu_5/\varphi^5 + \mu_4/\varphi^4$ mentioned above
to reproduce the quadratic potential derived in
Ref.~\cite{Jaekel:2006me}. On the other hand, they are less trivial
for the other possible expressions for $B$ quoted above, such as $B =
\lambda_9 s/ |\varphi|^9$ or $B = \lambda_1/(|\varphi| s)$.
However, they may still be easily checked at the lowest
post-Newtonian level, where the hyperbolicity conditions reduce to
(a2), (b2) and (c2) of Sec.~\ref{Disformal}, and where all terms
proportional to $u_*^\mu \partial_\mu\varphi$ may be neglected.]

The final consistency check is to show that the scalar field takes
indeed the form $\varphi \approx -\alpha GM_\odot/rc^2$ in the
solar system, since we assumed this expression to build our
anomalous potential. One just writes the scalar field equation
within matter, $2\nabla^*_\mu(\tilde f' \nabla_*^\mu\varphi) =
\partial\tilde f/\partial\varphi$, where $\tilde f$ is defined in
Eq.~(\ref{ftilde}). If the Sun were an isolated body, the scalar
field would be generated by it only, and one would have
$u_*^\mu\partial_\mu\varphi = 0$ strictly (up to a negligible time
variation of the scalar field possibly imposed by the cosmological
evolution). In such a case, the model reduces to Brans-Dicke
theory, and one gets thus obviously $\varphi \approx -\alpha
GM_\odot/rc^2$. When considering other bodies in the solar system,
the largest effects due to the disformal coupling function $B$
occur at large distances from the Sun. One finds that the coupling
constant $\alpha$ is then renormalized into $\alpha + \mathcal{O}(r
v^2/GM_\odot) k_1 r/\alpha$, where $k_1 \approx (10^6
\text{AU})^{-1}$ is the coefficient (\ref{k1}) of the linear
anomalous potential we added to the radial metric. The coefficient
$\mathcal{O}(r v^2/GM_\odot)$ is of order unity for planets because
of the virial theorem, and solar-system tests impose $\alpha^2 <
10^{-5}$. For the largest allowed value of $\alpha$, we can thus
conclude that our anomalous potential does not contribute
significantly to the scalar field equation up to $r \sim 10\,
\text{AU}$ (i.e., Saturn's orbit). For outer planets, $\alpha$ may
be increased by a factor $\alt 4$, but since their masses are
negligible with respect to that of the Sun, the approximate value
of the scalar field near the Pioneer spacecrafts remains anyway
$\varphi \approx -\alpha GM_\odot/rc^2$. The largest modification
of the matter-scalar coupling constant occurs for the Pioneer
spacecrafts themselves (or any object on a hyperbolic orbit at
similar distances from the Sun), since $r v^2/GM_\odot \approx 10$
for $r = 70\,\text{AU}$ and the spacecrafts' velocity $v \approx
1.2\times 10^4 \text{ m}.\text{s}^{-1}$. This means that this
coupling constant is increased by a factor $\sim 80$ at such
distances, explaining incidentally why these spacecrafts are
sensitive to the scalar field. But the local perturbation of
$\varphi$ they induce does not act on their own trajectory, and it
remains negligibly small for other bodies (as if we were
considering 80 spacecrafts instead of one, but weakly coupled to
the scalar field as in Brans-Dicke theory).

In conclusion, although the above model
(\ref{actionNonminST})-(\ref{DisformalPioneer}) is not justified
by any underlying symmetry principle, and should be considered
more as a fit of Pioneer data than as a predictive theory, it
proves that it is possible to account for this anomalous
acceleration in a consistent field theory, while satisfying all
other solar-system data (and binary-pulsar tests too, since the
model reduces to Brans-Dicke theory in vacuum). Let us recall
that there is no reason to trust Eq.~(\ref{DisformalPioneer}) in
a different regime than the one relevant to the Pioneer
spacecrafts, and notably in cosmology. Note also that there exist
many different ways to reproduce the same phenomenology, as
exhibited by various expressions of the disformal coupling
function $B$ above. Our only addition with respect to
Refs.~\cite{Jaekel:2005qe,Jaekel:2005qz,Jaekel:2006me} is that a
varying post-Newtonian parameter $\gamma$ is indeed possible
within a stable model admitting a well-posed Cauchy problem.

\section{Conclusions}\label{Concl}

One of the aims of this article was to clarify the mathematical
consistency of various field theories proposed in the literature
to reproduce the MOND phenomenology. We underlined that besides
experimental constraints which need to be satisfied, including
new ones that we discussed (notably binary-pulsar tests), there
are also several basic requirements that a field theory needs to
meet: stability (i.e., boundedness by below of its Hamiltonian)
and well-posedness of its Cauchy problem (i.e., notably the fact
that all field equations must be hyperbolic). In our opinion, the
TeVeS model \cite{Bekenstein:2004ne,Bekenstein:2004ca} is
presently the most promising, but it also presents several
serious difficulties discussed in Sec.~\ref{Section4}, and cannot
be considered yet as a fully consistent theory. We examined
another possible route to reproduce the MOND phenomenology,
within the class of the already studied RAQUAL models
\cite{Bekenstein:1984tv}, but with a different spirit. Its
originality is that the theory is particularly simple in vacuum
(pure general relativity or Brans-Dicke theory). However, the
analysis of the field equations \textit{within} matter exhibited
a deadly inconsistency. Although this framework did not provide
any serious alternative to TeVeS yet, we anyway believe further
work along its line might do so. Its interest also lies in its
relative simplicity. The model proposed in
Sec.~\ref{MatterSpin2}, assuming a nonminimal coupling of matter
to the spacetime curvature, is for instance an excellent toy
model of other MOND-like field theories, as its exhibits their
generic difficulties: non-predictiveness, fine-tuning, and above
all instability.

It should be underlined that other experimental problems also
need to be addressed by any field theory of MOND, although we did
not discuss them in the present article. First of all, weak
lensing observations have already exhibited several evidences of
dark matter not located around clusters of baryonic matter,
notably the famous bullet cluster\footnote{On the other hand, the
cluster Abell 520 \cite{Mahdavi:2007yp} exhibits a dark core which
might be more natural to explain within MOND than as dark
matter~\cite{Fort}.} \cite{Clowe:2003tk} and the very recent dark
matter ring \cite{Jee:2007nx}. The basic idea of the MOND dynamics
seems therefore to be contradicted. However, several studies
\cite{Angus:2006qy,Angus:2006ev,Brownstein:2007sr,Angus:2007qj}
have argued that such observations do not rule out MOND-like
field theories, whose predictions can differ significantly from
the original MOND proposal in non-spherical and dynamical
situations. Another generic problem of models avoiding the dark
matter hypothesis is to predict the right relative heights of the
second and third acoustic peaks of the CMB spectrum. Indeed, in
absence of collisionless dark matter, their heights should
decrease monotonically by Silk damping, whereas WMAP data
\cite{Spergel:2003cb,Spergel:2006hy} confirm that the second and
third peaks have similar heights (consistent with the dark matter
paradigm). Reference~\cite{Skordis:2005xk} analyzed the CMB
spectrum within the TeVeS model, and found that it can fit
experimental data provided their is a present density $\Omega_\nu
\approx 0.17$ of massive neutrinos. This number is not far from
the needed dark matter density $\Omega_\text{DM} \approx 0.24$ in
the standard $\Lambda$CDM (cosmological constant plus cold dark
matter) cosmological model, and illustrates that the existence of
dark matter (as massive neutrinos) would anyway be needed in the
TeVeS model, although the MOND dynamics was initially devised to
avoid the dark matter hypothesis. However, it is already known
that some amount of dark matter is anyway needed to account for
the cluster rotation curves, within the MOND model, and massive
neutrinos would be perfect candidates since they can cluster on
such scales although they are too light to cluster on galaxy
scales. Therefore, a MOND-like field theory remains a
possibility, provided there is a large enough density of massive
neutrinos --- or any other dark matter candidate with similar
properties. In the model \cite{Sanders:2005vd}, closely related
to TeVeS, one of the two scalar degrees of freedom is assumed to
play the role of this dark matter. A third difficulty of
MOND-like field theories is also to explain the numerical
coincidence $a_0 \approx c H_0/6$ between the MOND acceleration
constant and the Hubble constant (i.e., the present expansion
rate of the Universe)~\cite{Milgrom:1983ca}. Some very promising
ideas have been developed notably in \cite{Sanders:2005vd}, but
it cannot be considered as a prediction. For instance, it may
happen that $a_0$ is actually related to the cosmological
constant instead of the Hubble expansion rate, \textit{via} a
relation like $a_0 \approx(\sqrt{\Lambda})c^2/8$. In such a case,
it would not vary with time, and one would ``only'' have to
explain why it happens to take this precise value. On the other
hand, a constant value of $a_0$ would probably be ruled out by
the observed CMB spectrum, although this obviously depends on the
precise features of the model. A fourth and related difficulty is
to explain structure formation without cold dark matter, while
remaining consistent with the tiny amplitude of CMB fluctuations.
As far as we are aware, there is at present no obvious solution to this
crucial problem, and a larger value of $a_0$ at early cosmological
times might be necessary. The modification of gravity, the possible
renormalization of Newton's constant $G$ at large distances, and the
contribution of the various fields entering the model might also
contribute to a solution. A fifth difficulty is more theoretical than
experimental. The action-reaction principle implies that if a mass
$m_1$ feels a force $(\sqrt{Gm_2a_0})m_1/r$ caused by a second mass
$m_2$, in the MOND regime, then the latter must feel the opposite
force and thereby undergo an acceleration $(\sqrt{Ga_0/m_2})m_1/r$.
Since this expression diverges as $m_2 \rightarrow 0$, any light enough
particle should thus be infinitely accelerated by distant objects
feeling its MOND potential. Actually, the self-energy of such
particles must be properly renormalized, and it has been proven
in \cite{Bekenstein:1984tv} that the RAQUAL model does not suffer
from any inconsistency due to the action-reaction principle.
Indeed, the force felt by mass $m_1$ takes the MOND form
$(\sqrt{Gm_2a_0})m_1/r$ only when $m_2 \gg m_1$. The exact
expression of the forces has been derived in
\cite{Milgrom:1997gx}, $m_1 a_1 = m_2 a_2 = (2\sqrt{Ga_0}/3r)
[(m_1+m_2)^{3/2}- m_1^{3/2} - m_2^{3/2}]$, so that the
acceleration $a_2 = \sqrt{Gm_1a_0}/r + \mathcal{O}(\sqrt{m_2})$
remains finite even when $m_2 \rightarrow 0$. Although we expect
that similar arguments would apply for any field theory deriving
from an action principle, it remains to explicitly prove so for
new models that one may consider, for instance for the class of
nonminimally coupled fields studied in Sec.~\ref{Nonmin} above.

All these difficulties, in addition to those discussed in the
present paper, underline that the construction of a consistent
field theory reproducing the MOND dynamics and the right light
deflection is far from being obvious. Although the considered
actions depend on several free functions allowing us to
\textit{fit} different kinds of data, there are so many
theoretical and experimental constraints that no model passes all
of them at present. The conclusion of our analysis seems thus to
be in favor of the dark matter paradigm. On the other hand, the
Tully-Fisher law, i.e., the existence of a universal acceleration
scale $a_0$, has not yet been derived in dark matter models. More
generally, dark matter profiles seem to be tightly correlated to
baryonic ones \cite{McGaugh:2005er} (up to apparent
counterexamples such as the bullet cluster
\cite{Clowe:2003tk,Jee:2007nx}), and this lacks any explanation
in CDM whereas this is an obvious prediction of MOND. Therefore,
the dark and baryonic matter clustering still needs to be
understood in more detail, and the analysis of modified gravity
theories remains an interesting alternative to the standard
$\Lambda$CDM paradigm.

Identifying theoretical difficulties and finding ways to solve
them often provides new possible interpretations of experimental
data. For instance, we saw in Sec.~\ref{MONDRAQUAL} that the
RAQUAL kinetic term needed to be slightly modified for
vanishingly small accelerations, in order for the field theory to
remain consistent. The resulting model predicts that gravity
becomes Newtonian again at very large distances (after a
transition by the MOND regime), but with a much larger
gravitational constant. Both the cosmological predictions and our
interpretation of cosmological data would be fully changed within
such a model. This may lead to a new understanding of both the
``local'' mass discrepancies (within galaxies and clusters) and
of the cosmological dark matter. Another bonus of our study was
also to provide an example of a consistent field theory
reproducing the Pioneer anomaly without spoiling other
predictions of general relativity. But although it passes
theoretical and experimental constraints, it does not pass
``esthetical'' ones (as defined in the Introduction), i.e., it is
actually \textit{tuned} to account for Pioneer data. It would
remain to find an underlying symmetry principle which could
\textit{predict} this phenomenology.

\acknowledgments
We wish to thank many colleagues for discussions about dark
matter and MOND, notably L.~Blanchet, T.~Damour, C.~Deffayet,
\'E.~Flanagan, B.~Fort, A.~Lue, G.~Mamon, J.~Moffat, Y.~Mellier,
M.~Milgrom, R.~Sanders, \hbox{J.-P.}~Uzan, K.~Van Acoleyen and
R.~Woodard.



\end{document}